\documentclass[12pt, onecolumn, amsmath, amssymb, a4paper]{aastex631}
\usepackage{times}
\usepackage{amsmath, bm}
\usepackage{graphicx}
\usepackage{wrapfig}
\usepackage{verbatim}
\usepackage{tikz}
\usepackage{color}

\usepackage{enumitem}

\usepackage[utf8]{inputenc}
\usepackage{newunicodechar}
\usepackage{tcolorbox}
\newunicodechar{−}{-}

\usepackage[T1]{fontenc}
\usepackage{aecompl}
\usepackage{calligra}
\DeclareMathAlphabet{\mathcalligra}{T1}{calligra}{m}{n}
\DeclareFontShape{T1}{calligra}{m}{n}{<->s*[2.2]callig15}{}

\usepackage{hyperref}
\hypersetup{
    pdfnewwindow=true,      % links in new window
    colorlinks=true,       % false: boxed links; true: colored links
    linkcolor=cyan,          % color of internal links
    citecolor=cyan,        % color of links to bibliography
    filecolor=cyan,      % color of file links
    urlcolor=cyan           % color of external links
}

\def\apjl{ApJL}               % Astrophysical Journal, Letters
\def\apjs{ApJS}               % Astrophysical Journal, Supplement
\def\aap{A\&A}                % Astronomy and Astrophysics
\newcommand{\Ms}{{\rm M}_\odot}
\newcommand{\Rs}{{\rm R}_\odot}

\newcommand{\km}{{\rm km}}
\newcommand{\yr}{{\rm yr}}
\newcommand{\Mpc}{{\rm Mpc}}
\newcommand{\Gpc}{{\rm Gpc}}
\newcommand{\mstar}{{\rm M}_{\star}}
\newcommand{\rstar}{{\rm R}_{\star}}
\newcommand\sbullet[1][.5]{\mathbin{\vcenter{\hbox{\scalebox{#1}{$\bullet$}}}}}
\newcommand{\mbh}{{\rm M}_{\sbullet[0.7]}}
\newcommand{\maxmbh}{\widehat{{\rm M}}_{\sbullet[0.7]}}
\newcommand{\rtidal}{r_{\rm t}}

\newcommand{\bcri}{\beta_{\rm cri}}
\newcommand{\physRt}{\mathcal{R}_{\rm t}}
\newcommand{\erg}{{\rm erg}}
\newcommand{\second}{{\rm s}}
\newcommand{\cm}{{\rm cm}}
\usepackage[normalem]{ulem}

\DeclareMathAlphabet{\mathcalligra}{T1}{calligra}{m}{n}
\DeclareFontShape{T1}{calligra}{m}{n}{<->s*[2.2]callig15}{}

\begin{document}
\title[]{Multi-messenger astronomy with black holes: tidal disruption events}
\author[0000-0000-0000-0000]{Thomas Wevers}
%\altaffiliation{Both authors contributed equally to this work}
\email{Email: twevers@stsci.edu}
\affiliation{Space Telescope Science Institute, 3700 San Martin Drive, Baltimore, MD 21218, USA}
\affiliation{European Southern Observatory, Alonso de Córdova 3107, Vitacura, Santiago, Chile}
\author[0000-0003-2012-5217]{Taeho Ryu}
%\altaffiliation{Both authors contributed equally to this work}
\email{Email: tryu@mpa-garching.mpg.de}
\affiliation{Max Planck Institute for Astrophysics, Karl-Schwarzschild-Str.~1, 85748 Garching, Germany}

\begin{abstract}
    This chapter provides an overview of tidal disruption events, aiming to provide an overview of both the theoretical and the observational state of the field, with the overarching goal of introducing them as tools to indirectly observe massive black holes in the Universe. We start by introducing the relevant theoretical concepts, physical scales and timescales with an emphasis on the {\it classical framework} and how this has been (and continues to be) improved since the inception of the field. We then cover the current and future prospects of observing TDEs through a variety of messengers, including photons across the electromagnetic spectrum, as well as gravitational waves and neutrino particles. More recent advancements in the field, including repeating TDEs as well as TDEs by stellar-mass black holes, are also highlighted. \\

    \hspace{1.5in}\textit{The authors contributed equally to this work}

\end{abstract}

\section{Introduction}
At the centers of massive galaxies, at least one supermassive black hole ({SMBH}) with mass $\gtrsim 10^{5}\Ms$ is thought to reside, surrounded by a nuclear star cluster or bulge \citep{2013ARA&A..51..511K}. Stars in such dense environments undergo two-body scattering with other stars, which sometimes takes them on nearly radial orbits around the {SMBH}. If the pericenter distance is so small that the stellar self-gravity can not hold the star against the tidal force of the {SMBH}, it is tidally disrupted, generating a burst of radiation luminous enough to outshine entire host galaxies. Such events are called tidal disruption events ({TDEs}).

The basic theory of {TDEs} was first developed in the '80s \citep{1982ApJ...262..120L,1988Natur.331..687H,1988Natur.333..523R} for {SMBH}s at the centers of galaxies. Since then, major progress has been made in theoretical modelling of TDEs and observing TDE candidates with searches in wide-field transient surveys. Theoretically, it has been proposed that TDEs could be astrophysical sources of neutrinos \citep{2021NatAs...5..510S} and gravitational waves \citep{2020MNRAS.498..507T}, making TDEs promising multi-messenger sources. For example, TDEs as multi-messenger sources have recently drawn a great deal of attention because they have been suggested to produce some of the recently detected high-energy neutrinos \citep[e.g.][]{2019TNSTR.615....1N,2021ApJ...920...56F}.

Although {TDEs} were originally suggested as nuclear transients by the central {SMBH}s, {BH}s of all mass scales can in principle disrupt stars and generate radiation. Since the mid 2010s, serious attention has been paid to {TDEs} by stellar-mass black holes  \citep[or micro-TDEs,][]{2016ApJ...823..113P} as a possible outcome of dynamical encounters between stars and compact objects in clusters. Such events have also been considered as a potential explanation for highly energetic flares observed in clusters, such as ultra-long gamma ray bursts \citep{2016ApJ...823..113P} and fast blue optical transients \citep{2021ApJ...911..104K}. In addition, TDEs created in encounters involving binary star systems can have astrophysical implications for the formation history of star clusters \citep{2019PhRvD.100d3010S}, BH spins \citep{2019ApJ...877...56L} and the electromagnetic (EM) counterpart of gravitational wave emission \citep{2022MNRAS.516.2204R} of binary black holes. 

In this part of the chapter, we briefly describe the conventional picture of {TDE} and theoretical improvement beyond that (\S\ref{sec:TDEdescription}). Then we review some recent developments in the theory and observation of TDEs, focusing on discussing TDEs as multi-messenger sources (\S\ref{sec:multimessenger}). After briefly discussing repeating nuclear transients and their relation to partial TDEs (\S\ref{sec:ptdes}), we close this chapter with a summary of recent studies of micro-TDEs (\S\ref{sec:microTDE}) and open questions and summary (\S\ref{sec:summary}).

\begin{figure}
	\centering
	\includegraphics[width=12.cm]{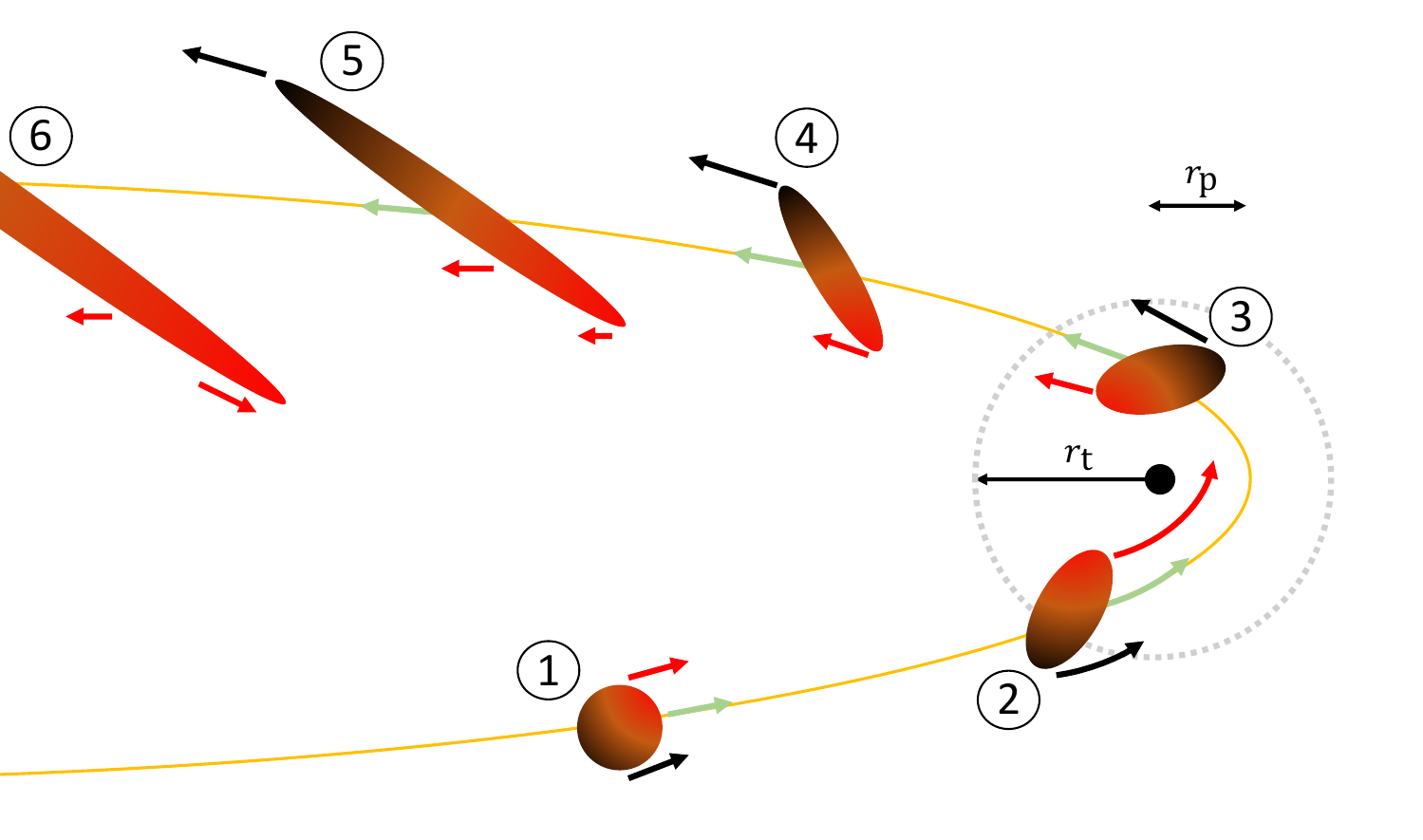}\\
        \includegraphics[width=12.cm]{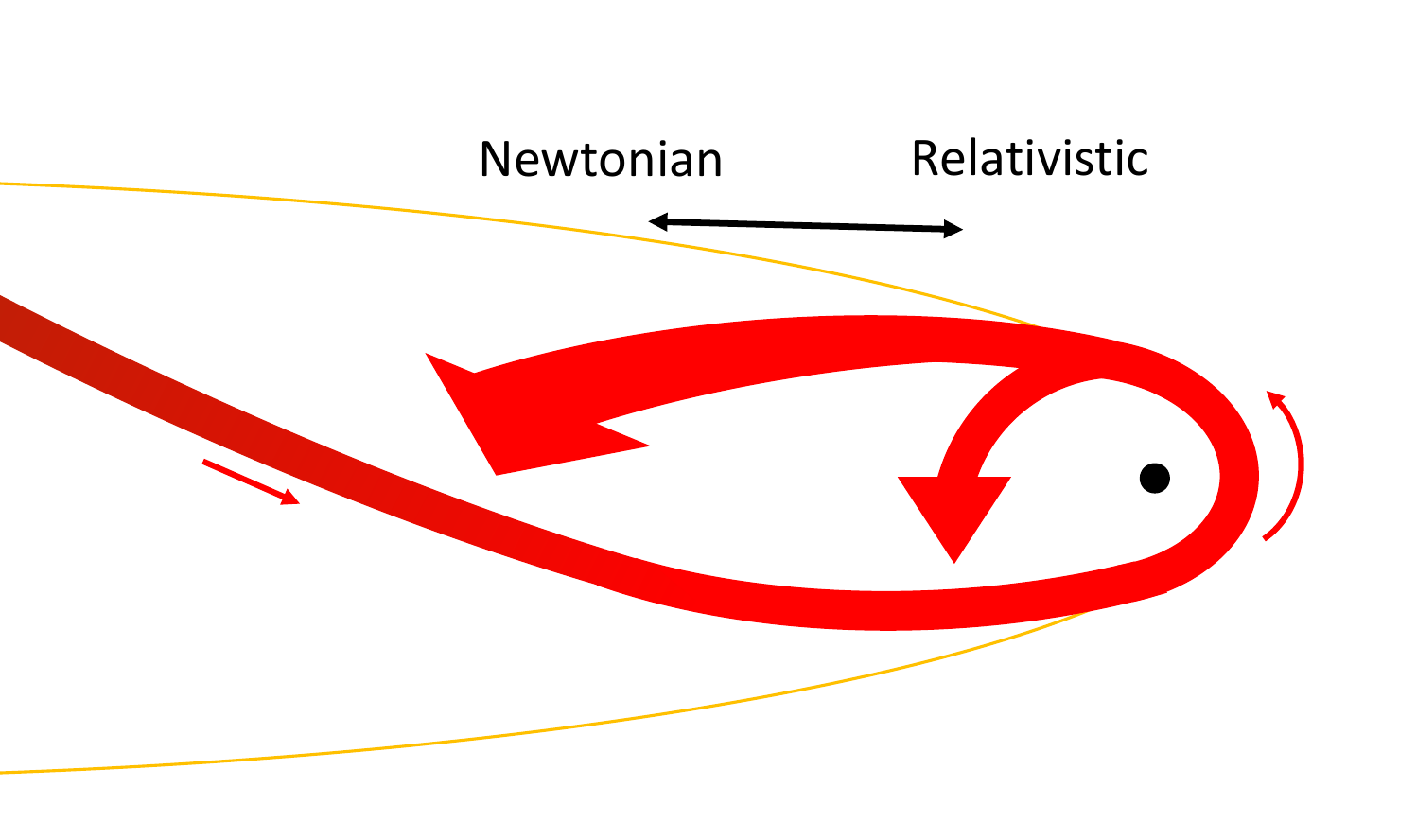}  
\caption{A schematic diagram showing a disruption event of a star. The \textit{top} panel depicts successive moments of the disruption from  the initial approach of the star (1) on a parabolic obit (counterclockwise) with pericenter distance $r_{\rm p} < r_{\rm t}$ around a black hole (black dot), to its disruption at the pericenter (2, 3), to an outward advance of stellar debris (4, 5, 6). The gas bound (unbound) to the black hole is colored in red (black). The arrows roughly indicate the velocity vector of bound (red), unbound (black) gas, and debris' center of mass (green). In the \textit{bottom} panel, the most bound debris returns (corresponding to the tip of the bound part of 6th debris in the \textit{left} panel), turns around the black hole, and collides with a newly returning stream. The collision occurs at shorter distances as relativistic apsidal precession becomes stronger.   }
	\label{fig:schematic2}
\end{figure}

\section{Description of tidal disruption event}\label{sec:TDEdescription}
  
  \subsection{Conventional picture}\label{subsec:conven_picture}

A star can be tidally disrupted if the BH's tidal forces exceed the self-gravity of the star. 
The distance below which a star of mass $\mstar$ and radius $\rstar$ is tidally disrupted by a {SMBH} of mass $\mbh$ is the so-called tidal radius \citet{1988Natur.331..687H}, estimated as
  \begin{align}\label{eq:rt}
      \rtidal&= \left(\frac{\mbh}{\mstar}\right)^{1/3}\rstar,\\\label{eq:rt2}
      &\simeq 47~r_{\rm g} \left(\frac{\mbh}{10^{6}\Ms}\right)^{-2/3}\left(\frac{\mstar}{1\Ms}\right)^{-1/3}\left(\frac{\rstar}{1\Rs}\right),
  \end{align}
  where $r_{\rm g}= G\mbh/c^{2}$ is the gravitational radius. Equation~\ref{eq:rt} can be derived by equating the order-of-magnitude estimates for the Newtonian self-gravity ($\simeq G\mstar/\rstar^{2}$) and the tidal force ($\simeq G\mbh \rstar/r^{3}$).
  
  We can consider two asymptotic values of $\rtidal$. First, as $\mbh$ decreases, $\rtidal$ asymptotes to $\rstar$ (Equation~\ref{eq:rt}). In the meantime, TDEs by less massive BHs occur at distances further from the horizon ($\rtidal/r_{\rm g}\sim O(10^{4}-10^{5})$ for stellar-mass black holes, Equation~\ref{eq:rt2}), meaning that relativistic effects become less important. On the other hand, as $\mbh$ increases, $\rtidal/r_{\rm g}$ falls, meaning that relativistic effects become more important for larger $\mbh$. The second limit constrains the maximum {BH} mass $\maxmbh$ capable of disrupting stars outside of the event horizon.  $\maxmbh$ for which $\rtidal\leq\alpha r_{\rm g}$,
  \begin{align}\label{eq:mmax}
      \maxmbh\gtrsim 10^{8}\Ms\left(\frac{2}{\alpha}\right)^{3/2}\left(\frac{\mstar}{1\Ms}\right)^{-1/2} \left(\frac{\rstar}{1\Rs}\right)^{3/2}.
  \end{align}
 In fact, although $\alpha=2$ (or the Schwarzschild radius) has been used frequently in literature, a star on Schwarzschild geodesic falling from infinity with orbital angular momentum $\leq4r_{\rm g}c$ (corresponding to $\alpha=4$)\footnote{The relativistic angular momentum $L$ in Schwarzschild space time is related to $r_{\rm p}$ by,
 \begin{align}
 L^{2}=(r_{\rm g}c)^{2}\frac{2(r_{\rm p}/r_{\rm g})^{2}}{r_{\rm p}/r_{\rm g}-2}.
 \end{align}} would directly fall into the SMBH without generating any flares.

As depicted in Figure~\ref{fig:schematic2}, once the star is disrupted, the gas continues to move on ballistic orbits with the orbital energy and angular momentum determined during the pericenter passage. This approximation, called ``the frozen-in approximation'', is  made based on the argument that the star remains essentially intact until it enters the tidal sphere and the energy is determined by the relative depth of a mass element in the {BH}'s potential well.  It then follows that the mass distribution of the specific orbital energy ($\epsilon$) of the debris is flat, that is, $dM/d\epsilon \simeq M_{\star}/2\Delta\epsilon$, within a characteristic energy width $\Delta \epsilon$ around $\epsilon \sim 0 $ \citep{1982ApJ...262..120L,1988Natur.333..523R,2013MNRAS.435.1809S}. $\Delta \epsilon$ is defined as,
  \begin{align}\label{eq:epsilon}
      \Delta\epsilon = \frac{G\mbh\rstar}{\rtidal^{2}}.
  \end{align}
Roughly speaking, the mass within the radius from the SMBH to the stellar center of mass is bound to the SMBH ($-\Delta\epsilon\lesssim\epsilon<0$) whereas the mass on the farther side is unbound ($0<\epsilon\lesssim\Delta\epsilon$).
The unbound debris escapes from the SMBH to large distances while the bound debris reaches its apocenter and returns to the SMBH. The most bound debris with $\epsilon\simeq -\Delta\epsilon$ has a semimajor axis
  \begin{align}\label{eq:amin}
      a &= \frac{G\mbh}{2\Delta\epsilon}\simeq 50 \rtidal\left(\frac{\mbh}{10^{6}\Ms}\right)^{1/3}\left(\frac{\mstar}{1\Ms}\right)^{-1/3}.
  \end{align} and a high eccentricity,
  \begin{align}\label{eq:ecc}
      e \simeq 1 - 0.02 \left(\frac{\mbh}{10^{6}\Ms}\right)^{-1/3}\left(\frac{\mstar}{1\Ms}\right)^{1/3},
  \end{align}
  from which it follows that the apocenter distance is roughly $(1+e)a\simeq 2a_{\rm \Delta\epsilon}$. The bound debris falls back to the {SMBH} at a rate $\dot{M}_{\rm fb}$,
\begin{align}\label{eq:mpeak}
\dot{M}_{\rm fb}=\frac{dM}{d\epsilon}\left|\frac{d\epsilon}{dt}\right|=\frac{(2\pi G \mbh)^{2/3}}{3}\frac{dM}{d\epsilon}t^{-5/3}.
\end{align}
The fallback rate peaks when the debris with $\epsilon=-\Delta\epsilon$ returns at $t=t_{\rm peak}$,
\begin{align}
 t_{\rm peak}& =\frac{\pi}{\sqrt{2}} \frac{G \mbh} {{\Delta \epsilon}^{3/2}}, \nonumber \\
 & \simeq 0.11\yr ~ \left(\frac{\mstar}{1\Ms}\right)^{-1}\left(\frac{R_{\star}}{1\Rs}\right)^{3/2}\left(\frac{\mbh}{10^{6}\Ms}\right)^{1/2},
\label{eq:peak_t}
\end{align}

and the mass return rate at peak $\dot{M}_{\rm peak}(t = t_{\rm peak})$ is,
\begin{align}
\label{eq:peak_mdot}
\dot{M}_{\rm peak}&\simeq \frac{M_{\star}}{3t_{\rm peak}},\nonumber\\
&\simeq1.49\Ms \yr^{-1}\left(\frac{\mstar}{1\Ms}\right)^{2}\left(\frac{R_{\star}}{1\Rs}\right)^{-3/2}\left(\frac{\mbh}{10^{6}\Ms}\right)^{-1/2},\\\label{eq:mdot_medd}
&\simeq 134\dot{M}_{\rm Edd}\left(\frac{\eta}{0.1}\right) \left(\frac{\mstar}{1\Ms}\right)^{2}\left(\frac{R_{\star}}{1\Rs}\right)^{-3/2}\left(\frac{\mbh}{10^{6}\Ms}\right)^{-3/2},
\end{align}
where $\dot{M}_{\rm Edd}=L_{\rm Edd}/\eta c^{2}$ is the accretion limit derived from the Eddington luminosity $L_{\rm Edd}$ with radiative efficiency $\eta$. As Equation~\ref{eq:mdot_medd} shows, the mass return rate can be highly super-Eddington, which implies that a bright flare can be generated in the process.

\subsection{Improvements beyond the conventional picture}\label{sec:development}

There have been significant improvements in modeling TDEs using advanced numerical methods \citep{2020SSRv..216...88K}. This section focuses on some major developments in estimating the two key quantities, tidal radius and energy width, that constitute the conventional picture.

\subsubsection{Determination of Tidal Radius}\label{subsec:tidalradius}
The tidal radius is the key quantity that readers might encounter first in the TDE literature. This is one of the most fundamental variables which defines the maximum distance at which a star is fully disrupted. The other key quantities (e.g., Equations~\ref{eq:mmax}, \ref{eq:epsilon}, \ref{eq:peak_t}, \ref{eq:peak_mdot}) are affected directly or indirectly by the tidal radius. 

It is necessary to appreciate what $\rtidal$ physically means. It is derived based on the order-of-magnitude argument for the balance between the Newtonian tidal gravity and self-gravity at the stellar surface. As a result, $\rtidal$ becomes a function of the average density $\bar{\rho}_{\star}$ ($\equiv \Ms/[4\pi \Rs^{3}/3]$) of the star, $\rtidal\simeq(\mbh/\bar{\rho}_{\star})^{1/3}$. Comparing the two forces at the stellar surface makes $\rtidal$ an approximate estimate for the characteristic pericenter distance below which the star loses some mass during the pericenter passage and survives (partial disruption events). 

While $\rtidal$ provides the right order-of-magnitude level estimate, it has to be corrected for more accurate modeling of TDEs. These corrections pertain to at least two determining factors, namely internal stellar structure and relativistic effects. We denote the maximum pericenter distance for full disruptions corrected by those factors by $\physRt$. From now on, we distinguish $\physRt$ from the nominal tidal radius $\rtidal$. Full disruptions occur only when the tidal force is greater than the self-gravity at the densest region of the star, i.e., its core. Then it naturally follows that the maximum pericenter distance for full disruptions should be more related to the core structure, especially the central density $\rho_{\rm c}$ \citep{2002ApJ...576..753L,2020ApJ...905..141L,2020ApJ...904...99R},
\begin{align}
    \physRt\simeq\left(\frac{\mbh}{\rho_{\rm c}}\right)^{1/3}.
\end{align}
In fact, \citet{2020ApJ...904...99R} analytically showed that $\physRt\simeq(\mbh/\rho_{\rm c})^{1/3}$ and, hence, $\physRt/\rtidal\simeq(\bar{\rho}/\rho_{\rm c})^{1/3}$, both of which were confirmed using their relativistic hydrodynamics simulations. Furthermore, because relativistic effects become stronger as $\mbh$ increases, relativistic corrections to $\rtidal$ become important.

There have been enormous efforts to more accurately measure $\physRt$ since the 1980's\footnote{The so-called Roche limit has the same functional form with $\rtidal$. But the classical problem of Roche considers the case where a homogeneous satellite moves in a stationary tidal field, i.e., on a circular Keplerian orbit about a central rigid spherical mass, which is not the case for TDEs.}. However, most of the early efforts were devoted to understanding the impact of the internal stellar structure on $\physRt$. Here, we define the penetration factor $\beta = r_{\rm t}/r_{\rm p}$, where $r_{\rm p}$ is the pericenter distance, and $\bcri=\rtidal/\physRt$ at $r_{\rm p}=\physRt$. Physically, larger $\bcri$ means full disruptions occur at smaller pericenter distances for a given black hole and star. If $\bcri>1$, a star is fully disrupted at $r_{\rm p}$, smaller than what is conventionally predicted. In other words, the star survives after passing though pericenter at the conventional tidal radius $\rtidal$. 

Many early works studied TDEs in the analytic framework of the so-called affine model \citep{1976ApJ...210..549L,1985MNRAS.212...23C}. In this model, a star evolves assuming that the density contours are homologous ellipsoids. This analytic model has proved extremely useful for capturing the basic aspects of TDEs. Using this model, \citet{1986ApJS...61..219L} analytically found that $\bcri\sim 0.66-0.67$ for incompressible stars (i.e., with a fixed volume) and $\bcri\simeq0.52$ for compressible stars modeled as a polytrope with polytropic index $\gamma=5/3$. By including non-linear effects associated with the excitation of Kelvin modes in the affine model, \citet{1992MNRAS.258..715K} found that $\bcri\simeq 0.6$ for incompressible stars and \citet{1995MNRAS.275..498D} found  $\bcri\simeq 0.67$ for compressible polytropic stars with $\gamma=4/3$ and $\simeq 1.1$ for $\gamma=5/3$. \citet{1989IAUS..136..543P} introduced a correction factor to $\rtidal$ that depends on the internal structure of stars, expressed as the ratio of the constant of apsidal motion to the nondimensional binding energy. This gives $\bcri$ ranging from 1.92 for fully radiative stars to 1.22 for fully convective stars. Although \citet{1989IAUS..136..543P}'s correction factor remains an order-of-magnitude estimate, it indeed predicts the trend that $\bcri$ is larger for more massive (i.e. centrally concentrated) stars. 

The critical penetration factor has been measured more accurately using numerical simulations since the 2010s. Most of those hydrodynamics simulations were Newtonian and approximated main-sequence (MS) stars by polytropic models. \citet{2013ApJ...767...25G} investigated the dependence of the mass fallback rate of the debris on the impact parameter using hydrodynamics simulations in which they considered a $1\Ms$ polytropic star with $\gamma=4/3$ and $5/3$. They found that no self-bound stellar remnant is produced at $\beta > \bcri =1.85 $ for $\gamma=4/3$ and $\bcri=0.9$ for $\gamma=5/3$. \citet{2017A&A...600A.124M} specifically focused on measuring $\bcri$ using three different numerical techniques (mesh-free finite mass, smoothed particle, and grid-based hydrodynamics simulations). Also assuming polytropes with $\gamma=4/3$
 and $5/3$, they measured $\bcri\sim 2.01$ for $\gamma=4/3$ and $\simeq 0.92$ for $\gamma=5/4$, which are similar with those found in \citet{2013ApJ...767...25G}.
    
More recently, numerical studies used a stellar evolution code (e.g., {\sc MESA}, \cite{2011ApJS..192....3P}) to create more realistic initial stellar models. Performing hydrodynamics simulations for TDEs of a $1\Ms$ zero-age MS star with a moving-mesh hydrodynamics code, \citet{2019MNRAS.487..981G} measured $\bcri\simeq 2$, which is close to that for the polytropic model with $\gamma=4/3$ in \citet{2013ApJ...767...25G,2017A&A...600A.124M}.
Later, some studies began to consider realistic MS models with a wide range of $\mstar$ \citep{2019ApJ...882L..25L,2020ApJ...905..141L,2020ApJ...904...98R,2020ApJ...904...99R,2020ApJ...904..101R}. This allows for studying the $\mstar$-dependence of $\bcri$, which can not be achieved with polytropic models. Those studies commonly found that $\bcri$ is larger for more concentrated stars (or larger $\rho_{\rm c}/\bar{\rho})$, although some scatter remains for the values of $\bcri$ in different studies: in general, $\bcri\lesssim 1$ for $\Ms\lesssim 0.3\Ms$, $\bcri\gtrsim 2$ for $\Ms\gtrsim 1\Ms$, and there is a relatively sharp transition in between associated with the stellar internal structure.

Several relativistic simulations were performed to investigate the dependence of TDE outcomes on the penetration factor and black hole mass, with a variety of approximations of relativity and self-gravity calculations and stellar models (e.g., \citet{2006A&A...448..843I,2012PhRvD..85b4037K,2019MNRAS.487.4790G} using polytropic models and \citet{2020ApJ...904...98R,2020ApJ...904...99R,2020ApJ...904..101R} using realistic stellar models). A quantity that comes naturally from such studies is the critical penetration factor as a function of the $\mbh$. A general trend found in the studies is that $\bcri$ decreases as $\mbh$ increases. The reason for the negative correlation between  $\bcri$ and  $\mbh$ is due to the fact that relativistic tidal stresses become increasingly more destructive than the Newtonian tidal forces for higher $\mbh$. Most recently, using fully relativistic simulations for TDEs of realistic MS stars taken from {\small MESA},  \citet{2020ApJ...904...98R,2020ApJ...904...99R,2020ApJ...904..101R} performed a systematic study for the $\mstar$ and $\mbh$-dependence of TDE outcomes by considering a wide range of $\Ms$, $\mbh$ and $\beta$. They found that the $\mstar$ and $\mbh$-dependence of $\bcri$ are separable and $\bcri$ can be reduced by up to a factor of 3 as $\mbh$ increases between $10^{5}\Ms$ and $5\times10^{7}\Ms$ for a fixed $\mstar$.

\subsubsection{Determination of the energy width and fallback rate}

The energy distribution of debris is another key quantity which determines the debris' subsequent orbit. Detailed numerical simulations showed that the overall shape of the energy distribution is not very different from a top-hat function. However, the details depend on several factors including stellar internal structure (so stellar mass, age, and spin) and relativistic effects, which sometimes lead to a change in the peak mass return rate and time by an order of magnitude. 

The dependence on the internal structure of the star was first studied by comparing the debris' energy spread from the disruption of polytropic stars with $\gamma$. The first hydrodynamics simulations with reasonable resolution were performed by \citet{1989ApJ...346L..13E} who found that $dM/dE$ for a full disruption of a $1\Ms$ polytropic star with $\gamma=5/3$ just inside the critical tidal radius is nearly flat within $\pm 0.9\Delta\epsilon$, which is in very good agreement with the conventional prediction.  \citet{2009MNRAS.392..332L} revisited the arguments for a flat energy distribution and the corresponding time evolution of mass fall back rate using smoothed particle hydrodynamics simulations with $1\Ms$ polytropic stars with different values of $\gamma$. They showed that the energy distribution deviates more significantly from a flat distribution for more centrally concentrated stars (i.e., smaller $\gamma$). More specifically, the distribution becomes wide beyond $\Delta \epsilon$ as $\gamma$ decreases, indicating some fraction of debris is more bound or unbound than what is expected from a flat distribution. As a result, the mass fallback curve reveals a smoother peak at an earlier time for more centrally concentrated stars. \citet{2013ApJ...767...25G} confirmed this trend from simulations for disruptions of $1\Ms$ polytropes with two different values of $\gamma$, $4/3$ and $5/3$. In addition, they showed that the energy width for full disruptions occurring close to the tidal radius does not strongly depend on the pericenter distance. 

Beyond the polytropic approximation, numerical investigation with stellar internal structure taken from a stellar evolution code has led to a more realistic picture of TDEs. \citet{2019MNRAS.487..981G} simulated disruptions of $1\Ms$ zero-age main-sequence stars, generated with the stellar evolution code {\small MESA} \citet{2011ApJS..192....3P}, at different pericenter distances. They found that the energy distribution has wings that extend beyond $\Delta\epsilon$ and decline less steeply than a top-hat function. This leads to a higher peak mass return rate of a full disruption than the conventional prediction by a factor of $\gtrsim 10$ and smaller peak mass return time also by a factor of $\gtrsim 10$, which is not strongly dependent on the penetration factor. \citet{2019ApJ...882L..26G} investigated the mass fallback rate from TDEs of $0.3$, $1$, $3\Ms$ main-sequence stars at different ages initially on parabolic orbits with $\beta=3$. They found that the peak mass fallback rate for a full disruption can be larger by up to a factor of $\sim10$ than that for polytropic stars with $4/3\leq\gamma\leq5/3$ and the difference is greater for more massive stars. Exploring a wider range of $\Ms$ ($0.1 - 10\Ms$) and ages (zero-age to terminal-age), \citet{2020ApJ...905..141L} confirmed that the energy distribution becomes wider for more centrally concentrated stars (i.e., more massive main sequence stars closer to their terminal age) for a given mass. 

The energy distribution and mass fallback rate are also affected by the spin magnitude and rotation axis of the star before disruption. \citet{2019ApJ...872..163G} simulated TDEs of initially rotating $1\Ms$ polytropic stars with $\gamma=5/3$ with the angular frequency up to 20\% of the breakup speed for prograde and retrograde encounters. They found a trend that the spin in the prograde (retrograde) direction effectively increases (reduces) the energy spread. This is because by the time the star is disrupted each mass element has effectively more (less) kinetic energy when the spin vector is (anti-)aligned with the orbital axis. This results in the mass return rate rising at an earlier (later) time and peaking at a higher (lower) value for the prograde (retrograde) case than the non-spinning case. However, the impact of stellar spin is rather small for the parameter space considered: a factor of $\sim 2$ change in the fallback rate relative to the non-spinning case. 

There have been efforts to study the impact of relativity on the shape of the energy distribution and fallback rate. Using relativistic hydrodynamics simulations with self-gravity calculated in a frame defined with Fermi normal coordinates, \citet{2014PhRvD..90f4020C} studied the disruption events of a $1\Ms$ polytropic star with $\gamma=5/3$ by BHs of varying masses between $10^{5} - 10^{7}\Ms$, which corresponds to pericenter distances of $220 - 10 r_{\rm g}$ for $\beta = 1$. They showed that relativistic effects shrink the energy spread when measured in units of $\Delta \epsilon$, which results in a shift of the peak mass return time to a larger value relative to the Newtonian value. But the difference was not significant, only by $\simeq 10\%$ even for the most relativistic case considered (i.e., $\mbh=10^{7}\Ms$). Later,  \citet{2020ApJ...904..101R} showed that the characteristic width of the energy distribution, relative to $\Delta \epsilon$, is smaller for TDEs by more massive {BH}s, confirming the results by \citet{2014PhRvD..90f4020C}. But \citet{2020ApJ...904..101R} found stronger relativistic effects on the shape of the distribution and, therefore, the mass fallback rate than \citet{2014PhRvD..90f4020C}: a decrease by a factor of $\lesssim 3$ in the peak mass fallback time from $\mbh=10^{5}\Ms$ to $5\times 10^{7}\Ms$.

The spread in angular momentum is negligible \citep{2014PhRvD..90f4020C,2020ApJ...904...99R}. This means that the pericenter distance of the bound debris is close to that of the original star.

\section{TDEs as multi-messenger sources}\label{sec:multimessenger}
The soon to be destroyed star goes through several phases after being perturbed onto an orbit intersecting the tidal radius of its host SMBH. These phases span a wide range of timescales and physical conditions\footnote{For example, the ratio of the stellar radius, one of the smallest characteristic length scales, to the apocenter distance of the most bound debris, one of the longest characteristic length scales, is $R_{\star}/a\simeq 10^{-4} \left(\frac{\mbh}{10^{6}}\right)^{-2/3}\left(\frac{\mstar}{1\Ms}\right)^{2/3}$.}, making it challenging to perform numerical simulations of the entire process self-consistently. We start this section by providing an overview of the different phases of a TDE and the expected and/or observed emission. This includes the now ubiquitously observed (although not necessarily understood) EM components, as well as neutrino (for which evidence is now emerging), cosmic ray (where no robust associations have been made) and gravitational wave emission (which, as we will describe, may require space-based detectors in the more distant future than the first generation space observatory, LISA).\\ 

When the star approaches the tidal radius of the SMBH, it is often assumed that it exists in a slightly perturbed hydrostatic equilibrium, based on the strong distance-dependence of tidal force, $\propto r^{-3}$. Passing the tidal radius, tidal forces become dominant over the stellar self gravity and the process of spaghettification (i.e. stretching of the star in the orbital plane of motion) begins in earnest. Upon reaching pericentre, strong compression in the direction orthogonal to the direction of motion occurs until the build up of internal pressure leads to a rebound in the form of shock waves. This compression leads to heating of the gas up to temperature of several keV, i.e. $\sim 10^{7}$K \citep{2004ApJ...615..855K}. The large optical depth of the star will suppress the emission of radiation, although at the surface of the gas some emission can leak out, leading to a double-peaked burst of X-ray emission; first the prompt emission from the surface, followed after a small delay by emission that diffuses out from the outer layers of the material \citet{2009ApJ...705..844G}. This X-ray shock breakout is expected to last for only $\sim$100 seconds, and has yet to be observed. These initial phases also represent the best opportunities to observe gravitational wave (GW) emission (described in more detail in Section \ref{sec:gw}). 

Following pericentre passage, relativistic apsidal precession will eventually lead to the outgoing and returning debris streams to intersect, resulting in shocks (e.g.  \citet{1989ApJ...346L..13E, 1990ApJ...351...38C, 1999ApJ...519..647K, 2013MNRAS.434..909H, 2015ApJ...804...85S}). Hydrodynamical simulations and analytic treatment have shown that the assembly of this material into an accretion flow can be impeded by $\sim$10s of orbital periods (i.e. the debris can orbit the black hole multiple times before it loses enough angular momentum to form a compact accretion disk) by several factors \citep{2015ApJ...804...85S}, including inefficient shocks \citep{2015ApJ...804...85S} and Lense-thirring (i.e. out of plane) precession \citep{2015ApJ...809..166G}, but ultimately it is expected that stream-stream collisions will remove orbital energy from the debris (\textit{top} panel of Figure~\ref{fig:schematic2}) and lead to the creation of an accretion disk. Motivated by the hydrodynamics simulation for a TDE by an intermediate-mass {BH} by \citet{2015ApJ...804...85S}, \citet{2015ApJ...806..164P} were the first to quantify the expected emission originating from stream self-intersections, finding that a significant amount of UV/optical emission can be released in the process. The result of stream self-intersection is the formation of a compact accretion flow, potentially retaining a significant amount of eccentricity \citep{2022A&A...666A...6W}. If the flow remains highly eccentric, this can significantly alter the emission (for example, the radiative efficiency; \citep{2015ApJ...806..164P, 2017MNRAS.467.1426S, 2020ApJ...904...73R}).

The accretion flow will emit primarily at X-ray wavelengths, although depending on the SMBH mass and the disk geometry, this emission can extend into the UV/optical regime. The expected mass fall-back rate (and hence mass accretion rate, assuming that the viscous timescale in the disk is short compared to the fallback timescale) are near- or super-Eddington, potentially giving rise to a disk wind / outflow of material driven by radiation pressure. If the fraction of the solid angle covered by this outflow is significant, it can intercept and downgrade some of the X-ray radiation towards UV/optical wavelengths \citep{2014ApJ...783...23G,2016MNRAS.455..859S, 2018ApJ...859L..20D}.

The formation of an accretion disk may also be accompanied by the launching of a (potentially relativistic) jet, which may give rise to hard X-ray (when beamed along our line of sight; e.g. \citet{2011Sci...333..203B, 2011Natur.476..425Z, 2015MNRAS.452.4297B}) and/or radio emission (for a review, see \citet{2020SSRv..216...81A}). The base of such a jet is also considered to be a prime location for the production of neutrinos \citep{2020ApJ...902..108M, 2021NatAs...5..510S}, and may also act as an efficient particle accelerator for high energy cosmic rays \citep{2009ApJ...693..329F, 2014arXiv1411.0704F}. 

Finally, if the SMBH is surrounded by circumnuclear gas/dust (on scales of $\sim$0.1--1 pc), the sudden release of a burst of highly energetic photons may give rise to an echo at IR wavelengths due to the down-processing of UV/optical light into IR photons by circum-nuclear dust \citep{2016MNRAS.458..575L}.

\subsection{Electromagnetic emission}
\begin{figure}
    \centering
    \includegraphics[width=\textwidth]{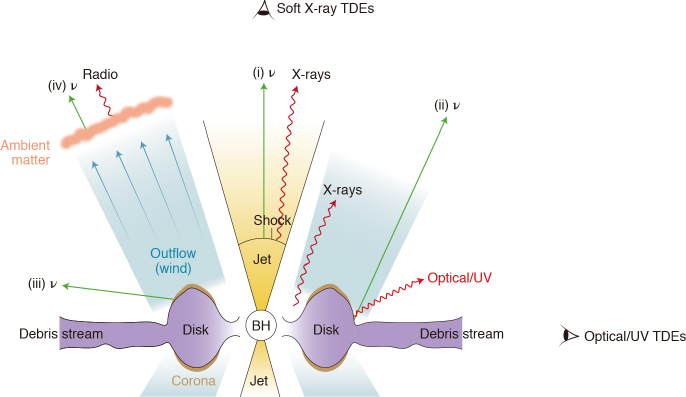}
    \caption{The various emission components envisaged to occur following a TDE, and their physical location. Adapted from \citet{2021NatAs...5..436H}.}
    \label{fig:emission}
\end{figure}

Tidal disruption events are a class of truly multi-wavelength events, with emission ranging from radio waves up to very hard X-rays. In recent years, many excellent reviews (e.g., \citet{2021SSRv..217...16B}) that explore the EM emission of TDEs in great detail have been written. For this reason, we limit ourselves here to a concise overview of the observed properties of TDEs in the various wavelength ranges, and encourage the interested reader to delve into the details of the various reviews that are referenced throughout the text. 
The physical picture that is emerging is shown in Figure \ref{fig:emission}, and indicates the various physical locations of each emission component.

\subsubsection{X-ray emission}
Following the pioneering theoretical work in the late seventies and eighties, the first observational evidence for TDEs appeared in the mid nineties and was facilitated primarily by the ROSAT soft X-ray all sky survey \citep{1984PhST....7..209T}. Several TDE candidates were identified by selecting for high amplitude, high luminosity (10$^{43-44}$ erg s$^{-1}$) transient X-ray flares in galaxies that were spectroscopically classified as inactive \citep{1996A&A...309L..35B, 1999A&A...350L..31G, 1999A&A...349L..45K, 2000A&A...362L..25G}. These flares conformed to the theoretical expectation of very soft X-ray spectra near peak luminosity (blackbody temperatures $\sim$50--100 eV) and a lightcurve that declined in time as a power law $\propto t^{-5/3}$. Later on, other X-ray observatories including Chandra \citep{2000SPIE.4012....2W}, Swift \citep{2004ApJ...611.1005G} and XMM-Newton (chiefly through its slew survey, \citep{2008A&A...480..611S}) also started contributing to the sample of X-ray TDEs. More recently, the eROSITA instrument aboard the Rontgen Gamma Telescope provided a step change in our capability to detect TDEs at X-ray wavelengths \citep{2021MNRAS.508.3820S}, although valuable follow-up observations are still provided by the workhorse missions of X-ray astronomy (Swift, Chandra, XMM-Newton and NICER, \citealt{2012SPIE.8443E..13G}). 
In addition to these X-ray selected TDE candidates, follow-up observations of UV/optically discovered TDEs are providing another way to identify X-ray bright TDEs, although the exact link between the X-ray and UV/optical brightness of TDEs remains an active topic of study.

The X-ray properties of TDEs are broadly consistent with theoretical expectations of compact accretion disk emission: very soft X-ray spectra around peak light and blackbody temperatures of kT $\sim$ 10s -- 100s eV consistent with black holes with mass $\lesssim 10^8$ M$_{\odot}$ (see \citet{2021SSRv..217...18S} for an excellent review on the history and properties of X-ray TDEs discovered to date).
A comprehensive analysis of X-ray TDE lightcurves shows that the power law indices are somewhat shallower than t$^{-5/3}$, with the majority of TDEs behaving as L$_X \sim t^{-1 - -0.5}$ \citep{2017ApJ...838..149A}.%; this is consistent with a number of predictions for accretion disk evolution \citet{}.

Some early studies found that the X-ray emission can significantly harden over time and evolve to become dominated by non-thermal, rather than thermal emission \citep{2004ApJ...603L..17K, 2014ApJ...792L..29M}. This is often seen in accreting stellar mass black holes when the accretion rate decreases below a transition value (thought to be around 1\% of Eddington, e.g. \citet{2006ARA&A..44...49R}). As a population, TDEs seem to exhibit a similar evolution \citep{2020MNRAS.497L...1W}; such changes imply a different dominant production mechanism for the X-rays, and are called accretion state transitions. It is thought that the accretion disk transitions from a radiatively efficient, optically thick, geometrically thin configuration into a radiatively inefficient, optically thin state. Such changes have been confirmed in TDEs by several other studies \citep{2020ApJ...889..166J, 2021ApJ...912..151W, 2021MNRAS.508.3820S}, highlighting the importance of continued X-ray monitoring over long timescales. Such observations are now becoming standard practice, and are enabling studies of the post-peak X-ray evolution in great detail, providing further evidence of the accretion state transitions \citep{2021ApJ...912..151W} but also uncovering diverse and unexpected behaviour such as the emergence of X-ray emission long after the UV/optical peak \citep{2017ApJ...851L..47G, 2020A&A...639A.100K, 2020ApJ...889..166J}, intermittent flaring on short and long timescales \citep{2012A&A...541A.106S, 2021ApJ...912..151W, 2021ApJ...908....4V}, and the rebrightening of emission following extended X-ray dark periods \citep{Wevers:23,Liu:23}.

A small sub-sample of X-ray TDEs exhibit very high X-ray luminosities ($10^{45-48}$ erg s$^{-1}$, \citet{2011Sci...333..203B, 2011Sci...333..199L, 2012ApJ...753...77C, 2011Natur.476..421B, 2022NatAs.tmp..252P}), and recalling that a solar mass star will be swallowed whole if $\mbh > 10^8~\Ms$  this implies emission in excess of the Eddington limit regardless of the exact black hole mass. Rather than being thermal in nature, these X-ray photons are thought to be produced by internal dissipation in highly collimated relativistic jets with Lorentz factors $\Gamma \sim 1--10$ and opening angles $\theta \lesssim 1/\Gamma$ = 0.1  (e.g. \citet{2011Sci...333..203B,2012ApJ...753...77C}). This can explain i) the X-ray spectral shape, which is indicative of non-thermal emission produced by (for example) inverse Compton
upscattering of the accretion disk photons, internal dissipation within the jet itself, and/or photons
from structures external to the accretion disk, and ii) the highly super-Eddington luminosity, which can be explained by a relativistically boosted jet pointed (nearly) along our line of sight. Unlike gamma-ray bursts (lasting up to at most several 100 seconds), these jets can persist for 100s of days, and therefore provide the exciting opportunity to study jet physics in great detail.

By exploiting the growing sample of X-ray TDEs, it is becoming possible to constrain the X-ray luminosity function (LF), and by integration over the LF also the TDE rate. Analysis of the first sizeable homogeneously selected sample of X-ray TDEs discovered by eROSITA shows that the X-ray LF scales $\propto$L$^{-0.6}$ \citep{2021MNRAS.508.3820S}, although it may become somewhat steeper when excluding the nearest and faintest event ($\propto$L$^{-0.8}$). This functional form is similar to that observed for UV/optical TDEs, but significantly shallower ($\sim$L$^{-2.3}$; e.g. \citet{2018ApJ...852...72V}).  This LF implies an X-ray TDE rate of 1.1$\pm$0.5 $\times 10^{-5}$ per galaxy, which amounts to $\sim$10 per cent of the UV/optical TDE rate estimates. This could indicate that geometrical factors (similar to the AGN unification scheme, where the orientation with respect to the observer line of sight largely determines the X-ray properties) play an important role in the detection of TDE X-ray emission, which in turn may help to understand the complex behaviour of TDEs at other wavelengths (as described in the next sections). 

\subsubsection{UV/optical emission}
UV and optical imaging surveys were not exploited to search for TDEs until $\sim$ 10 years after the first TDE candidates were reported at X-ray wavelengths. The first TDE candidates at UV/optical wavelengths were discovered in the GALEX medium deep survey (in the UV; \citet{2006ApJ...653L..25G}). This and similar archival searches (see also \citet{2011ApJ...741...73V} for results from SDSS stripe 82 multi-epoch optical imaging) focused primarily on deselecting flares consistent with AGN variability, resulting in several transient (but non-repeating) nuclear events exhibiting a high ($\sim$50\,000 K) and near-constant blackbody temperature for at least several months. Neither supernovae nor AGN flares (the most common contaminants in TDE searches) display such behaviour, making it a telltale signature for TDE identification\footnote{It should be noted that although empirically a blackbody fit provides a reasonable description of the UV/optical SED, it remains unclear whether this interpretation is physical. For example, the common assumption of spherical symmetry of the emitting region is unlikely to be correct in some cases.}.

These pioneering early searches laid the groundwork for the near real-time photometric identification and spectroscopic follow-up observations of TDE candidates, first using the Pan-STARRS survey \citep{2002SPIE.4836..154K} and gradually expanding to include nearly all major time domain surveys (e.g. ZTF \citep{2019PASP..131a8002B}, ASAS-SN \citep{2014ApJ...788...48S}, OGLE \citep{2004AcA....54..129S}). Astrometric constraints on the nuclear nature of new transients can help to optimise spectroscopic follow-up resources. The availability of Gaia photometric science alerts data (which provides single epoch astrometry with an accuracy of 55 mas \citet{2021A&A...652A..76H})%, corresponding to xx parsec at a typical redshift of z=0.05) 
can provide very tight constraints; unfortunately, its cadence is low (30 days on average). With an ever-increasing torrent of transient candidates from today's large photometric surveys, and exacerbated by the sparsity of access to the UV part of the wavelength spectrum (accessible with the required scheduling flexibility almost exclusively through the Swift UVOT telescope), optical spectroscopic follow-up observations have taken over as the most efficient and accurate form of TDE identification. 

As a group the sample of UV/optical selected TDEs exhibits several common properties (see the reviews by \citet{2020SSRv..216..124V} and \citet{2021ARA&A..59...21G} for more in-depth discussions). These events display a blackbody temperature of 10\,000 -- 50\,000 K that persists throughout the peak and decline of the lightcurve (over several weeks/months), and potentially a mild temperature increase at later times \citep{2019ApJ...878...82V}. The UV/optical lightcurves broadly decay over time following a power law ($\propto t^{\gamma}$, where $\gamma$ varies between --1 and --3, \citet{2022arXiv220301461H}) to flatten out to a near constant level at very late times (100s--1000s of days, e.g. \citet{2019ApJ...878...82V}), although there is some heterogeneity among the sample. For example, many TDEs exhibit signs of a double-peaked lightcurve \citep[e.g.][]{2016NatAs...1E...2L,2019MNRAS.488.4816W}. \citet{2020ApJ...894L..10H} reported a relation between the peak luminosity and decay timescale, similar to type Ia supernovae, exemplified by several faint and fast TDEs (e.g. iPTF16--fnl \citet{2017ApJ...844...46B} and AT2019qiz \citet{2020MNRAS.499..482N}). These differences may relate to the stellar structure as well as the depth (i.e. impact parameter $\beta$) of the encounter \citep{2013ApJ...767...25G,2020ApJ...905..141L,2020ApJ...904...98R,2020ApJ...904...99R}. Correlations have also been found between the black hole mass and the lightcurve decline rate \citep[e.g.][]{2017MNRAS.467.1426S, 2021ApJ...908....4V}, as well as the late time UV luminosity \citep{2021arXiv210406212M}.

The main spectroscopic signatures of TDEs are a hot, blue continuum with very broad (5\,000 -- 30\,000 km s$^{-1}$) Gaussian emission lines of the H Balmer series (most notably H$\alpha$ and H$\beta$) and He\,\textsc{ii} (most notably the $4686\mathring{{\rm A}}\ $ transition). He\,\textsc{i} features are also seen on occasion, as are O\,\textsc{iii} and N\,\textsc{iii} ($\lambda\lambda$ 4100, 4640) Bowen fluorescence lines \citep{2019ApJ...873...92B, 2019ApJ...887..218L} and, on rarer occasions, Fe\,\textsc{ii} lines \citep{2019MNRAS.488.4816W, 2021MNRAS.504..792C}. Given the width of the lines several features are expected to overlap in the $4600\mathring{{\rm A}}\ $ region (home to He\,\textsc{ii}, N\,\textsc{iii} and also H$\beta$). As a result, the Bowen and Fe\,\textsc{ii} lines weren't unambiguously identified until recently. The width of these emission lines does not correlate with the black hole mass, nor is their evolution consistent with a kinematic origin. Instead, optical polarimetric observations indicate that electron scattering sets the line widths \citep{2022NatAs...6.1193L}. This implies the existence around peak light of a high covering factor, optically thick envelope of material engulfing the inner accretion disk. This material can become optically thin and reveal the accretion disk at late times \citep{2022MNRAS.515..138P}.

Although the spectral content of TDEs is heterogeneous and can change depending on the phase of the lightcurve, \citet{2021ApJ...908....4V} have proposed a classification scheme based on the emission line content. Three classes of TDEs have been identified: TDE-H, which show broad H Balmer (primarily H$\alpha$ and H$\beta$); TDE-H+He, which show both signatures of H Balmer, He\,\textsc{i} and/or He\,\textsc{ii} lines--the majority of these sources also show N\,\textsc{iii}/O\,\textsc{iii} Bowen fluorescence lines, and a small subsample show Fe\,\textsc{ii} lines; and TDE-He, which show only broad He\,\textsc{ii} emission. In the TDE-H+He class, the emission lines do not necessarily emerge simultaneously. Both an evolution of H-rich to He-rich (e.g. \citet{2019MNRAS.488.1878N}) and vice versa (e.g. \citet{2022A&A...666A...6W}) have been observed.  
We note that due to this behaviour, while single epoch spectroscopy is sufficient to classify a new transient as a TDE, it is likely insufficient to robustly classify a TDE in one of the three classes.
Correlations exist between the photometric (e.g. blackbody) properties of TDEs and their spectroscopic appearance \citep{2021ApJ...908....4V,2022A&A...659A..34C}. For example, the H$\alpha$ luminosity is found to correlate linearly with the blackbody radius, and the TDE-H+He class has smaller blackbody radii than the TDE-H class.

A small but interesting sub-class of TDEs display double-peaked emission line profiles in H$\alpha$, H$\beta$ \citep{2019ApJ...880..120H,2020MNRAS.498.4119S, 2020ApJ...903...31H} and/or He\,\textsc{ii} \citep{2022A&A...666A...6W} at various points in their evolution. These events are of particular interest because of the possibility to model their emission features with a relativistic, elliptical accretion disk to constrain their properties. This allows a direct determination of the disk geometry and inclination (and potentially even the black hole spin, \citet{2022A&A...666A...6W}), properties for which a significant body of theoretical literature predictions exist but are not directly testable. Areas of particular interest include verifying whether or not the tidal debris retains significant eccentricity or can form a circular disk efficiently, as well as directly testing the hypothesis that the TDE line content as well as X-ray properties are viewing angle dependent. 

An independent spectroscopic signature that has been suggested to be related to TDEs are narrow, high excitation ($>$100 eV) coronal emission lines such as Fe\,\textsc{x,xi,xiv}, Ar\,\textsc{xiv} and S\,\textsc{xii} \citep{2008ApJ...678L..13K, 2011ApJ...740...85W, 2012ApJ...749..115W}. In contrast with the other spectroscopic features described earlier, such lines are hypothesised to be excited after the TDE radiation interacts with a clumpy, pre-existing medium at larger scales ($\sim$10$^{17-18}$ cm). These lines are therefore expected to be delayed with respect to the peak of the UV/optical flare. While most such sources have been discovered in archival searches, more recently this association was confirmed through the detection of coronal lines following the TDE AT2017gge \citep{2022MNRAS.517...76O}.

UV spectroscopy of TDEs is more rare, with existing datasets providing coverage of only a handful of sources. These spectra tend to show both emission (e.g. Ly$\alpha$, C\,\textsc{iv} and N\,\textsc{v}, \citet{2016ApJ...818L..32C, 2018MNRAS.473.1130B, 2019ApJ...873...92B}) and broad absorption lines (BALs; e.g. C\,\textsc{iv}, N\,\textsc{v}, S\,\textsc{iv}, e.g. \citet{2017ApJ...843..106B,2019ApJ...879..119H}) that tend to have blueshifts of 1000--10\,000 km s$^{-1}$ \citep{2019ApJ...879..119H}, similar to those observed in broad absorption line quasars -- albeit with some peculiarities. For example, TDEs do not show the ubiquitously observed Mg\,\textsc{ii} doublet at $2796,2803 \mathring{{\rm A}}$. Furthermore, the ratio of nitrogen to carbon is much higher than observed in QSOs \citep{2016ApJ...818L..32C}. The broad, blue-shifted absorption lines may be evidence for the presence of fast outflows (e.g. \citet{2020MNRAS.494.4914P}), although the driving mechanism remains unclear.

The heterogeneity of discovery surveys (in terms of cadence, depth, sky coverage, etc.) make accurate rate estimates based on the UV/optical sample challenging. Early calculations based on small samples yielded estimates of order $\dot{N}_{TDE} \sim$10$^{-5}$ per galaxy per year (e.g. \citet{2008ApJ...676..944G, 2014ApJ...792...53V}), discrepant by an order of magnitude with conservative theoretical expectations of $\dot{N}_{TDE} \sim 10^{-4}$ \citep{1999MNRAS.309..447M, 2004ApJ...600..149W}. \citet{2018ApJ...852...72V} provided the first detailed treatment of selection effects of a sample of UV/optically discovered TDEs, leading to the first measurement of their LF. The observed rest frame g-band LF follows a steep power law, $\frac{dN}{dL_g} \propto L_g^{-2.5}$. Integrating over this LF for the observed range of peak luminosities results in a TDE rate of $\approx 10^{-4}$ per galaxy per year (for a mean galaxy stellar mass of 10$^{10}$M$_{\odot}$), consistent with theoretical expectations. The growing sample of homogeneously selected UV/optical TDEs with ZTF \citep{2022arXiv220301461H} has also been used to constrain the LF, with results that are largely consistent in terms of the LF slope ($\propto L^{-2.3}$, \citep{2022arXiv220912913C,2022arXiv221014950L}) but with significant differences in the resulting TDE rate (up to an order of magnitude lower). 

In the near future, large and deep sky surveys (e.g. the legacy survey of space and time on the Vera Rubin observatory, \citet{2019ApJ...873..111I,2020ApJ...890...73B}) will help to further enlarge the TDE population and enable studies of large, homogeneously selected samples.

\subsubsection{IR emission}
The luminous and energetic radiation that is emitted during a TDE has the power to reveal and alter the black hole's circum-nuclear environment on sub-pc scales. Dust particles residing in the immediate vicinity of the SMBH will efficiently absorb the X-ray/UV/optical radiation, heating them to high temperatures. If this temperature is higher than the dust sublimation temperature (which depends on the dust composition, particle size and geometry, among other factors), it will lead to the dust evaporating up to the so-called sublimation radius, the smallest radius where dust particles can survive given a ionising luminosity, dust temperature and particle size. Because the SMBH is generally inactive before the TDE, dust can exist in close proximity to the black hole (unlike the situation for AGNs, which are continuously radiating photons that will evaporate the dust). The luminous TDE flare will then carve out a dust-free region with size R$_{\rm sub}$. For the typical case of graphite particles and in the assumption that the dust absorption efficiency is $\approx 1$, the sublimation radius is given by \citep{2016ApJ...829...19V} 
\begin{equation}
\label{eq:dust}
 \rm    R_{sub} = 0.15 \left(\frac{L_{45}}{a_{0.1}^2~T_{1850}^{5.8}}\right)^{1/2}~ \rm pc
\end{equation}
Here L$_{45}$ is the luminosity normalised to 10$^{45}$ erg s$^{-1}$, a$_{0.1}$ is the particle size normalised to 0.1 $\mu m$ (because the dust emissivity scales $\propto a^2$ it will be dominated by the largest grains, and 0.1 $\mu m$ corresponds to the largest dust grain size observed in our Galaxy), and T$_{1850}$ is the dust temperature normalised to 1850 Kelvin (the expected sublimation temperature of graphite-like dust).
The surviving dust particles will reradiate the absorbed energy at IR wavelengths -- where the precise wavelength is determined by the particle size, composition, geometry and temperature. The TDE light echo will be dominated by the largest particles able to exist at R$_{\rm sub}$, as these will reradiate the absorbed energy in the shortest timeframe. 

The IR lightcurve will evolve with a time delay with respect to the UV/optical flare. Assuming a single shell of material is responsible for the dust absorption and re-emission, once the photons of the UV/optical peak have been reprocessed an abrupt drop in IR emission will occur. Measuring this break in the IR lightcurve therefore provides a direct estimate of the approximate radius R$_{\rm sub}$ of the reprocessing shell. With R$_{\rm sub}$ measured from the observations, solving Equation \ref{eq:dust} for L$_{45}$ provides an independent measurement of the bolometric luminosity of the TDE required to power the observed dust echo. 

% observations, review chapter 2021SSRv..217...63V
The first measurement of the IR luminosity of a TDE candidate was reported in \citet{2009ApJ...701..105K}, who observed it in the mid-IR with the Spitzer space telescope. The (surprising) result was a very large mid-IR (10--20 $\mu m$) luminosity $\approx 10^{43.5}$ erg s$^{-1}$. In the assumption of blackbody emission, this implies an emitting region size of 0.5 pc. WISE imaging observations taken $\sim$2 years later showed a decrease in luminosity by a factor of $\sim$2, confirming that this MIR emission is transient and not associated to star formation in the host.

Other results are based mainly on the reactivated WISE mission (NEOWISE, \citet{2014ApJ...792...30M}), which is performing an all sky IR survey with a cadence of $\sim 6$ months. Several authors reported on both theoretical expectations based on 1D radiative transfer models \citep{2016MNRAS.458..575L}, and observational detections of IR echoes \citep{2016ApJ...829...19V, 2016ApJ...832..188D, 2016ApJ...828L..14J}. These observations indicate light delays of the IR peak brightness of 100--200 days with respect to the UV/optical flares, and generally small amplitudes/luminosities (L$_{\rm IR} \sim 10^{41-42}$ erg s$^{-1}$, $\delta m \approx 0.1$ mag). Due to the small amplitudes, a careful analysis and high quality data are required to robustly detect the IR echoes. By comparing the UV/optical blackbody luminosity to the bolometric luminosity implied by the dust echo, \citet{2016ApJ...829...19V} infer a bolometric correction factor of $\sim 10$, i.e. $L_{\rm BB}/L_{\rm bol}$ = 0.1. This could suggest that a large fraction of the available energy is radiated at wavelengths that are not directly accessible (e.g. in the EUV band).

The most comprehensive study to date is presented by \citet{2021ApJ...911...31J}, who find similarly modest L$_{\rm IR}$ values and, by equating the energy radiated in the IR with the energy able to heat up the dust, very low dust covering fractions f$_c = E_{\rm IR}/E_{\rm abs} \sim$ 0.01 (or similar upper limits in the case of non-detections). This is in stark contrast with AGN studies, where the torus covering factor is typically $\sim$0.1 or higher.

Recently launched and planned IR space observatories (e.g. JWST and the Nancy Grace Roman telescope) will enable, through spectroscopic and time-domain observations, in-depth studies of the circumnuclear dust and gas content and properties of quiescent galaxies beyond the local group, something that is otherwise very hard to do with existing ground-based (and space) observatories.

\subsubsection{Radio emission}
Relativistic electrons gyrating in magnetic fields will emit synchrotron radiation, which for the magnetic fields found near SMBHs will peak at radio wavelengths. 
Following a TDE, there are three main physical mechanisms that can produce radio emission. 

One of these is associated to the (previously ignored) unbound tail of the stellar debris. This debris will move away from the SMBH at high velocities and eventually interact with the ambient circum-nuclear medium, creating a bow shock along the leading edge of the
debris stream (e.g. \citet{2016ApJ...827..127K}). The radio brightness will depend on the ambient medium density and the cross-section of the debris stream. Depending on the location of the ambient material with respect to the SMBH, there may be a delay in the onset of radio emission related to the travel time from the disruption site to the circum-nuclear material.

Another way to produce the relativistic electrons necessary to produce radio synchrotron radiation is through external shocks (e.g. \citet{2013ApJ...767..152Z}). In this scenario, an outflow or (sub-)relativistic jet --produced following the accretion of material, and therefore unrelated to the previously mentioned unbound debris-- is launched from the SMBH at high velocity, and while moving outwards eventually plows into pre-existing circum-nuclear material to produce a forward (as well as a reverse) shock. This shock subsequently propagates through the medium, producing relativistic electrons which emit synchrotron photons when accelerated in the medium's magnetic field. Although this emission is not produced directly by the jet, modelling the radio lightcurves can be used to constrain the internal jet structure. For example, if the jet is composed of an ultra-relativistic core and a slower surrounding sheath (for example, produced by jet precession), a multi-peaked light curve should be observed \citep{2015MNRAS.450.2824M}. 
Radio variability can also be produced by synchrotron self-absorption, if the self-absorption frequency moves through the observed frequency bands \citep{2012MNRAS.420.3528M}; by changes in the slope of the circum-nuclear medium density profile the decelerating jet encounters \citep[although these changes are more modest][]{2017MNRAS.464.2481G}; and through inverse Compton cooling of the electrons on the jet X-rays, the rate of which depends on the collimation of the jet \citep{2013MNRAS.434.3078K}.

The third production mechanism for radio emission is distinct from the previous two scenarios in that it is internal to the jet (i.e. co-spatial with the X-rays; the relativistic electrons can be provided by collimation shocks, turbulent electromagnetic fields or magnetic reconnection), and does not depend on the presence of dense circum-nuclear material. 
A freely (adiabatically) expanding conical jet model has successfully been applied to compact (extragalactic) radio cores and X-ray binaries, and has also been proposed to power the radio emission of ASASSN--14li, a canonical example of a TDE \citep{2018ApJ...856....1P}.
A key property of this model is the prediction of correlated behaviour between emission in different bands; in particular, if the jet power is coupled to the mass accretion rate through the accretion disk, a correlation between radio (from the jet) and X-ray (from the accretion disk) emission is expected (and observed, \citet{2018ApJ...856....1P}).\\

Radio follow-up observations of TDEs have uncovered two broad classes \citep{2013A&A...552A...5V}: radio-loud and radio-quiet TDEs, although the dividing line is somewhat arbitrary. Radio-loud TDEs stand out due to their association to extremely bright X-ray emission, and are thought to be produced when a relativistic jet is fortuitously pointed along our line of sight. Such relativistic jets can be detected out to cosmological distances due to Doppler boosting caused by the relativistic bulk motion of the material -- the current record-holder is AT2022cmc, a TDE at $z=1.19$ (luminosity distance of 8.45 Gpc) -- although their observed rate (based on the 4 known events) is only $\sim$1\% of UV/optical and X-ray TDEs, $\sim$ 0.03 Gpc$^{-3}$ yr$^{-1}$ \citep{2015MNRAS.452.4297B,2022arXiv221116530A}. 

The sources with less luminous radio emission could in part be those with off-axis relativistic jets, which due to the absence of Doppler boosting will not reach very high peak luminosities. Such systems may become detectable only at later times, when the jet has decelerated. Deep radio limits suggest that the radio-quiet TDEs may in fact launch intrinsically less powerful jets or outflows. The radio luminosity in such systems is governed by a wide variety of properties, including the magnetic field strength, the density profile of the circum-nuclear medium \citep{2011MNRAS.416.2102G, 2017MNRAS.464.2481G}, as well as the degree of equipartition of energy between the electrons and the magnetic field (e.g. \citet{2020NewAR..8901538D, 2022NatAs.tmp..252P}).

\subsection{High energy neutrino and cosmic rays}

Cosmic rays are high-energy ($10 - 10^{20}$GeV) charged hadrons (e.g., protons and atomic nuclei) that impinge on Earth from outer space. In particular, those with energies larger than $10^{19}$ GeV are referred to as ultrahigh-energy cosmic rays (UHECRs). Upon interactions with ambient baryonic matter and radiation fields, these cosmic rays produce secondary particles such as gamma rays and hadrons. One of the secondary particles is the neutrino, a subatomic neutral particle with a very small mass that interacts only via weak force and gravity. Because of the very weak gravitational interaction and the weak force having a short range, they so weakly interact with ordinary matter that it makes them incredibly hard to detect, even though they are the second most common elementary particle in nature \citep{1972gcpa.book.....W}. However, this also means that neutrinos have an extremely long mean free path such that they carry information about very distant astrophysical phenomena, including the origin of cosmic rays. A few known possible astrophysical sites of neutrino production include the Earth's atmosphere, the solar atmosphere, the Galactic disk, and the cores of massive stars (when they explode as supernovae). See review papers for cosmic rays \citep{2011ARAA..49..119K}
 and neutrinos \citep{2017ARNPS..67...45M}.

Tidal disruption events have been proposed as possible sources of UHECRs \citep{2009ApJ...693..329F,2014arXiv1411.0704F} and neutrinos \citep{2011PhRvD..84h1301W,2016PhRvD..93h3005W,2017MNRAS.469.1354D,2017ApJ...838....3S,2018NatSR...810828B,2018A&A...616A.179G}. There are two main possible production zones. 
\begin{enumerate}
    \item \textit{Relativistic jet}:   relativistic jets in TDEs could be a potential production site of cosmic rays \citep{2009ApJ...693..329F}.
    The detection of jetted TDEs (Swift J1644+57 \citet{2011Sci...333..203B,2011Natur.476..421B},  Swift J1112.2-8238 \citet{2015MNRAS.452.4297B} and Swift J1112-82 \citet{2012ApJ...753...77C}) and high neutrino flux at $\sim 30$ TeV by IceCube \citep{2013PhRvL.111b1103A} exceeding the Fermi isotropic gamma-ray background at 10-100 GeV, triggered the research on the cosmic ray and neutrino production in the context of jetted TDEs.  It was suggested that in relativistic TDE jets, protons can be accelerated via internal shocks created due to the collision of rapidly moving ejecta and slower ejecta \citep{2014arXiv1411.0704F}. \citet{2011PhRvD..84h1301W} analytically confirmed this idea that accelerated protons by the internal shock can produce $0.1-10$ PeV neutrinos via photomeson interactions with X-ray photons but also found that their flux is not sufficient to account for the observed flux of UHECRs. It was further noted by \citet{2017PhRvD..96f3007Z} that nuclei can be significantly disintegrated in internal shocks, although they could survive for low-luminosity TDE jets. So alternative scenarios, such as  UHECR and neutrino production in external forward and reverse shocks created by the interaction of the jet with circumnuclear medium \citep{2016PhRvD..93h3005W,2017PhRvD..96f3007Z}, or by the inner part of jet \citep{2018A&A...616A.179G}, were examined. However, it appears that the tension between the detection rate of neutrinos and cosmic rays by IceCube and the jetted TDE rate has not been fully resolved. \citet{2018NatSR...810828B} revisited the internal shock scenario and found that the disintegration of nuclei found by \citet{2017PhRvD..96f3007Z}, in fact, should be frequent to produce more neutrinos and, if so, abundant TDEs of white dwarfs can explain the observed neutrinos and cosmic rays. On the other hand, other studies \citep{2017MNRAS.469.1354D,2017ApJ...838....3S,2018A&A...616A.179G} argued that TDEs can not be the dominant sources of IceCube neutrinos because the observationally inferred upper limit on the rate of jetted TDEs that happen to be pointed towards us is much lower than the lower limit of the production rate of neutrinos inferred from the IceCube data.

    \item \textit{Accretion disk and corona}: the protons can be accelerated in the disk and corona that can form in TDEs. \citet{2019ApJ...886..114H} found that the stochastic acceleration of protons is efficient when magnetic fields are strong in super-Eddington accretion flows. They further showed that the acceleration could be efficient, independent of the strength of magnetic fields, when the disk becomes radiatively inefficient and optically thin because of the heat generated by turbulent viscosity and transported inward with accretion, which was supported by \citet{2020ApJ...902..108M}. In addition, \citet{2020ApJ...902..108M}, taking the spectral template for AGNs, showed that efficient neutrino production in a hot corona above a disk is possible. High-energy gamma-rays produced from the decay of neutrinos are then likely to be absorbed in the disk and the corona via $\gamma\gamma\rightarrow e^{-}e^{+}$ pair production, the latter of which are subsequently reprocessed to lower energies $\simeq O(10)$ MeV. However, as \citet{2020ApJ...902..108M} noted, the spectral features for TDEs are not well understood and may be different from those for AGNs.

\end{enumerate}
So far, there are two TDE candidates claimed to be associated with neutrino observations. AT2019dsg \citep{2019TNSTR.615....1N} was detected by ZTF in optical/UV, X-rays and Radio. The optical/UV emission is well described by a blackbody spectrum of $10^{4.59\pm0.02}$ K and radius of $10^{14.59\pm0.03}\cm$ \citep{2021ApJ...908....4V}. The peak luminosity is estimated to be $10^{44.54\pm0.08} \erg~\second^{-1}$. The first detection of X-rays from AT2019dsg occurred 37 days after discovery. The detected X-ray fluxes were soft, and declined very rapidly, by more than a factor of 50 in flux over $\simeq 160$ days \citep{2021MNRAS.504..792C}. Radio follow-up observations revealed synchrotron emission from non-thermal electrons. This has been interpreted as an indication for a mildly relativistic outflow \citep{2021ApJ...919..127C,2022MNRAS.511.5085M} in which particles can be accelerated. AT2019dsg also displayed a strong dust echo signal \citep{2021arXiv211109391V}. Later, it was suggested as the astrophysical origin of the detected 0.2PeV neutrino IC191001A 
\citep{2020ApJ...902..108M,2021NatAs...5..472W,2022ApJ...927...74M}. However, the AT2019dsg-IC191001A association has not been confirmed. \citet{2021ApJ...919..127C} casts doubt on the claimed association of this event with a high-energy neutrino based on that even a mildly relativistic outflow requires too narrow opening angle. 

AT2019fdr is a TDE candidate in a Narrow-Line Seyfert 1 active galaxy \citep{2021ApJ...920...56F} possibly in association with the IceCube high-energy neutrino IC200530A \citep{2022PhRvL.128v1101R}. It lies within the reported 90\% localization region of the neutrino source. The EM emission at multi-wavelength from this event was detected. It was first discovered by ZTF \citep{2019TNSTR.771....1N} as a very luminous and long-lived event with the peak optical/UV luminosity of $\simeq (2-3)\times10^{44}\erg~\second^{-1}$. X-ray emission from the candidate was only detected in the 0.3-2.0 keV band roughly 590 days since discovery. AT2019fdr was further detected at mid-infrared wavelengths. The peak MIR luminosity of $1.9 \times 10^{44}\erg\second^{-1}$ was observed over one year after the optical/UV peak.

\subsection{Gravitational wave emission}
\label{sec:gw}
Fundamentally, the tidal disruption of a star is a two body encounter between massive objects accelerating under the influence of each other's gravitational fields. Such encounters will produce GW emission, chiefly due to the changing mass quadrupole moment of the BH-star system (but see the remainder of this section for other GW emission mechanisms). The main difference between the now well-known compact object binaries and TDEs is the mass ratio ($q$) of the system; while typically $q > 0.1$ for the former, TDEs can have $10^{-8} < q < 10^2$ depending on the mass of the black hole (typically assumed to be 10$^{4-8} M_{\odot}$) and the type of star involved (ranging from low mass main sequence stars to white dwarfs as well as more massive stars). Because the BH tends to be (quasi-)stationary while the low mass component moves in an orbit around it, such systems are also called extreme mass ratio inspirals (EMRIs; $10^{-8} < q < 10^{-5}$) or intermediate mass ratio inspirals (IMRIS; $10^{-5} < q < 10^{-2}$); among these systems one can distinguish between full disruptions (where a star is completely destroyed upon pericentre passage) and stellar captures (where a stellar remnant or compact object occupies a stable, bound orbit around the MBH that evolves mainly due to GW radiation). In this section we will provide a brief overview of the main mechanisms that can produce gravitational waves during and after TDEs, their expected observational properties, and the prospects of detecting such signals with current and future detectors. We discuss these scenarios in order of decreasing expected strain amplitude and hence, detectability.

\subsubsection{The first pericentre passage}
\label{sec:gwburst}
Stars that are fully disrupted after passing within the tidal radius will be transformed into a cloud of tidal debris. This cloud becomes sufficiently diffuse (compared to the density of the initial star) that such an encounter will produce a single, burst-like gravitational wave signal (see Section \ref{sec:gwemri} for details about partial disruptions and long-lived stellar captures).

In the simplifying assumption that the star is disrupted instantaneously at pericentre (the {\it impulse} approximation), we can estimate the characteristic frequency of this signal from Kepler's third law (evaluated at r$_{\rm p}$), recalling that $r_{\rm t} = \rstar \left( \frac{\mbh}{\mstar} \right)^{1/3}$ and $\beta = \frac{r_{\rm t}}{r_{\rm p}}$:
\begin{equation}
\label{eq:gwburstfreq}
    f_{\rm GW} \approx \frac{\beta^{3/2}}{2 \pi} \left( \frac{G \mbh}{r_{\rm t}^3} \right)^{1/2} \approx 10^{-4} \beta^{-3/2} \left( \frac{\mstar}{\Ms} \right)^{1/2} \left( \frac{\rstar}{\Rs} \right)^{3/2} {\rm Hz}
\end{equation}
This relatively low frequency signal is beyond the reach of current ground-based GW detectors, but is within range of space-based laser antennas such as LISA \citep{2017arXiv170200786A}.

To estimate the strength of the gravitational waves emitted by the BH-star system, we start from the quadrupole approximation to the Einstein field equations (we refer the reader to \citet{1998bhrs.conf...41T} for a more detailed discussion of the underlying assumptions):
\begin{equation}
\label{eq:quadapprox}
    h \simeq \frac{G}{c^4} \frac{1}{d} 4 E_{\rm kin}
\end{equation}
where $d$ is the distance to the source and $E_{\rm kin}$ is the kinetic energy of the star, evaluated at pericentre for the TDE scenario (given the large mass ratio, the SMBH is assumed to be at rest in the system's centre of mass frame):
\begin{equation}
\label{eq:ekin}
    E_{\rm kin} = \frac{G \mbh \mstar}{r_{\rm p}}
\end{equation}
Combining equations \ref{eq:quadapprox} and \ref{eq:ekin}, substituting the definitions of $r_{\rm t}$ and $\beta$, and renormalising for a 1 M$_{\odot}$ star being disrupted by a 10$^6\Ms$ SMBH at the distance of the Virgo cluster (the closest galaxy cluster at $\approx$ 20 Mpc) we obtain the strain of the GW emission (e.g. \citet{2004ApJ...615..855K}):
\begin{equation}
\label{eq:gwburststrain}
    h \approx 4\times10^{-22}~ \beta~ \left( \frac{d}{20~\rm Mpc} \right)^{-1} \left( \frac{\rstar}{\Rs} \right)^{-1} \left( \frac{\mstar}{\Ms} \right)^{4/3} \left( \frac{\mbh}{10^6\Ms} \right)^{2/3}
\end{equation}
The dependence on $\mstar$, $\rstar$ and $\mbh$ indicates that more compact stars (such as WDs) will produce stronger GW signals. However, the parameter space for TDEs of WDs to produce both a GW and an EM signal is limited by the requirement that the tidal radius must be larger than the Schwarzschild radius.

\subsubsection{Strong vertical compression of the star near pericentre}
\label{sec:gwcompression}
Upon pericentre approach, the unlucky star will experience differential tidal forces at its leading and trailing edges, resulting in the star being stretched in the direction of orbital motion (sometimes called {\it spaghettification}). In addition, it experiences strong one dimensional compression orthogonal to its trajectory of motion \citep[also called {\it pancaking}][]{1983A&A...121...97C}. During this phase the material is in tidal free fall, and this pancaking effect scales strongly with the penetration factor of the encounter. This spaghettification and pancaking lead to the star being strongly asymmetrically deformed, resulting in a variable quadrupole moment and hence GW emission \citep{2009ApJ...705..844G, 2013MNRAS.435.1809S}. The build-up of internal pressure will eventually overcome the tidal field, leading to outward propagation of the flow, which is likely accompanied by strong shock waves that can lead to an observable EM signal (X-ray shock breakout, \citet{2004ApJ...615..855K, 2009ApJ...705..844G}). 

At the moment of maximum compression, the vertical extent of the star is $\sim \beta^{-3}\rstar$ \citep{1986ApJS...61..219L}. Approximating the second derivative of the moment of the inertia tensor by assuming that the pericentre passage time is $T_{\rm p} \approx$ $\frac{v_{\rm p}}{r_{\rm p}} \approx$  $\frac{G\mbh}{r_{\rm p}^3}$ leads (e.g. \citet{2009ApJ...705..844G}) to the following expression
\begin{equation}
    \ddot{I}^{ij} \propto \frac{\mstar \rstar^2}{T_{\rm p}^2} \approx \mstar \rstar^2 \frac{G\mstar}{(\rstar / \beta)^3} = G~\frac{\mstar^2 \beta^3}{\rstar} 
\end{equation}
This results in a GW strain
\begin{equation}
    h \approx \frac{G}{c^4 d} \frac{\mstar \rstar^2}{T_{\rm p}^2} = \frac{G^2}{c^4 d} ~\frac{\beta^3 \mstar^2}{\rstar} \approx 5\times10^{-27}~ \beta^3 ~ \left( \frac{\mstar}{\Ms} \right)^2 ~\left( \frac{\rstar}{\Rs} \right)^{-1} \left( \frac{d}{20~\rm Mpc} \right)^{-1}
\end{equation}
The dependence of the strain h $\propto \beta^3$ indicates that a significant signal will occur only for very deep plunges (for reference, $\beta = 10$ and a solar type star at 20 Mpc leads to h $\approx 5\times 10^{-24}$, while for $\beta = 25$, h $\approx 8\times 10^{-23}$). 
With the assumption that the signal is emitted during the entire pericentre passage, the frequency of this signal is given by Equation \ref{eq:gwburstfreq}.\\

An alternative approach is to calculate the components of the moment of inertia tensor explicitly \citep{2013MNRAS.435.1809S}. The characteristic frequency from the timescale of maximum compression $\tau_c$ scales as
\begin{equation}
    \tau_c = 8.5 \beta^4 \tau_\star
\end{equation}
where $\tau_\star \propto \frac{1}{\sqrt{G \bar{\rho}_{\star}}}$, which quantifies the compressibility of the star \citep{1986ApJS...61..219L, 2008A&A...481..259B, 2013MNRAS.435.1809S}.
This leads to a frequency for the GW signal of 
\begin{equation}
    f_{\rm GW} \approx 4\times10^{-5}~ \beta^4 \left(\frac{\mstar}{\Ms}\right)^{\frac{1}{2}} \left(\frac{\rstar}{\Rs}\right)^{-\frac{3}{2}} \rm Hz
\end{equation}
For deeply plunging encounters with $\beta = 10~ (25)$, this results in $f_{\rm GW} \approx 0.4~ (15)$ Hz.

From the explicit calculation of $\ddot{I}^{ij}$, \citet{2013MNRAS.435.1809S} find that the diagonal elements scale as $\ddot{I}^{ij} \propto \beta^2$, leading to an expression of the strain
\begin{equation}
    h \approx 10^{-27} \beta^2 \left( \frac{\mstar}{\Ms} \right)^2 ~\left( \frac{\rstar}{\Rs} \right)^{-1} \left( \frac{d}{20~\rm Mpc} \right)^{-1}
\end{equation}

For the same deeply plunging encounters as envisaged before with $\beta = 10~(25)$, this leads to strain values of $h\sim10^{-25} (10^{-24})$.

The off-diagonal elements of $\ddot{I}^{ij}$ scale as $\propto \beta^5$ -- but these elements have coefficients that reduce to 0 if the system has reflection symmetry about the principle  axes and orbital plane (which occurs, for example, in the Schwarzschild metric). This may no longer be true for inclined orbits (with respect to the black hole spin plane) in Kerr space. 

More important, however, is the argument of \citet{2013MNRAS.435.1809S} that in reality, the finite size of the star will lead to {\it desynchronization} of the GW emission (which becomes significant for $\beta>6$), i.e. the production of GW emission from the leading and trailing edges of the star occurs non-simultaneously, and is smeared out over the entire pericentre passage time (rather than emitted as a coherent signal over the maximum compression timescale). Because GW emission from stellar compression scales strongly with $\beta$, the parameter space of high $\beta$ would in principle provide the best opportunity to observe this component. However, desynchronization leads to a suppression in the GW strain in particular for these high $\beta$ events, and the likely result is that the GW emission from this mechanism is well below the detection threshold of future GW detectors. 

\subsubsection{Instabilities in the post-disruption accretion disk}
\label{sec:gwinstability}
An additional mechanism to produce GW emission in TDEs relates to the (potential) formation of a massive, geometrically thick disk \citep{2011PhRvL.106y1102K,2019MNRAS.489..699T}. Such a structure may arise if the radiative cooling timescale in the disk is larger than the viscous timescale. \citet{2018MNRAS.474.1737N} showed through numerical simulations (see also \citet{2018MNRAS.475..108B}) that under certain conditions such a disk can form following a TDE, and is unstable to a global non-axisymmetric instability (the Papaloizou-Pringle instability or PPI, \citet{1986MNRAS.220..593P}).

\citet{2019MNRAS.489..699T} treat this problem both analytically and numerically, assuming a 1 M$_{\odot}$ accretion disk surrounding a 10$^6$ M$_{\odot}$ SMBH. The analytical treatment follows from the results of \citet{2018MNRAS.474.1737N}, who find that the PPI occurs at approximately the Keplerian orbital frequency of the torus, which is assumed to be located at $2r_{\rm p}$. Hence the frequency of the signal as well as the maximum strain are similar to Equations \ref{eq:gwburstfreq} and \ref{eq:gwburststrain}. The main difference with the scenario described in Section \ref{sec:gwburst} is that the instability occurs over multiple orbital timescales, and the total (characteristic) strain (denoted $h_c$) will be larger by a factor $\sqrt{N_c}$, where $N_c$ is the number of cycles spent within the detector bandwidth \citep{maggiore2018}. \citet{2019MNRAS.489..699T} estimate that $N_c \sim $ several 10s of orbits, so the characteristic strain will be correspondingly larger by a factor of a few ($<$10). 
The numerical treatment of this problem results in significant differences, due to several assumptions that are unlikely to be valid (notably the point-like nature of the disk, as well as the assumption that the entire disk undergoes the PPI, rather than only part of the 1 M$_{\odot}$ available). This leads to estimates of the strain that are a factor of $\sim$100 lower ($\sim$10$^{-24}$) than the analytical estimate (Equation \ref{eq:gwburststrain}). Similar to the GW emission from stellar compression, the prospects for observing emission from accretion disk instabilities in the near future remain bleak.

\subsubsection{Partial disruptions, direct plunges, and extreme mass ratio inspirals}
\label{sec:gwemri}
Lastly, we briefly touch upon scenarios that are closely related to TDEs, although they do not constitute a {\it classical}, full tidal disruption -- envisaged here as a single pericentre passage with observed EM emission.

Stars in galactic nuclei are thought to be diffusing through orbital angular momentum space through weak stellar encounters that represent small perturbations to their orbital angular momentum / energy \citep{1976ApJ...209..214B,1976MNRAS.176..633F,1977ApJ...211..244L}. One can then imagine a non-fatal interaction between said stars and the SMBH if they pass close to, but not within, the tidal radius (i.e. $\beta < 1$). In this scenario, dubbed a partial disruption event, some mass may be stripped from the stellar envelope but the core will survive to live another orbit (or orbits; e.g. \citet{2005ApJ...624L..25D,2013ApJ...767...25G,2020ApJ...904..100R}), possibly leading to repeated close encounters \citet{Wevers:23,Liu:23}. With each subsequent close passage, more orbital angular momentum will be removed from the star and it will eventually spiral into the SMBH to be completely destroyed \citep{2003ApJ...590L..29A, 2005ApJ...629..362H}. Such a scenario can produce multiple bursts of GW (and EM) emission, although as we saw earlier, the GW strain scales with $\beta$ in all cases, hence the GW signal will be weak.

Alternatively, a direct plunge (where the stellar orbit directly impacts the SMBH without being able to pass through pericentre) provides a scenario where the GW strain will be highest (due to the very high $\beta$ value of such an encounter), but little or no EM emission will be visible (as in this case, the material cannot reach pericentre but rather is swallowed directly by the black hole). 

Finally, there is a limited parameter space in which a star may be captured on a bound orbit around the SMBH, e.g. through direct capture (e.g. \citet{1993MNRAS.264..388K, 1995ApJ...445L...7H, 1997MNRAS.284..318S}), tidal capture (e.g., \citet{1992MNRAS.255..276N}), tidal stripping of stars \citet{2001ApJ...551L..37D}, disk migration \citet{2022ApJ...926..101M,2023MNRAS.521.4522D}, or the disruption of a binary star system. If the pericenter distance of a star is not sufficiently close to the BH, the star is tidally deformed without losing its mass at the pericenter passage during which its orbital energy is converted into heat energy \citep{1977ApJ...213..183P}. Through this process, an initially unbound star can be captured by the BH \citep{1976MNRAS.176..633F,1992MNRAS.255..276N}. If the pericenter is between the ``tidal capture'' radius and full disruption radius, the star loses some fraction, but not all, of its mass. Depending on which one is the dominant effect between tidal excitation and heating, and asymmetric mass loss, the partially disrupted star can be bound \citep{2005Icar..175..248F,2013PhRvD..87j4010C,2020ApJ...904..100R}.  The last scenario will lead to one star being flung out at high velocity, while the other component of the binary can remain gravitationally bound in a stable orbit the \citet{1988Natur.331..687H} mechanism. Although the rate of bound systems produced via the Hills mechanism is uncertain\footnote{\citet{2022ApJ...926..101M} made an order-of-magnitude estimate for the rate to be $10^{-5}-10^{-7}$ $\yr^{-1}$ per Milky Way-like galaxies, which would be significantly affected by the binary fraction and the shape of the potential (e.g., spherical or trixial).}, such systems will occupy orbits that eventually evolve primarily due to the emission of GWs -- at which point they are designated extreme mass ratio inspirals (see \S\ref{EMRI} or e.g. \citet{2007CQGra..24R.113A} for a review).

Given that the dynamical time scale of a solar type star at the tidal radius of a 10$^6$ M$_{\odot}$ black hole is $\sim$10$^4$ seconds, the GW frequency of emission in this scenario will be $\sim$10$^{-4}$ Hz\footnote{This can range up to $\sim$10$^{-2}$ Hz if one also considers NSs or stellar mass BHs, which can orbit within the tidal radius as defined for a solar type star \citep{1997MNRAS.284..318S}}, and hence within the frequency sensitivity of space-based laser antennae. The GW signal of such systems will be relatively weak, but it will be boosted by the fact that many cycles will occur within the detector bandpass. Coherent integration of the GW signal for $> 10^4$ cycles with a frequency in the LISA band would be required for detecting this signal \citep{2007CQGra..24R.113A}.\\

The prospects of observing this radiation with future detectors remains a poorly explored topic. 
Given the burst-like nature of the strongest predicted GW strain (due to the varying quadrupole moment of the BH-star system), detecting such signals requires significant signal-to-noise ratio to confidently distinguish it from the noise characteristics. Besides the noise, there will also be the background signal from the various sources of GW emission within the LISA band. 

\citet{2020MNRAS.498..507T} study the background signal of TDEs of MS and WD stars, and find its spectral shape scales as h$_c \propto$ f$^{\frac{-4\gamma+5}{6}}$, where $\gamma$ represents the distribution of $\beta$ values (i.e. $\frac{d\ddot{N}}{d\beta} \propto \beta^{-\gamma}$, with a typical value for $\gamma = 2$ \citet{2016MNRAS.455..859S}). These authors also found that the background signal for WD disruptions around IMBHs may be detectable by future generation space arrays (although the uncertainties remain large due to unknown factors such as the IMBH occupation fraction of globular clusters), but that the background signal for MS disruptions of SMBHs will likely remain undetectable, not due to sensitivity issues, but rather because the lower frequency will be outside of detectors such as DECIGO \citep{2006CQGra..23S.125K} and BBO \citep{2006CQGra..23.4887H}.

\begin{figure}
    \centering
\includegraphics[width=\textwidth]{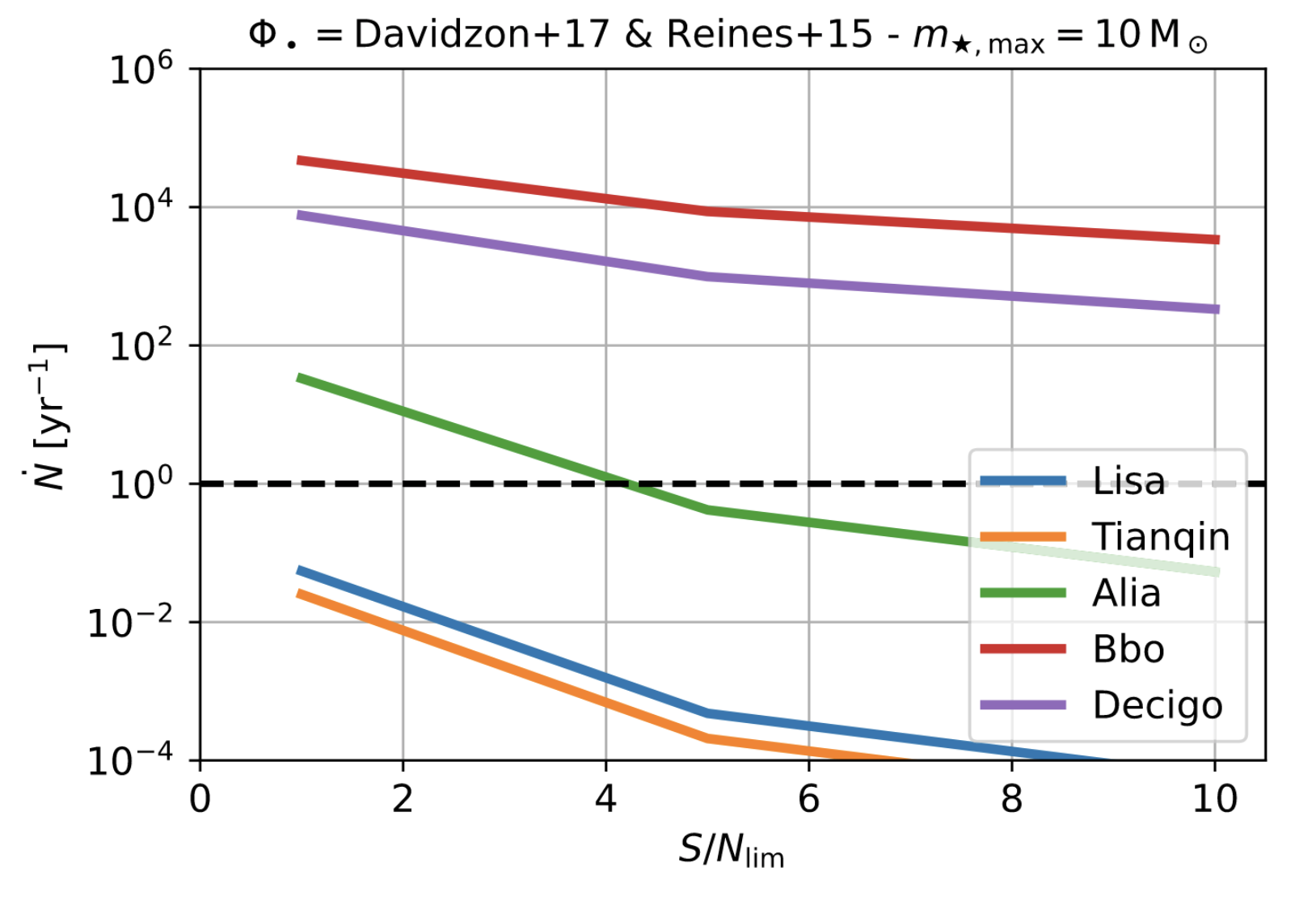}
    \caption{Calculation of the expected TDE rate for future space-based gravitational wave detectors (LISA, Tianqin, Alia, Bbo, Decigo), as a function the signal-to-noise ratio. For this estimate, they adopted Kroupa stellar mass function between $0.08\Ms\leq \mstar \leq 10\Ms$ and a black hole mass function combined by the galaxy stellar mass function from \citet{2017A&A...605A..70D} and the BH mass-galaxy mass relation from \citet{2015ApJ...813...82R}. Figure adapted from \citet{2022MNRAS.510.2025P}.}
    \label{fig:gw}
\end{figure}

Regarding the detection rate of TDE signals with current and future generation detectors, \citet{2022MNRAS.510.2025P} perform the most comprehensive study to date, by combining the expected GW and EM signals with TDE rate estimates. As illustrated in Figure~\ref{fig:gw}, the take-away message is that it is highly unlikely for LISA and TIANQIN \citep{2016CQGra..33c5010L} to detect individual TDEs (nor, as previously discussed, the TDE background). Future generation detectors, on the other hand, will detect 100s -- 1000s of TDEs over a large redshift range, and the planned EM facilities will enable true multi-messenger TDE astronomy by concurrently detecting their optical/UV/X-ray counterparts.

\section{Repeating nuclear transients and partial tidal disruption events}
\label{sec:ptdes}
In the previous Sections we have touched upon the concept of non-fatal encounters between stars and SMBHs. Here, because the rates of the partial tidal disruption events (pTDEs) are expected to comparable to or more frequent than full disruption events, primarily due to larger encounter cross sections \citep{2020ApJ...904...68K,2022ApJ...933...96Z,2023MNRAS.524.3026B} and they may have a potential connection with the origin of observed repeated bursts, we digress briefly into the observed phenomenology that has been linked with pTDEs. 
On the one hand, there are the subtle variations in the predictions for the behaviour of the fall-back rate (which is expected to follow a steeper power law rather than t$^{-5/3}$, e.g. \citet{2013ApJ...767...25G, 2019ApJ...883L..17C,2019ApJ...872..163G,2019MNRAS.487..981G,2020ApJ...904..100R}), which is generally assumed to translate into differences in light curve evolution (see \citet{2020arXiv200512528R} for an excellent review). \citet{2023ApJ...942....9H} analysed a sample of 30 TDEs, and modelling their light curves with a power-law decay found decay indices ranging from --0.8 to --3, hence encompassing the range of both full and partial TDEs. 
More detailed light curve modelling has also led to some claims for candidate pTDEs, usually based on inference of the impact parameter $\beta$ being lower than required for a full disruption (e.g. \citet{2020MNRAS.497.1925G,2022MNRAS.515.5604N}). However, the intrinsic light curve scatter, the possible dependence on the wavelengths sampled by the data, and the exact coupling between the fallback rate and the light curve behaviour complicate the interpretation of these results.

On the other hand, one can envisage a less subtle consequence of a pTDE if the surviving stellar core settles into a bound orbit around the SMBH. When the stellar remnant returns to pericentre again, it will experience another close encounter with the SMBH, leading to a partial or full disruption (e.g., \citet{2017A&A...600A.124M}). Nominally, when a single star on a near parabolic initial trajectory experiences a pTDE, the energy of the new orbit is set by tidal dissipation, which can be at most the binding energy of the star (assuming that the initial orbital energy is equal to 0, which corresponds to a parabolic orbit) but is much smaller in practice (\citet{2023MNRAS.520L..38C}; of order a few per cent of the binding energy) and asymmetric mass loss, which tends to exert a momentum kick on the remnant away from the SMBH. Depending on which effects dominate, the remnant can be bound or unbound \citep{2020ApJ...904..100R}. Typically, the orbital period of the remnant is likely to be very long ($\gtrsim$1\,000 years). One can immediately see the practical problem in trying to identify repeating pTDEs: it requires monitoring on very long timescales. 

Nevertheless, favourable circumstances exist that can dramatically shorten the orbital period of the remnant following a pTDE. The most efficient of these occurs\footnote{We note that another potential mechanism to shorten the orbital period is GW emission. However, \citet{2023MNRAS.520L..38C} show that given the large expected orbital apocenters (of order 1-10 pc) it is highy likely that such an orbit would experience further perturbations within the nuclear star cluster, and hence not lead to repeated pTDEs.} when the disrupted star lives in a binary star system, rather than in isolation. In this case, the companion star is imparted the bulk of the orbital energy during the encounter. The resulting \citet{1975AJ.....80..809H} mechanism leads to one so-called hyper-velocity star (the companion that received the energy kick and is flung out at high velocity) and a star that is tightly bound to the SMBH. It is this latter component that will be subject to repeated tidal stripping, which due to the much tighter orbital configuration can occur on timescales of months--years. The new orbital period depends on the impact parameter, the initial binary separation, the SMBH--star mass ratio and the stellar structure; in practice, orbits with periods as short as hours and as long as years are plausible outcomes.
With the advent of large sky surveys, and increasing attention by astronomers, identifying repetitive behaviour on these timescales is becoming possible. 

Observational evidence is mounting to support the existence of systems which could be powered by repeating partial TDEs, although alternative explanations exist and their nature remains a topic of intense debate. \citet{Miniutti:19} reported the existence of quasi-periodic X-ray eruptions from the nucleus of a nearby galaxy with a recurrence time of $\sim$ a day; additional systems which repeat over timescales as short as 2 hours and as long as 30 years have also been discovered \citep{Campana:15, Giustini:20, Arcodia:21, Payne:21, Wevers:23, Liu:23, Evans:23, Guolo:23, Malyali:23} at X-ray and optical wavelengths. Several of these show a distinct evolution where the recurrence period decreases over time, which could indicate energy dissipation through GW emission (although this is unlikely to be the complete explanation, \citet{Payne:21, 2023MNRAS.520L..38C}) as discussed in the previous Section and Chapter. The prospect of identifying electromagnetic counterparts to sources of GW emission will undoubtedly fuel additional efforts to fully characterize the largely unexplored parameter space of repeating transients in the nuclei of galaxies, taking advantage of the data that will be provided by new large scale surveys at optical and X-ray wavelengths.

\section{Tidal disruption events by stellar-mass black holes}\label{sec:microTDE}

Stellar-mass {BH}s are the natural end product of massive stars. Despite several mechanisms for the ejection of BHs, recent simulation studies suggest that a number of stellar-mass BHs can be retained in clusters \citep{2015ApJ...800....9M,2016PhRvD..93h4029R,2018MNRAS.479.4652A,2020ApJ...898..162W,2009ApJ...692..917M,2021Symm...13.1678M,2021Symm...13.1678M}. In dense stellar clusters such as globular clusters or nuclear stellar clusters, mass segregation will bring them to the central part of the host where they frequently undergo close encounters and collisions with more abundant normal stars. When stars pass too close to the stellar-mass {BH}s, form a triple system with the {BH}s \citep{2016arXiv161000593M,2017PhRvL.118o1101S,2019MNRAS.489..727F}, or undergo dynamical three-body interactions with one or two BHs \citep{2019ApJ...877...56L,2022MNRAS.516.2204R,2023MNRAS.519.5787R,2023MNRAS.tmp.1885R,2023arXiv230703097R}, they can be tidally disrupted \citep[also called ``micro-''TDEs][]{2016ApJ...823..113P}.

Although the overall debris evolution is qualitatively the same in micro-TDEs and TDEs by SMBHs, a few quantitative differences result in different observational signatures. First, the evolution time scales of micro-TDEs are shorter as the dynamical time scales of gas around the BH become shorter for smaller 
 $\mbh$. For example, the peak mass return time (Equation~\ref{eq:peak_t}) is hours for stellar-mass BHs, whereas that is weeks to months for SMBHs. 
Second, forming an accretion disk would be more efficient in micro-TDEs. What primarily determines the pace of the accretion disk formation is how much orbital energy debris should lose to circularize and how efficiently the orbital energy is dissipated. Unlike very eccentric orbits of the returning debris in TDEs by SMBHs (Equation~\ref{eq:ecc}), the bound debris in micro-TDEs is already on an orbit close to circular, which would require a relatively small amount of energy loss for its full circularization. If the incoming stellar orbit is parabolic\footnote{Assuming the typical kinetic energy of stars in clusters is $\simeq 0.5\sigma^{2}$, one can estimate that the eccentricity of two-body encounters with $r_{\rm p}\simeq \rtidal$ with velocity dispersion $\sigma$ is $|1-e| \simeq 10^{-4} ( \sigma /15 \km ~ {\rm s}^{-1})^{2} (\mbh/20\Ms)^{-2/3}(\mstar/1\Ms)^{-1/3}(\rstar/1\Rs)$ for $\mbh \gg \mstar$.}, $1-e \simeq 0.7 \beta^{-1}(\mbh /20 \Ms)^{-1/3}(\mstar /1\Ms)^{1/3}$.
Furthermore, the shorter peak mass return time for micro-TDEs indicates a more rapid accumulation of gas around the BH; the peak mass return rate ($\propto t_{\rm peak}^{-1}$) for micro-TDEs is $O(10^{2})\Ms\yr^{-1}$ whereas that is $O(1)\Ms\yr^{-1}$ for ordinary TDEs. The larger return rate indicates that interactions between gas streams would be more violent and dissipative on a shorter time scale. Lastly, unlike SMBH-driven TDEs, micro-TDEs can give a significant momentum kick to the motion of the stellar-mass BHs \citep{2022ApJ...933..203K}, resulting in active runaway BHs \citep{2023MNRAS.tmp...76R}. For TDEs in close interactions between compact object binaries and a star, such a momentum kick can affect the orbit of the binaries significantly \citep{2022MNRAS.516.2204R}.

Given the differences in the evolution of debris, the observational signatures from micro-TDEs would be distinctive from ordinary TDEs. Assuming the accretion time is set by the viscous time scale for a geometrically thick disk that rapidly forms, \citet{2016ApJ...823..113P} expected that gamma-ray or X-ray flares, depending on the radiation efficiency of accretion, may be generated in micro-TDEs over tens of minutes to hours, resembling ultra-long gamma-ray bursts/X-ray flashes. Building upon the order-of-magnitude estimates, \citet{2019ApJ...881...75K} carried out a calculation for the radiation hydrodynamic evolution of winds from hyper-Eddington accretion, and found that ejected mass can generate a optical/UV flare with a peak luminosity above Eddington lasting roughly 10 - 100 days. Later, \citet{2022arXiv220112368K} performed a suite of smoothed particle hydrodynamics simulations for collisions and tidal disruption events of stars by stellar mass black holes. Because accretion and radiation are not included in their calculations, they predicted possible electromagnetic signatures by analyzing the structure of the material bound to the black holes. Assuming the accretion rate is set by the infall of bound material on a viscous timescale and depends on the radius from the black hole, they expected a peak luminosity of $\sim10^{46\pm1}\erg/\second$ which will mostly come out in the X-ray band. However, more detailed studies for the light curves of micro-TDEs are necessary to reliably identify them among other transients. 

There have been a handful of attempts to estimate the rate of micro-TDEs for different types of star clusters. \citet{2016ApJ...823..113P} analytically estimated a lower limit of the TDE rate for BH $\simeq$ a few $10^{-6}$ $\yr^{-1}$ per Milky Way (MW) galaxy using typical globular cluster parameters. Assuming the MW galaxy density of one per $(4.4\Mpc)^{3}$, the rate becomes roughly $\simeq10\Gpc^{-3}\yr^{-1}$. Using Monte-Carlo simulations, \citet{2019ApJ...881...75K} found a similar rate ($\simeq 3\Gpc^{-3}\yr^{-1}$) for typical globular clusters in the local universe, which inclines as redshift increases to $25-70\Gpc^{-3}\yr^{-1}$ until $z\simeq4$, then declines. The rate for BHs in nuclear stellar clusters is found to be roughly comparable ($1-10\Gpc^{-3}\yr^{-1}$ or $10^{-7}-10^{-6}\yr^{-1}$ per galaxy, see e.g. \cite{2019MNRAS.489..727F}), but that in young stellar clusters may be larger ($\simeq 200\Gpc^{-3}\yr^{-1}$, \cite{2021ApJ...911..104K}).

Recently, micro-TDEs in AGN disks have attracted a great deal of attention as a transient factory that can create transient phenomena, including gravitational wave emission and micro-TDEs \footnote{TDEs by SMBHs in AGNs are also possible, and their observational signatures are likely to be different from those that happen in a vacuum (e.g., \citet{2019ApJ...881..113C,2022MNRAS.514.4102M})}. AGN disks can impact the dynamics of stellar-mass objects as their orbital energy is dissipated every time they cross \citep{2005ApJ...619...30M}. The angular torques by the disk make their orbits less eccentric (relative to the SMBH) and inclined relative to the disk. They are eventually entrained into the disk if their orbits become fully circularized \citep{1995MNRAS.275..628R}, after which they migrate inwards. As the stellar-mass objects accumulate in the inner AGN disk, stars can be disrupted by stellar-mass black holes through close encounters. Alternatively, TDEs can occur in star-BH binaries which could be eccentric at birth and/or become eccentric via gas-binary interactions \citep{2021ApJ...909L..13Z,2021ApJ...914L..21D}. Detailed hydrodynamics calculations for TDEs in AGN disks will provide more accurate estimates of their observables. \citet{2021arXiv210502342Y} estimated that the rate of micro-TDEs in AGN disks is $\sim170\Gpc^{-3}\yr^{-1}$.

\section{ Summary}\label{sec:summary}

Tidal disruption events provide an excellent tool to detect and study otherwise quiescent supermassive black holes and their close environments, both in the local Universe and at cosmological distances. Their optical/UV/X-ray light curves can be modeled to constrain the fundamental properties of these black holes. Furthermore, X-ray spectral fitting of sophisticated accretion disk models can provide an independent measurement of these quantities, making TDEs a promising tool to study SMBH demographics in the era of large surveys such as the Legacy Survey for Space and Time and the eROSITA mission, when supplemented by the next generation of X-ray telescopes such as Athena and other mission concepts currently being reviewed (e.g. the High Energy X-ray Probe or HEX-P).

Our theoretical understanding of these events at the first pericenter passage is relatively well established. A number of numerical simulations dedicated to simulating the disruption of stars in the early stages made significant improvement beyond the conventional picture. However, the long-term evolution of debris, which is directly related to the observables of these events, remains yet to be fully understood. Several open questions such as the main source that powers optical/UV TDEs, magnetic field structure and enhancement in debris in relation to jetted TDEs, the role of recombination in shaping the debris structure and its long-term effects on disk formation, can be only properly answered via global magnetohydrodynamics simulations with astrophysically realistic initial conditions. Although it is still notoriously challenging, novel numerical techniques \citep{2020SSRv..216...88K} seem to show the possibility of such simulations with reasonable computational costs. 

Studies on TDEs by stellar-mass BHs have been gaining momentum recently. Such events can naturally occur in close multi-body encounters between stars and compact objects in star clusters or AGN disks. Unlike TDEs by SMBHs, multi-body interactions involving binaries can produce TDEs that can constitute a non-negligible fraction of the entire micro-TDE population. Given the chaotic nature of three-body encounters, TDEs by stellar-mass BHs in such interactions would reveal unique observational signatures that can not be described by TDEs in two-body interactions. 

We are living in an exciting era where ongoing and future transient surveys will detect TDEs at unprecedented rates. The impending growth of TDE candidates will provide an amazing opportunity for us to deepen our understanding of TDEs as observations will confirm or confront the existing theoretical models. In addition, observations of various signals 
 of disparate nature will bring up new aspects of TDEs as multi-messenger sources. In the mean time, to grasp such opportunities, it is important to develop reliable automated identification algorithms for TDEs to facilitate the multi-wavelength follow-up observations necessary to drive theoretical advances in the field. 

\section*{Acknowledgments}
This work is part of an invited review chapter for a book entitled {\it Black holes in the era of multi-messenger astronomy}. We warmly thank the editors for their effort in putting together this resource. 

\bibliographystyle{apj} 
\bibliography{bib.bib,bib2.bib}

\begin{thebibliography}{}
\expandafter\ifx\csname natexlab\endcsname\relax\def\natexlab#1{#1}\fi

\bibitem[{198(????)}]{1984PhST....7..209T}
 ????

\bibitem[{{Aartsen} {et~al.}(2013){Aartsen}, {Abbasi}, {Abdou}, {Ackermann},
  {Adams}, {Aguilar}, {Ahlers}, {Altmann}, {Auffenberg}, {Bai}, {Baker},
  {Barwick}, {Baum}, {Bay}, {Beatty}, {Bechet}, {Becker Tjus}, {Becker},
  {Bell}, {Benabderrahmane}, {BenZvi}, {Berdermann}, {Berghaus}, {Berley},
  {Bernardini}, {Bernhard}, {Bertrand}, {Besson}, {Binder}, {Bindig}, {Bissok},
  {Blaufuss}, {Blumenthal}, {Boersma}, {Bohaichuk}, {Bohm}, {Bose},
  {B{\"o}ser}, {Botner}, {Brayeur}, {Bretz}, {Brown}, {Bruijn}, {Brunner},
  {Carson}, {Casey}, {Casier}, {Chirkin}, {Christov}, {Christy}, {Clark},
  {Clevermann}, {Coenders}, {Cohen}, {Cowen}, {Cruz Silva}, {Danninger},
  {Daughhetee}, {Davis}, {De Clercq}, {De Ridder}, {Desiati}, {de With},
  {DeYoung}, {D{\'\i}az-V{\'e}lez}, {Dunkman}, {Eagan}, {Eberhardt}, {Eisch},
  {Ellsworth}, {Euler}, {Evenson}, {Fadiran}, {Fazely}, {Fedynitch},
  {Feintzeig}, {Feusels}, {Filimonov}, {Finley}, {Fischer-Wasels}, {Flis},
  {Franckowiak}, {Franke}, {Frantzen}, {Fuchs}, {Gaisser}, {Gallagher},
  {Gerhardt}, {Gladstone}, {Gl{\"u}senkamp}, {Goldschmidt}, {Golup},
  {Gonzalez}, {Goodman}, {G{\'o}ra}, {Grant}, {Gro{\ss}}, {Gurtner}, {Ha}, {Haj
  Ismail}, {Hallen}, {Hallgren}, {Halzen}, {Hanson}, {Heereman}, {Heinen},
  {Helbing}, {Hellauer}, {Hickford}, {Hill}, {Hoffman}, {Hoffmann}, {Homeier},
  {Hoshina}, {Huelsnitz}, {Hulth}, {Hultqvist}, {Hussain}, {Ishihara},
  {Jacobi}, {Jacobsen}, {Jagielski}, {Japaridze}, {Jero}, {Jlelati},
  {Kaminsky}, {Kappes}, {Karg}, {Karle}, {Kelley}, {Kiryluk}, {Kislat},
  {Kl{\"a}s}, {Klein}, {K{\"o}hne}, {Kohnen}, {Kolanoski}, {K{\"o}pke},
  {Kopper}, {Kopper}, {Koskinen}, {Kowalski}, {Krasberg}, {Krings}, {Kroll},
  {Kunnen}, {Kurahashi}, {Kuwabara}, {Labare}, {Landsman}, {Larson},
  {Lesiak-Bzdak}, {Leuermann}, {Leute}, {L{\"u}nemann}, {Madsen}, {Maruyama},
  {Mase}, {Matis}, {McNally}, {Meagher}, {Merck}, {M{\'e}sz{\'a}ros}, {Meures},
  {Miarecki}, {Middell}, {Milke}, {Miller}, {Mohrmann}, {Montaruli}, {Morse},
  {Nahnhauer}, {Naumann}, {Niederhausen}, {Nowicki}, {Nygren}, {Obertacke},
  {Odrowski}, {Olivas}, {Olivo}, {O'Murchadha}, {Paul}, {Pepper}, {P{\'e}rez de
  los Heros}, {Pfendner}, {Pieloth}, {Pinat}, {Pirk}, {Posselt}, {Price},
  {Przybylski}, {R{\"a}del}, {Rameez}, {Rawlins}, {Redl}, {Reimann}, {Resconi},
  {Rhode}, {Ribordy}, {Richman}, {Riedel}, {Rodrigues}, {Rott}, {Ruhe},
  {Ruzybayev}, {Ryckbosch}, {Saba}, {Salameh}, {Sander}, {Santander}, {Sarkar},
  {Schatto}, {Scheel}, {Scheriau}, {Schmidt}, {Schmitz}, {Schoenen},
  {Sch{\"o}neberg}, {Sch{\"o}nwald}, {Schukraft}, {Schulte}, {Schulz},
  {Seckel}, {Sestayo}, {Seunarine}, {Sheremata}, {Smith}, {Soiron}, {Soldin},
  {Spiczak}, {Spiering}, {Stamatikos}, {Stanev}, {Stasik}, {Stezelberger},
  {Stokstad}, {St{\"o}{\ss}l}, {Strahler}, {Str{\"o}m}, {Sullivan}, {Taavola},
  {Taboada}, {Tamburro}, {Ter-Antonyan}, {Te{\v{s}}i{\'c}}, {Tilav}, {Toale},
  {Toscano}, {Usner}, {van der Drift}, {van Eijndhoven}, {Van Overloop}, {van
  Santen}, {Vehring}, {Voge}, {Vraeghe}, {Walck}, {Waldenmaier}, {Wallraff},
  {Wasserman}, {Weaver}, {Wellons}, {Wendt}, {Westerhoff}, {Whitehorn},
  {Wiebe}, {Wiebusch}, {Williams}, {Wissing}, {Wolf}, {Wood}, {Woschnagg},
  {Xu}, {Xu}, {Xu}, {Yanez}, {Yodh}, {Yoshida}, {Zarzhitsky}, {Ziemann},
  {Zierke}, {Zilles}, \& {Zoll}}]{2013PhRvL.111b1103A}
{Aartsen}, M.~G., {Abbasi}, R., {Abdou}, Y., {et~al.} 2013, \prl, 111, 021103

\bibitem[{{Alexander} {et~al.}(2020){Alexander}, {van Velzen}, {Horesh}, \&
  {Zauderer}}]{2020SSRv..216...81A}
{Alexander}, K.~D., {van Velzen}, S., {Horesh}, A., \& {Zauderer}, B.~A. 2020,
  \ssr, 216, 81

\bibitem[{{Alexander} \& {Hopman}(2003)}]{2003ApJ...590L..29A}
{Alexander}, T., \& {Hopman}, C. 2003, \apjl, 590, L29

\bibitem[{{Amaro-Seoane} {et~al.}(2007){Amaro-Seoane}, {Gair}, {Freitag},
  {Miller}, {Mandel}, {Cutler}, \& {Babak}}]{2007CQGra..24R.113A}
{Amaro-Seoane}, P., {Gair}, J.~R., {Freitag}, M., {et~al.} 2007, Classical and
  Quantum Gravity, 24, R113

\bibitem[{{Amaro-Seoane} {et~al.}(2017){Amaro-Seoane}, {Audley}, {Babak},
  {Baker}, {Barausse}, {Bender}, {Berti}, {Binetruy}, {Born}, {Bortoluzzi},
  {Camp}, {Caprini}, {Cardoso}, {Colpi}, {Conklin}, {Cornish}, {Cutler},
  {Danzmann}, {Dolesi}, {Ferraioli}, {Ferroni}, {Fitzsimons}, {Gair}, {Gesa
  Bote}, {Giardini}, {Gibert}, {Grimani}, {Halloin}, {Heinzel}, {Hertog},
  {Hewitson}, {Holley-Bockelmann}, {Hollington}, {Hueller}, {Inchauspe},
  {Jetzer}, {Karnesis}, {Killow}, {Klein}, {Klipstein}, {Korsakova}, {Larson},
  {Livas}, {Lloro}, {Man}, {Mance}, {Martino}, {Mateos}, {McKenzie},
  {McWilliams}, {Miller}, {Mueller}, {Nardini}, {Nelemans}, {Nofrarias},
  {Petiteau}, {Pivato}, {Plagnol}, {Porter}, {Reiche}, {Robertson},
  {Robertson}, {Rossi}, {Russano}, {Schutz}, {Sesana}, {Shoemaker}, {Slutsky},
  {Sopuerta}, {Sumner}, {Tamanini}, {Thorpe}, {Troebs}, {Vallisneri},
  {Vecchio}, {Vetrugno}, {Vitale}, {Volonteri}, {Wanner}, {Ward}, {Wass},
  {Weber}, {Ziemer}, \& {Zweifel}}]{2017arXiv170200786A}
{Amaro-Seoane}, P., {Audley}, H., {Babak}, S., {et~al.} 2017, arXiv e-prints,
  arXiv:1702.00786

\bibitem[{{Andreoni} {et~al.}(2022){Andreoni}, {Coughlin}, {Perley}, {Yao},
  {Lu}, {Cenko}, {Kumar}, {Anand}, {Ho}, {Kasliwal}, {de Ugarte Postigo},
  {Sagu{\'e}s-Carracedo}, {Schulze}, {Kann}, {Kulkarni}, {Sollerman}, {Tanvir},
  {Rest}, {Izzo}, {Somalwar}, {Kaplan}, {Ahumada}, {Anupama}, {Auchettl},
  {Barway}, {Bellm}, {Bhalerao}, {Bloom}, {Bremer}, {Bulla}, {Burns},
  {Campana}, {Chandra}, {Charalampopoulos}, {Cooke}, {D'Elia}, {Das}, {Dobie},
  {Ag{\"u}{\'\i} Fern{\'a}ndez}, {Freeburn}, {Fremling}, {Gezari}, {Goode},
  {Graham}, {Hammerstein}, {Karambelkar}, {Kilpatrick}, {Kool}, {Krips},
  {Laher}, {Leloudas}, {Levan}, {Lundquist}, {Mahabal}, {Medford}, {Miller},
  {M{\"o}ller}, {Mooley}, {Nayana}, {Nir}, {Pang}, {Paraskeva}, {Perley},
  {Petitpas}, {Pursiainen}, {Ravi}, {Ridden-Harper}, {Riddle}, {Rigault},
  {Rodriguez}, {Rusholme}, {Sharma}, {Smith}, {Stein}, {Th{\"o}ne},
  {Tohuvavohu}, {Valdes}, {van Roestel}, {Vergani}, {Wang}, \&
  {Zhang}}]{2022arXiv221116530A}
{Andreoni}, I., {Coughlin}, M.~W., {Perley}, D.~A., {et~al.} 2022, \nat, 612,
  430

\bibitem[{{Arca Sedda} {et~al.}(2018){Arca Sedda}, {Askar}, \&
  {Giersz}}]{2018MNRAS.479.4652A}
{Arca Sedda}, M., {Askar}, A., \& {Giersz}, M. 2018, \mnras, 479, 4652

\bibitem[{{Arcodia} {et~al.}(2021){Arcodia}, {Merloni}, {Nandra}, {Buchner},
  {Salvato}, {Pasham}, {Remillard}, {Comparat}, {Lamer}, {Ponti}, {Malyali},
  {Wolf}, {Arzoumanian}, {Bogensberger}, {Buckley}, {Gendreau}, {Gromadzki},
  {Kara}, {Krumpe}, {Markwardt}, {Ramos-Ceja}, {Rau}, {Schramm}, \&
  {Schwope}}]{Arcodia:21}
{Arcodia}, R., {Merloni}, A., {Nandra}, K., {et~al.} 2021, \nat, 592, 704

\bibitem[{{Auchettl} {et~al.}(2017){Auchettl}, {Guillochon}, \&
  {Ramirez-Ruiz}}]{2017ApJ...838..149A}
{Auchettl}, K., {Guillochon}, J., \& {Ramirez-Ruiz}, E. 2017, \apj, 838, 149

\bibitem[{{Bade} {et~al.}(1996){Bade}, {Komossa}, \&
  {Dahlem}}]{1996A&A...309L..35B}
{Bade}, N., {Komossa}, S., \& {Dahlem}, M. 1996, \aap, 309, L35

\bibitem[{{Bahcall} \& {Wolf}(1976)}]{1976ApJ...209..214B}
{Bahcall}, J.~N., \& {Wolf}, R.~A. 1976, \apj, 209, 214

\bibitem[{{Bellm} {et~al.}(2019){Bellm}, {Kulkarni}, {Graham}, {Dekany},
  {Smith}, {Riddle}, {Masci}, {Helou}, {Prince}, {Adams}, {Barbarino},
  {Barlow}, {Bauer}, {Beck}, {Belicki}, {Biswas}, {Blagorodnova}, {Bodewits},
  {Bolin}, {Brinnel}, {Brooke}, {Bue}, {Bulla}, {Burruss}, {Cenko}, {Chang},
  {Connolly}, {Coughlin}, {Cromer}, {Cunningham}, {De}, {Delacroix}, {Desai},
  {Duev}, {Eadie}, {Farnham}, {Feeney}, {Feindt}, {Flynn}, {Franckowiak},
  {Frederick}, {Fremling}, {Gal-Yam}, {Gezari}, {Giomi}, {Goldstein},
  {Golkhou}, {Goobar}, {Groom}, {Hacopians}, {Hale}, {Henning}, {Ho}, {Hover},
  {Howell}, {Hung}, {Huppenkothen}, {Imel}, {Ip}, {Ivezi{\'c}}, {Jackson},
  {Jones}, {Juric}, {Kasliwal}, {Kaspi}, {Kaye}, {Kelley}, {Kowalski},
  {Kramer}, {Kupfer}, {Landry}, {Laher}, {Lee}, {Lin}, {Lin}, {Lunnan},
  {Giomi}, {Mahabal}, {Mao}, {Miller}, {Monkewitz}, {Murphy}, {Ngeow},
  {Nordin}, {Nugent}, {Ofek}, {Patterson}, {Penprase}, {Porter}, {Rauch},
  {Rebbapragada}, {Reiley}, {Rigault}, {Rodriguez}, {van Roestel}, {Rusholme},
  {van Santen}, {Schulze}, {Shupe}, {Singer}, {Soumagnac}, {Stein}, {Surace},
  {Sollerman}, {Szkody}, {Taddia}, {Terek}, {Van Sistine}, {van Velzen},
  {Vestrand}, {Walters}, {Ward}, {Ye}, {Yu}, {Yan}, \&
  {Zolkower}}]{2019PASP..131a8002B}
{Bellm}, E.~C., {Kulkarni}, S.~R., {Graham}, M.~J., {et~al.} 2019, \pasp, 131,
  018002

\bibitem[{{Biehl} {et~al.}(2018){Biehl}, {Boncioli}, {Lunardini}, \&
  {Winter}}]{2018NatSR...810828B}
{Biehl}, D., {Boncioli}, D., {Lunardini}, C., \& {Winter}, W. 2018, Scientific
  Reports, 8, 10828

\bibitem[{{Blagorodnova} {et~al.}(2017){Blagorodnova}, {Gezari}, {Hung},
  {Kulkarni}, {Cenko}, {Pasham}, {Yan}, {Arcavi}, {Ben-Ami}, {Bue}, {Cantwell},
  {Cao}, {Castro-Tirado}, {Fender}, {Fremling}, {Gal-Yam}, {Ho}, {Horesh},
  {Hosseinzadeh}, {Kasliwal}, {Kong}, {Laher}, {Leloudas}, {Lunnan}, {Masci},
  {Mooley}, {Neill}, {Nugent}, {Powell}, {Valeev}, {Vreeswijk}, {Walters}, \&
  {Wozniak}}]{2017ApJ...844...46B}
{Blagorodnova}, N., {Gezari}, S., {Hung}, T., {et~al.} 2017, \apj, 844, 46

\bibitem[{{Blagorodnova} {et~al.}(2019){Blagorodnova}, {Cenko}, {Kulkarni},
  {Arcavi}, {Bloom}, {Duggan}, {Filippenko}, {Fremling}, {Horesh},
  {Hosseinzadeh}, {Karamehmetoglu}, {Levan}, {Masci}, {Nugent}, {Pasham},
  {Veilleux}, {Walters}, {Yan}, \& {Zheng}}]{2019ApJ...873...92B}
{Blagorodnova}, N., {Cenko}, S.~B., {Kulkarni}, S.~R., {et~al.} 2019, \apj,
  873, 92

\bibitem[{{Blanchard} {et~al.}(2017){Blanchard}, {Nicholl}, {Berger},
  {Guillochon}, {Margutti}, {Chornock}, {Alexander}, {Leja}, \&
  {Drout}}]{2017ApJ...843..106B}
{Blanchard}, P.~K., {Nicholl}, M., {Berger}, E., {et~al.} 2017, \apj, 843, 106

\bibitem[{{Bloom} {et~al.}(2011){Bloom}, {Giannios}, {Metzger}, {Cenko},
  {Perley}, {Butler}, {Tanvir}, {Levan}, {O'Brien}, {Strubbe}, {De Colle},
  {Ramirez-Ruiz}, {Lee}, {Nayakshin}, {Quataert}, {King}, {Cucchiara},
  {Guillochon}, {Bower}, {Fruchter}, {Morgan}, \& {van der
  Horst}}]{2011Sci...333..203B}
{Bloom}, J.~S., {Giannios}, D., {Metzger}, B.~D., {et~al.} 2011, Science, 333,
  203

\bibitem[{{Bonnerot} \& {Stone}(2021)}]{2021SSRv..217...16B}
{Bonnerot}, C., \& {Stone}, N.~C. 2021, \ssr, 217, 16

\bibitem[{{Bortolas} {et~al.}(2023){Bortolas}, {Ryu}, {Broggi}, \&
  {Sesana}}]{2023MNRAS.524.3026B}
{Bortolas}, E., {Ryu}, T., {Broggi}, L., \& {Sesana}, A. 2023, \mnras, 524,
  3026

\bibitem[{{Brassart} \& {Luminet}(2008)}]{2008A&A...481..259B}
{Brassart}, M., \& {Luminet}, J.~P. 2008, \aap, 481, 259

\bibitem[{{Bricman} \& {Gomboc}(2020)}]{2020ApJ...890...73B}
{Bricman}, K., \& {Gomboc}, A. 2020, \apj, 890, 73

\bibitem[{{Brown} {et~al.}(2015){Brown}, {Levan}, {Stanway}, {Tanvir}, {Cenko},
  {Berger}, {Chornock}, \& {Cucchiaria}}]{2015MNRAS.452.4297B}
{Brown}, G.~C., {Levan}, A.~J., {Stanway}, E.~R., {et~al.} 2015, \mnras, 452,
  4297

\bibitem[{{Brown} {et~al.}(2018){Brown}, {Kochanek}, {Holoien}, {Stanek},
  {Auchettl}, {Shappee}, {Prieto}, {Morrell}, {Falco}, {Strader}, {Chomiuk},
  {Post}, {Villanueva}, {Mathur}, {Dong}, {Chen}, \&
  {Bose}}]{2018MNRAS.473.1130B}
{Brown}, J.~S., {Kochanek}, C.~S., {Holoien}, T.~W.~S., {et~al.} 2018, \mnras,
  473, 1130

\bibitem[{{Bugli} {et~al.}(2018){Bugli}, {Guilet}, {M{\"u}ller}, {Del Zanna},
  {Bucciantini}, \& {Montero}}]{2018MNRAS.475..108B}
{Bugli}, M., {Guilet}, J., {M{\"u}ller}, E., {et~al.} 2018, \mnras, 475, 108

\bibitem[{{Burrows} {et~al.}(2011){Burrows}, {Kennea}, {Ghisellini}, {Mangano},
  {Zhang}, {Page}, {Eracleous}, {Romano}, {Sakamoto}, {Falcone}, {Osborne},
  {Campana}, {Beardmore}, {Breeveld}, {Chester}, {Corbet}, {Covino},
  {Cummings}, {D'Avanzo}, {D'Elia}, {Esposito}, {Evans}, {Fugazza}, {Gelbord},
  {Hiroi}, {Holland}, {Huang}, {Im}, {Israel}, {Jeon}, {Jeon}, {Jun}, {Kawai},
  {Kim}, {Krimm}, {Marshall}, {P. M{\'e}sz{\'a}ros}, {Negoro}, {Omodei},
  {Park}, {Perkins}, {Sugizaki}, {Sung}, {Tagliaferri}, {Troja}, {Ueda},
  {Urata}, {Usui}, {Antonelli}, {Barthelmy}, {Cusumano}, {Giommi}, {Melandri},
  {Perri}, {Racusin}, {Sbarufatti}, {Siegel}, \&
  {Gehrels}}]{2011Natur.476..421B}
{Burrows}, D.~N., {Kennea}, J.~A., {Ghisellini}, G., {et~al.} 2011, \nat, 476,
  421

\bibitem[{{Campana} {et~al.}(2015){Campana}, {Mainetti}, {Colpi}, {Lodato},
  {D'Avanzo}, {Evans}, \& {Moretti}}]{Campana:15}
{Campana}, S., {Mainetti}, D., {Colpi}, M., {et~al.} 2015, \aap, 581, A17

\bibitem[{{Cannizzaro} {et~al.}(2021){Cannizzaro}, {Wevers}, {Jonker},
  {P{\'e}rez-Torres}, {Moldon}, {Mata-S{\'a}nchez}, {Leloudas}, {Pasham},
  {Mattila}, {Arcavi}, {Decker French}, {Onori}, {Inserra}, {Nicholl},
  {Gromadzki}, {Chen}, {M{\"u}ller-Bravo}, {Short}, {Anderson}, {Young},
  {Gendreau}, {Arzoumanian}, {L{\"o}wenstein}, {Remillard}, {Roy}, \&
  {Hiramatsu}}]{2021MNRAS.504..792C}
{Cannizzaro}, G., {Wevers}, T., {Jonker}, P.~G., {et~al.} 2021, \mnras, 504,
  792

\bibitem[{{Cannizzo} {et~al.}(1990){Cannizzo}, {Lee}, \&
  {Goodman}}]{1990ApJ...351...38C}
{Cannizzo}, J.~K., {Lee}, H.~M., \& {Goodman}, J. 1990, \apj, 351, 38

\bibitem[{{Carter} \& {Luminet}(1983)}]{1983A&A...121...97C}
{Carter}, B., \& {Luminet}, J.~P. 1983, \aap, 121, 97

\bibitem[{{Carter} \& {Luminet}(1985)}]{1985MNRAS.212...23C}
---. 1985, \mnras, 212, 23

\bibitem[{{Cendes} {et~al.}(2021){Cendes}, {Alexander}, {Berger}, {Eftekhari},
  {Williams}, \& {Chornock}}]{2021ApJ...919..127C}
{Cendes}, Y., {Alexander}, K.~D., {Berger}, E., {et~al.} 2021, \apj, 919, 127

\bibitem[{{Cenko} {et~al.}(2012){Cenko}, {Krimm}, {Horesh}, {Rau}, {Frail},
  {Kennea}, {Levan}, {Holland}, {Butler}, {Quimby}, {Bloom}, {Filippenko},
  {Gal-Yam}, {Greiner}, {Kulkarni}, {Ofek}, {Olivares E.}, {Schady},
  {Silverman}, {Tanvir}, \& {Xu}}]{2012ApJ...753...77C}
{Cenko}, S.~B., {Krimm}, H.~A., {Horesh}, A., {et~al.} 2012, \apj, 753, 77

\bibitem[{{Cenko} {et~al.}(2016){Cenko}, {Cucchiara}, {Roth}, {Veilleux},
  {Prochaska}, {Yan}, {Guillochon}, {Maksym}, {Arcavi}, {Butler}, {Filippenko},
  {Fruchter}, {Gezari}, {Kasen}, {Levan}, {Miller}, {Pasham}, {Ramirez-Ruiz},
  {Strubbe}, {Tanvir}, \& {Tombesi}}]{2016ApJ...818L..32C}
{Cenko}, S.~B., {Cucchiara}, A., {Roth}, N., {et~al.} 2016, \apjl, 818, L32

\bibitem[{{Chan} {et~al.}(2019){Chan}, {Piran}, {Krolik}, \&
  {Saban}}]{2019ApJ...881..113C}
{Chan}, C.-H., {Piran}, T., {Krolik}, J.~H., \& {Saban}, D. 2019, \apj, 881,
  113

\bibitem[{{Charalampopoulos} {et~al.}(2023){Charalampopoulos}, {Pursiainen},
  {Leloudas}, {Arcavi}, {Newsome}, {Schulze}, {Burke}, \&
  {Nicholl}}]{2022arXiv220912913C}
{Charalampopoulos}, P., {Pursiainen}, M., {Leloudas}, G., {et~al.} 2023, \aap,
  673, A95

\bibitem[{{Charalampopoulos} {et~al.}(2022){Charalampopoulos}, {Leloudas},
  {Malesani}, {Wevers}, {Arcavi}, {Nicholl}, {Pursiainen}, {Lawrence},
  {Anderson}, {Benetti}, {Cannizzaro}, {Chen}, {Galbany}, {Gromadzki},
  {Guti{\'e}rrez}, {Inserra}, {Jonker}, {M{\"u}ller-Bravo}, {Onori}, {Short},
  {Sollerman}, \& {Young}}]{2022A&A...659A..34C}
{Charalampopoulos}, P., {Leloudas}, G., {Malesani}, D.~B., {et~al.} 2022, \aap,
  659, A34

\bibitem[{{Cheng} \& {Bogdanovi{\'c}}(2014)}]{2014PhRvD..90f4020C}
{Cheng}, R.~M., \& {Bogdanovi{\'c}}, T. 2014, \prd, 90, 064020

\bibitem[{{Cheng} \& {Evans}(2013)}]{2013PhRvD..87j4010C}
{Cheng}, R.~M., \& {Evans}, C.~R. 2013, \prd, 87, 104010

\bibitem[{{Coughlin} \& {Nixon}(2019)}]{2019ApJ...883L..17C}
{Coughlin}, E.~R., \& {Nixon}, C.~J. 2019, \apjl, 883, L17

\bibitem[{{Cufari} {et~al.}(2023){Cufari}, {Nixon}, \&
  {Coughlin}}]{2023MNRAS.520L..38C}
{Cufari}, M., {Nixon}, C.~J., \& {Coughlin}, E.~R. 2023, \mnras, 520, L38

\bibitem[{{Dai} \& {Fang}(2017)}]{2017MNRAS.469.1354D}
{Dai}, L., \& {Fang}, K. 2017, \mnras, 469, 1354

\bibitem[{{Dai} {et~al.}(2018){Dai}, {McKinney}, {Roth}, {Ramirez-Ruiz}, \&
  {Miller}}]{2018ApJ...859L..20D}
{Dai}, L., {McKinney}, J.~C., {Roth}, N., {Ramirez-Ruiz}, E., \& {Miller},
  M.~C. 2018, \apjl, 859, L20

\bibitem[{{Davidzon} {et~al.}(2017){Davidzon}, {Ilbert}, {Laigle}, {Coupon},
  {McCracken}, {Delvecchio}, {Masters}, {Capak}, {Hsieh}, {Le F{\`e}vre},
  {Tresse}, {Bethermin}, {Chang}, {Faisst}, {Le Floc'h}, {Steinhardt}, {Toft},
  {Aussel}, {Dubois}, {Hasinger}, {Salvato}, {Sanders}, {Scoville}, \&
  {Silverman}}]{2017A&A...605A..70D}
{Davidzon}, I., {Ilbert}, O., {Laigle}, C., {et~al.} 2017, \aap, 605

\bibitem[{{Davies} \& {King}(2005)}]{2005ApJ...624L..25D}
{Davies}, M.~B., \& {King}, A. 2005, \apjl, 624, L25

\bibitem[{{De Colle} \& {Lu}(2020)}]{2020NewAR..8901538D}
{De Colle}, F., \& {Lu}, W. 2020, \nar, 89, 101538

\bibitem[{{Di Stefano} {et~al.}(2001){Di Stefano}, {Greiner}, {Murray}, \&
  {Garcia}}]{2001ApJ...551L..37D}
{Di Stefano}, R., {Greiner}, J., {Murray}, S., \& {Garcia}, M. 2001, \apjl,
  551, L37

\bibitem[{{Diener} {et~al.}(1995){Diener}, {Kosovichev}, {Kotok}, {Novikov}, \&
  {Pethick}}]{1995MNRAS.275..498D}
{Diener}, P., {Kosovichev}, A.~G., {Kotok}, E.~V., {Novikov}, I.~D., \&
  {Pethick}, C.~J. 1995, \mnras, 275, 498

\bibitem[{{D'Orazio} \& {Duffell}(2021)}]{2021ApJ...914L..21D}
{D'Orazio}, D.~J., \& {Duffell}, P.~C. 2021, \apjl, 914, L21

\bibitem[{{Dou} {et~al.}(2016){Dou}, {Wang}, {Jiang}, {Yang}, {Lyu}, \&
  {Zhou}}]{2016ApJ...832..188D}
{Dou}, L., {Wang}, T.-g., {Jiang}, N., {et~al.} 2016, \apj, 832, 188

\bibitem[{{Evans} \& {Kochanek}(1989)}]{1989ApJ...346L..13E}
{Evans}, C.~R., \& {Kochanek}, C.~S. 1989, \apjl, 346, L13

\bibitem[{{Evans} {et~al.}(2023){Evans}, {Nixon}, {Campana},
  {Charalampopoulos}, {Perley}, {Breeveld}, {Page}, {Oates}, {Eyles-Ferris},
  {Malesani}, {Izzo}, {Goad}, {O'Brien}, {Osborne}, \& {Sbarufatti}}]{Evans:23}
{Evans}, P.~A., {Nixon}, C.~J., {Campana}, S., {et~al.} 2023, arXiv e-prints,
  arXiv:2309.02500

\bibitem[{{Faber} {et~al.}(2005){Faber}, {Rasio}, \&
  {Willems}}]{2005Icar..175..248F}
{Faber}, J.~A., {Rasio}, F.~A., \& {Willems}, B. 2005, \icarus, 175, 248

\bibitem[{{Farrar} \& {Gruzinov}(2009)}]{2009ApJ...693..329F}
{Farrar}, G.~R., \& {Gruzinov}, A. 2009, \apj, 693, 329

\bibitem[{{Farrar} \& {Piran}(2014)}]{2014arXiv1411.0704F}
{Farrar}, G.~R., \& {Piran}, T. 2014, arXiv e-prints, arXiv:1411.0704

\bibitem[{{Fragione} {et~al.}(2019){Fragione}, {Leigh}, {Perna}, \&
  {Kocsis}}]{2019MNRAS.489..727F}
{Fragione}, G., {Leigh}, N. W.~C., {Perna}, R., \& {Kocsis}, B. 2019, \mnras,
  489, 727

\bibitem[{{Frank} \& {Rees}(1976)}]{1976MNRAS.176..633F}
{Frank}, J., \& {Rees}, M.~J. 1976, \mnras, 176, 633

\bibitem[{{Frederick} {et~al.}(2021){Frederick}, {Gezari}, {Graham},
  {Sollerman}, {van Velzen}, {Perley}, {Stern}, {Ward}, {Hammerstein}, {Hung},
  {Yan}, {Andreoni}, {Bellm}, {Duev}, {Kowalski}, {Mahabal}, {Masci},
  {Medford}, {Rusholme}, {Smith}, \& {Walters}}]{2021ApJ...920...56F}
{Frederick}, S., {Gezari}, S., {Graham}, M.~J., {et~al.} 2021, \apj, 920, 56

\bibitem[{{Gafton} \& {Rosswog}(2019)}]{2019MNRAS.487.4790G}
{Gafton}, E., \& {Rosswog}, S. 2019, \mnras, 487, 4790

\bibitem[{{Gehrels} {et~al.}(2004){Gehrels}, {Chincarini}, {Giommi}, {Mason},
  {Nousek}, {Wells}, {White}, {Barthelmy}, {Burrows}, {Cominsky}, {Hurley},
  {Marshall}, {M{\'e}sz{\'a}ros}, {Roming}, {Angelini}, {Barbier}, {Belloni},
  {Campana}, {Caraveo}, {Chester}, {Citterio}, {Cline}, {Cropper}, {Cummings},
  {Dean}, {Feigelson}, {Fenimore}, {Frail}, {Fruchter}, {Garmire}, {Gendreau},
  {Ghisellini}, {Greiner}, {Hill}, {Hunsberger}, {Krimm}, {Kulkarni}, {Kumar},
  {Lebrun}, {Lloyd-Ronning}, {Markwardt}, {Mattson}, {Mushotzky}, {Norris},
  {Osborne}, {Paczynski}, {Palmer}, {Park}, {Parsons}, {Paul}, {Rees},
  {Reynolds}, {Rhoads}, {Sasseen}, {Schaefer}, {Short}, {Smale}, {Smith},
  {Stella}, {Tagliaferri}, {Takahashi}, {Tashiro}, {Townsley}, {Tueller},
  {Turner}, {Vietri}, {Voges}, {Ward}, {Willingale}, {Zerbi}, \&
  {Zhang}}]{2004ApJ...611.1005G}
{Gehrels}, N., {Chincarini}, G., {Giommi}, P., {et~al.} 2004, \apj, 611, 1005

\bibitem[{{Gendreau} {et~al.}(2012){Gendreau}, {Arzoumanian}, \&
  {Okajima}}]{2012SPIE.8443E..13G}
{Gendreau}, K.~C., {Arzoumanian}, Z., \& {Okajima}, T. 2012, in Society of
  Photo-Optical Instrumentation Engineers (SPIE) Conference Series, Vol. 8443,
  Space Telescopes and Instrumentation 2012: Ultraviolet to Gamma Ray, ed.
  T.~{Takahashi}, S.~S. {Murray}, \& J.-W.~A. {den Herder}, 844313

\bibitem[{{Generozov} {et~al.}(2017){Generozov}, {Mimica}, {Metzger}, {Stone},
  {Giannios}, \& {Aloy}}]{2017MNRAS.464.2481G}
{Generozov}, A., {Mimica}, P., {Metzger}, B.~D., {et~al.} 2017, \mnras, 464,
  2481

\bibitem[{{Gezari}(2021)}]{2021ARA&A..59...21G}
{Gezari}, S. 2021, \araa, 59, 21

\bibitem[{{Gezari} {et~al.}(2017){Gezari}, {Cenko}, \&
  {Arcavi}}]{2017ApJ...851L..47G}
{Gezari}, S., {Cenko}, S.~B., \& {Arcavi}, I. 2017, \apjl, 851, L47

\bibitem[{{Gezari} {et~al.}(2006){Gezari}, {Martin}, {Milliard}, {Basa},
  {Halpern}, {Forster}, {Friedman}, {Morrissey}, {Neff}, {Schiminovich},
  {Seibert}, {Small}, \& {Wyder}}]{2006ApJ...653L..25G}
{Gezari}, S., {Martin}, D.~C., {Milliard}, B., {et~al.} 2006, \apjl, 653, L25

\bibitem[{{Gezari} {et~al.}(2008){Gezari}, {Basa}, {Martin}, {Bazin},
  {Forster}, {Milliard}, {Halpern}, {Friedman}, {Morrissey}, {Neff},
  {Schiminovich}, {Seibert}, {Small}, \& {Wyder}}]{2008ApJ...676..944G}
{Gezari}, S., {Basa}, S., {Martin}, D.~C., {et~al.} 2008, \apj, 676, 944

\bibitem[{{Giannios} \& {Metzger}(2011)}]{2011MNRAS.416.2102G}
{Giannios}, D., \& {Metzger}, B.~D. 2011, \mnras, 416, 2102

\bibitem[{{Giustini} {et~al.}(2020){Giustini}, {Miniutti}, \&
  {Saxton}}]{Giustini:20}
{Giustini}, M., {Miniutti}, G., \& {Saxton}, R.~D. 2020, arXiv e-prints,
  arXiv:2002.08967

\bibitem[{{Goicovic} {et~al.}(2019){Goicovic}, {Springel}, {Ohlmann}, \&
  {Pakmor}}]{2019MNRAS.487..981G}
{Goicovic}, F.~G., {Springel}, V., {Ohlmann}, S.~T., \& {Pakmor}, R. 2019,
  \mnras, 487, 981

\bibitem[{{Golightly} {et~al.}(2019{\natexlab{a}}){Golightly}, {Coughlin}, \&
  {Nixon}}]{2019ApJ...872..163G}
{Golightly}, E. C.~A., {Coughlin}, E.~R., \& {Nixon}, C.~J. 2019{\natexlab{a}},
  \apj, 872, 163

\bibitem[{{Golightly} {et~al.}(2019{\natexlab{b}}){Golightly}, {Nixon}, \&
  {Coughlin}}]{2019ApJ...882L..26G}
{Golightly}, E.~C.~A., {Nixon}, C.~J., \& {Coughlin}, E.~R. 2019{\natexlab{b}},
  \apjl, 882, L26

\bibitem[{{Gomez} {et~al.}(2020){Gomez}, {Nicholl}, {Short}, {Margutti},
  {Alexander}, {Blanchard}, {Berger}, {Eftekhari}, {Schulze}, {Anderson},
  {Arcavi}, {Chornock}, {Cowperthwaite}, {Galbany}, {Herzog}, {Hiramatsu},
  {Hosseinzadeh}, {Laskar}, {M{\"u}ller Bravo}, {Patton}, \&
  {Terreran}}]{2020MNRAS.497.1925G}
{Gomez}, S., {Nicholl}, M., {Short}, P., {et~al.} 2020, \mnras, 497, 1925

\bibitem[{{Greiner} {et~al.}(2000){Greiner}, {Schwarz}, {Zharikov}, \&
  {Orio}}]{2000A&A...362L..25G}
{Greiner}, J., {Schwarz}, R., {Zharikov}, S., \& {Orio}, M. 2000, \aap, 362,
  L25

\bibitem[{{Grupe} {et~al.}(1999){Grupe}, {Thomas}, \&
  {Leighly}}]{1999A&A...350L..31G}
{Grupe}, D., {Thomas}, H.~C., \& {Leighly}, K.~M. 1999, \aap, 350, L31

\bibitem[{{Gu{\'e}pin} {et~al.}(2018){Gu{\'e}pin}, {Kotera}, {Barausse},
  {Fang}, \& {Murase}}]{2018A&A...616A.179G}
{Gu{\'e}pin}, C., {Kotera}, K., {Barausse}, E., {Fang}, K., \& {Murase}, K.
  2018, \aap, 616, A179

\bibitem[{{Guillochon} {et~al.}(2014){Guillochon}, {Manukian}, \&
  {Ramirez-Ruiz}}]{2014ApJ...783...23G}
{Guillochon}, J., {Manukian}, H., \& {Ramirez-Ruiz}, E. 2014, \apj, 783, 23

\bibitem[{{Guillochon} \& {Ramirez-Ruiz}(2013)}]{2013ApJ...767...25G}
{Guillochon}, J., \& {Ramirez-Ruiz}, E. 2013, \apj, 767, 25

\bibitem[{{Guillochon} \& {Ramirez-Ruiz}(2015)}]{2015ApJ...809..166G}
---. 2015, \apj, 809, 166

\bibitem[{{Guillochon} {et~al.}(2009){Guillochon}, {Ramirez-Ruiz}, {Rosswog},
  \& {Kasen}}]{2009ApJ...705..844G}
{Guillochon}, J., {Ramirez-Ruiz}, E., {Rosswog}, S., \& {Kasen}, D. 2009, \apj,
  705, 844

\bibitem[{{Guolo} {et~al.}(2023){Guolo}, {Pasham}, {Zaja{\v{c}}ek}, {Coughlin},
  {Gezari}, {Sukov{\'a}}, {Wevers}, {Witzany}, {Tombesi}, {van Velzen},
  {Alexander}, {Yao}, {Arcodia}, {Karas}, {Miller-Jones}, {Remillard},
  {Gendreau}, \& {Ferrara}}]{Guolo:23}
{Guolo}, M., {Pasham}, D.~R., {Zaja{\v{c}}ek}, M., {et~al.} 2023, arXiv
  e-prints, arXiv:2309.03011

\bibitem[{{Hammerstein} {et~al.}(2023{\natexlab{a}}){Hammerstein}, {van
  Velzen}, {Gezari}, {Cenko}, {Yao}, {Ward}, {Frederick}, {Villanueva},
  {Somalwar}, {Graham}, {Kulkarni}, {Stern}, {Andreoni}, {Bellm}, {Dekany},
  {Dhawan}, {Drake}, {Fremling}, {Gatkine}, {Groom}, {Ho}, {Kasliwal},
  {Karambelkar}, {Kool}, {Masci}, {Medford}, {Perley}, {Purdum}, {van Roestel},
  {Sharma}, {Sollerman}, {Taggart}, \& {Yan}}]{2022arXiv220301461H}
{Hammerstein}, E., {van Velzen}, S., {Gezari}, S., {et~al.} 2023{\natexlab{a}},
  \apj, 942, 9

\bibitem[{{Hammerstein} {et~al.}(2023{\natexlab{b}}){Hammerstein}, {van
  Velzen}, {Gezari}, {Cenko}, {Yao}, {Ward}, {Frederick}, {Villanueva},
  {Somalwar}, {Graham}, {Kulkarni}, {Stern}, {Andreoni}, {Bellm}, {Dekany},
  {Dhawan}, {Drake}, {Fremling}, {Gatkine}, {Groom}, {Ho}, {Kasliwal},
  {Karambelkar}, {Kool}, {Masci}, {Medford}, {Perley}, {Purdum}, {van Roestel},
  {Sharma}, {Sollerman}, {Taggart}, \& {Yan}}]{2023ApJ...942....9H}
---. 2023{\natexlab{b}}, \apj, 942, 9

\bibitem[{{Harry} {et~al.}(2006){Harry}, {Fritschel}, {Shaddock}, {Folkner}, \&
  {Phinney}}]{2006CQGra..23.4887H}
{Harry}, G.~M., {Fritschel}, P., {Shaddock}, D.~A., {Folkner}, W., \&
  {Phinney}, E.~S. 2006, Classical and Quantum Gravity, 23, 4887

\bibitem[{{Hayasaki}(2021)}]{2021NatAs...5..436H}
{Hayasaki}, K. 2021, Nature Astronomy, 5, 436

\bibitem[{{Hayasaki} {et~al.}(2013){Hayasaki}, {Stone}, \&
  {Loeb}}]{2013MNRAS.434..909H}
{Hayasaki}, K., {Stone}, N., \& {Loeb}, A. 2013, \mnras, 434, 909

\bibitem[{{Hayasaki} \& {Yamazaki}(2019)}]{2019ApJ...886..114H}
{Hayasaki}, K., \& {Yamazaki}, R. 2019, \apj, 886, 114

\bibitem[{{Hills}(1988)}]{1988Natur.331..687H}
{Hills}, J.~G. 1988, \nat, 331, 687

\bibitem[{{Hils} \& {Bender}(1995)}]{1995ApJ...445L...7H}
{Hils}, D., \& {Bender}, P.~L. 1995, \apjl, 445, L7

\bibitem[{{Hinkle} {et~al.}(2020){Hinkle}, {Holoien}, {Shappee}, {Auchettl},
  {Kochanek}, {Stanek}, {Payne}, \& {Thompson}}]{2020ApJ...894L..10H}
{Hinkle}, J.~T., {Holoien}, T. W.~S., {Shappee}, B.~J., {et~al.} 2020, \apjl,
  894, L10

\bibitem[{{Hodgkin} {et~al.}(2021){Hodgkin}, {Harrison}, {Breedt}, {Wevers},
  {Rixon}, {Delgado}, {Yoldas}, {Kostrzewa-Rutkowska}, {Wyrzykowski}, {van
  Leeuwen}, {Blagorodnova}, {Campbell}, {Eappachen}, {Fraser}, {Ihanec},
  {Koposov}, {Kruszy{\'n}ska}, {Marton}, {Rybicki}, {Brown}, {Burgess},
  {Busso}, {Cowell}, {De Angeli}, {Diener}, {Evans}, {Gilmore}, {Holland},
  {Jonker}, {van Leeuwen}, {Mignard}, {Osborne}, {Portell}, {Prusti},
  {Richards}, {Riello}, {Seabroke}, {Walton}, {{\'A}brah{\'a}m}, {Altavilla},
  {Baker}, {Bastian}, {O'Brien}, {de Bruijne}, {Butterley}, {Carrasco},
  {Casta{\~n}eda}, {Clark}, {Clementini}, {Copperwheat}, {Cropper},
  {Damljanovic}, {Davidson}, {Davis}, {Dennefeld}, {Dhillon}, {Dolding},
  {Dominik}, {Esquej}, {Eyer}, {Fabricius}, {Fridman}, {Froebrich}, {Garralda},
  {Gomboc}, {Gonz{\'a}lez-Vidal}, {Guerra}, {Hambly}, {Hardy}, {Holl},
  {Hourihane}, {Japelj}, {Kann}, {Kiss}, {Knigge}, {Kolb}, {Komossa},
  {K{\'o}sp{\'a}l}, {Kov{\'a}cs}, {Kun}, {Leto}, {Lewis}, {Littlefair},
  {Mahabal}, {Mundell}, {Nagy}, {Padeletti}, {Palaversa}, {Pigulski},
  {Pretorius}, {van Reeven}, {Ribeiro}, {Roelens}, {Rowell}, {Schartel},
  {Scholz}, {Schwope}, {Sip{\H{o}}cz}, {Smartt}, {Smith}, {Serraller},
  {Steeghs}, {Sullivan}, {Szabados}, {Szegedi-Elek}, {Tisserand}, {Tomasella},
  {van Velzen}, {Whitelock}, {Wilson}, \& {Young}}]{2021A&A...652A..76H}
{Hodgkin}, S.~T., {Harrison}, D.~L., {Breedt}, E., {et~al.} 2021, \aap, 652,
  A76

\bibitem[{{Holoien} {et~al.}(2019){Holoien}, {Huber}, {Shappee}, {Eracleous},
  {Auchettl}, {Brown}, {Tucker}, {Chambers}, {Kochanek}, {Stanek}, {Rest},
  {Bersier}, {Post}, {Aldering}, {Ponder}, {Simon}, {Kankare}, {Dong},
  {Hallinan}, {Reddy}, {Sanders}, {Topping}, {Pan-STARRS}, {Bulger}, {Lowe},
  {Magnier}, {Schultz}, {Waters}, {Willman}, {Wright}, {Young}, {ASAS-SN},
  {Dong}, {Prieto}, {Thompson}, {ATLAS}, {Denneau}, {Flewelling}, {Heinze},
  {Smartt}, {Smith}, {Stalder}, {Tonry}, \& {Weiland}}]{2019ApJ...880..120H}
{Holoien}, T.~W.~S., {Huber}, M.~E., {Shappee}, B.~J., {et~al.} 2019, \apj,
  880, 120

\bibitem[{{Hopman} \& {Alexander}(2005)}]{2005ApJ...629..362H}
{Hopman}, C., \& {Alexander}, T. 2005, \apj, 629, 362

\bibitem[{{Hung} {et~al.}(2019){Hung}, {Cenko}, {Roth}, {Gezari}, {Veilleux},
  {van Velzen}, {Gaskell}, {Foley}, {Blagorodnova}, {Yan}, {Graham}, {Brown},
  {Siebert}, {Frederick}, {Ward}, {Gatkine}, {Gal-Yam}, {Yang}, {Schulze},
  {Dimitriadis}, {Kupfer}, {Shupe}, {Rusholme}, {Masci}, {Riddle}, {Soumagnac},
  {van Roestel}, \& {Dekany}}]{2019ApJ...879..119H}
{Hung}, T., {Cenko}, S.~B., {Roth}, N., {et~al.} 2019, \apj, 879, 119

\bibitem[{{Hung} {et~al.}(2020){Hung}, {Foley}, {Ramirez-Ruiz}, {Dai},
  {Auchettl}, {Kilpatrick}, {Mockler}, {Brown}, {Coulter}, {Dimitriadis},
  {Holoien}, {Law-Smith}, {Piro}, {Rest}, {Rojas-Bravo}, \&
  {Siebert}}]{2020ApJ...903...31H}
{Hung}, T., {Foley}, R.~J., {Ramirez-Ruiz}, E., {et~al.} 2020, \apj, 903, 31

\bibitem[{{Ivanov} \& {Chernyakova}(2006)}]{2006A&A...448..843I}
{Ivanov}, P.~B., \& {Chernyakova}, M.~A. 2006, \aap, 448, 843

\bibitem[{{Ivezi{\'c}} {et~al.}(2019){Ivezi{\'c}}, {Kahn}, {Tyson}, {Abel},
  {Acosta}, {Allsman}, {Alonso}, {AlSayyad}, {Anderson}, {Andrew}, {Angel},
  {Angeli}, {Ansari}, {Antilogus}, {Araujo}, {Armstrong}, {Arndt}, {Astier},
  {Aubourg}, {Auza}, {Axelrod}, {Bard}, {Barr}, {Barrau}, {Bartlett}, {Bauer},
  {Bauman}, {Baumont}, {Bechtol}, {Bechtol}, {Becker}, {Becla}, {Beldica},
  {Bellavia}, {Bianco}, {Biswas}, {Blanc}, {Blazek}, {Blandford}, {Bloom},
  {Bogart}, {Bond}, {Booth}, {Borgland}, {Borne}, {Bosch}, {Boutigny},
  {Brackett}, {Bradshaw}, {Brandt}, {Brown}, {Bullock}, {Burchat}, {Burke},
  {Cagnoli}, {Calabrese}, {Callahan}, {Callen}, {Carlin}, {Carlson},
  {Chandrasekharan}, {Charles-Emerson}, {Chesley}, {Cheu}, {Chiang}, {Chiang},
  {Chirino}, {Chow}, {Ciardi}, {Claver}, {Cohen-Tanugi}, {Cockrum}, {Coles},
  {Connolly}, {Cook}, {Cooray}, {Covey}, {Cribbs}, {Cui}, {Cutri}, {Daly},
  {Daniel}, {Daruich}, {Daubard}, {Daues}, {Dawson}, {Delgado}, {Dellapenna},
  {de Peyster}, {de Val-Borro}, {Digel}, {Doherty}, {Dubois},
  {Dubois-Felsmann}, {Durech}, {Economou}, {Eifler}, {Eracleous}, {Emmons},
  {Fausti Neto}, {Ferguson}, {Figueroa}, {Fisher-Levine}, {Focke}, {Foss},
  {Frank}, {Freemon}, {Gangler}, {Gawiser}, {Geary}, {Gee}, {Geha}, {Gessner},
  {Gibson}, {Gilmore}, {Glanzman}, {Glick}, {Goldina}, {Goldstein}, {Goodenow},
  {Graham}, {Gressler}, {Gris}, {Guy}, {Guyonnet}, {Haller}, {Harris},
  {Hascall}, {Haupt}, {Hernandez}, {Herrmann}, {Hileman}, {Hoblitt}, {Hodgson},
  {Hogan}, {Howard}, {Huang}, {Huffer}, {Ingraham}, {Innes}, {Jacoby}, {Jain},
  {Jammes}, {Jee}, {Jenness}, {Jernigan}, {Jevremovi{\'c}}, {Johns}, {Johnson},
  {Johnson}, {Jones}, {Juramy-Gilles}, {Juri{\'c}}, {Kalirai}, {Kallivayalil},
  {Kalmbach}, {Kantor}, {Karst}, {Kasliwal}, {Kelly}, {Kessler}, {Kinnison},
  {Kirkby}, {Knox}, {Kotov}, {Krabbendam}, {Krughoff}, {Kub{\'a}nek},
  {Kuczewski}, {Kulkarni}, {Ku}, {Kurita}, {Lage}, {Lambert}, {Lange},
  {Langton}, {Le Guillou}, {Levine}, {Liang}, {Lim}, {Lintott}, {Long},
  {Lopez}, {Lotz}, {Lupton}, {Lust}, {MacArthur}, {Mahabal}, {Mandelbaum},
  {Markiewicz}, {Marsh}, {Marshall}, {Marshall}, {May}, {McKercher}, {McQueen},
  {Meyers}, {Migliore}, {Miller}, {Mills}, {Miraval}, {Moeyens}, {Moolekamp},
  {Monet}, {Moniez}, {Monkewitz}, {Montgomery}, {Morrison}, {Mueller},
  {Muller}, {Mu{\~n}oz Arancibia}, {Neill}, {Newbry}, {Nief}, {Nomerotski},
  {Nordby}, {O'Connor}, {Oliver}, {Olivier}, {Olsen}, {O'Mullane}, {Ortiz},
  {Osier}, {Owen}, {Pain}, {Palecek}, {Parejko}, {Parsons}, {Pease},
  {Peterson}, {Peterson}, {Petravick}, {Libby Petrick}, {Petry},
  {Pierfederici}, {Pietrowicz}, {Pike}, {Pinto}, {Plante}, {Plate}, {Plutchak},
  {Price}, {Prouza}, {Radeka}, {Rajagopal}, {Rasmussen}, {Regnault}, {Reil},
  {Reiss}, {Reuter}, {Ridgway}, {Riot}, {Ritz}, {Robinson}, {Roby}, {Roodman},
  {Rosing}, {Roucelle}, {Rumore}, {Russo}, {Saha}, {Sassolas}, {Schalk},
  {Schellart}, {Schindler}, {Schmidt}, {Schneider}, {Schneider}, {Schoening},
  {Schumacher}, {Schwamb}, {Sebag}, {Selvy}, {Sembroski}, {Seppala}, {Serio},
  {Serrano}, {Shaw}, {Shipsey}, {Sick}, {Silvestri}, {Slater}, {Smith},
  {Smith}, {Sobhani}, {Soldahl}, {Storrie-Lombardi}, {Stover}, {Strauss},
  {Street}, {Stubbs}, {Sullivan}, {Sweeney}, {Swinbank}, {Szalay}, {Takacs},
  {Tether}, {Thaler}, {Thayer}, {Thomas}, {Thornton}, {Thukral}, {Tice},
  {Trilling}, {Turri}, {Van Berg}, {Vanden Berk}, {Vetter}, {Virieux},
  {Vucina}, {Wahl}, {Walkowicz}, {Walsh}, {Walter}, {Wang}, {Wang}, {Warner},
  {Wiecha}, {Willman}, {Winters}, {Wittman}, {Wolff}, {Wood-Vasey}, {Wu},
  {Xin}, {Yoachim}, \& {Zhan}}]{2019ApJ...873..111I}
{Ivezi{\'c}}, {\v{Z}}., {Kahn}, S.~M., {Tyson}, J.~A., {et~al.} 2019, \apj,
  873, 111

\bibitem[{{Jiang} {et~al.}(2016){Jiang}, {Dou}, {Wang}, {Yang}, {Lyu}, \&
  {Zhou}}]{2016ApJ...828L..14J}
{Jiang}, N., {Dou}, L., {Wang}, T., {et~al.} 2016, \apjl, 828, L14

\bibitem[{{Jiang} {et~al.}(2021){Jiang}, {Wang}, {Hu}, {Sun}, {Dou}, \&
  {Xiao}}]{2021ApJ...911...31J}
{Jiang}, N., {Wang}, T., {Hu}, X., {et~al.} 2021, \apj, 911, 31

\bibitem[{{Jonker} {et~al.}(2020){Jonker}, {Stone}, {Generozov}, {van Velzen},
  \& {Metzger}}]{2020ApJ...889..166J}
{Jonker}, P.~G., {Stone}, N.~C., {Generozov}, A., {van Velzen}, S., \&
  {Metzger}, B. 2020, \apj, 889, 166

\bibitem[{{Kaiser} {et~al.}(2002){Kaiser}, {Aussel}, {Burke}, {Boesgaard},
  {Chambers}, {Chun}, {Heasley}, {Hodapp}, {Hunt}, {Jedicke}, {Jewitt},
  {Kudritzki}, {Luppino}, {Maberry}, {Magnier}, {Monet}, {Onaka}, {Pickles},
  {Rhoads}, {Simon}, {Szalay}, {Szapudi}, {Tholen}, {Tonry}, {Waterson}, \&
  {Wick}}]{2002SPIE.4836..154K}
{Kaiser}, N., {Aussel}, H., {Burke}, B.~E., {et~al.} 2002, in Society of
  Photo-Optical Instrumentation Engineers (SPIE) Conference Series, Vol. 4836,
  Survey and Other Telescope Technologies and Discoveries, ed. J.~A. {Tyson} \&
  S.~{Wolff}, 154--164

\bibitem[{{Kajava} {et~al.}(2020){Kajava}, {Giustini}, {Saxton}, \&
  {Miniutti}}]{2020A&A...639A.100K}
{Kajava}, J. J.~E., {Giustini}, M., {Saxton}, R.~D., \& {Miniutti}, G. 2020,
  \aap, 639, A100

\bibitem[{{Kawamura} {et~al.}(2006){Kawamura}, {Nakamura}, {Ando}, {Seto},
  {Tsubono}, {Numata}, {Takahashi}, {Nagano}, {Ishikawa}, {Musha}, {Ueda},
  {Sato}, {Hosokawa}, {Agatsuma}, {Akutsu}, {Aoyanagi}, {Arai}, {Araya},
  {Asada}, {Aso}, {Chiba}, {Ebisuzaki}, {Eriguchi}, {Fujimoto}, {Fukushima},
  {Futamase}, {Ganzu}, {Harada}, {Hashimoto}, {Hayama}, {Hikida}, {Himemoto},
  {Hirabayashi}, {Hiramatsu}, {Ichiki}, {Ikegami}, {Inoue}, {Ioka},
  {Ishidoshiro}, {Itoh}, {Kamagasako}, {Kanda}, {Kawashima}, {Kirihara},
  {Kiuchi}, {Kobayashi}, {Kohri}, {Kojima}, {Kokeyama}, {Kozai}, {Kudoh},
  {Kunimori}, {Kuroda}, {Maeda}, {Matsuhara}, {Mino}, {Miyakawa}, {Miyoki},
  {Mizusawa}, {Morisawa}, {Mukohyama}, {Naito}, {Nakagawa}, {Nakamura},
  {Nakano}, {Nakao}, {Nishizawa}, {Niwa}, {Nozawa}, {Ohashi}, {Ohishi},
  {Ohkawa}, {Okutomi}, {Oohara}, {Sago}, {Saijo}, {Sakagami}, {Sakata},
  {Sasaki}, {Sato}, {Shibata}, {Shinkai}, {Somiya}, {Sotani}, {Sugiyama},
  {Tagoshi}, {Takahashi}, {Takahashi}, {Takahashi}, {Takano}, {Tanaka},
  {Taniguchi}, {Taruya}, {Tashiro}, {Tokunari}, {Tsujikawa}, {Tsunesada},
  {Yamamoto}, {Yamazaki}, {Yokoyama}, {Yoo}, {Yoshida}, \&
  {Yoshino}}]{2006CQGra..23S.125K}
{Kawamura}, S., {Nakamura}, T., {Ando}, M., {et~al.} 2006, Classical and
  Quantum Gravity, 23, S125

\bibitem[{{Kesden}(2012)}]{2012PhRvD..85b4037K}
{Kesden}, M. 2012, \prd, 85, 024037

\bibitem[{{Kim} {et~al.}(1999){Kim}, {Park}, \& {Lee}}]{1999ApJ...519..647K}
{Kim}, S.~S., {Park}, M.-G., \& {Lee}, H.~M. 1999, \apj, 519, 647

\bibitem[{{King} \& {Done}(1993)}]{1993MNRAS.264..388K}
{King}, A.~R., \& {Done}, C. 1993, \mnras, 264, 388

\bibitem[{{Kiuchi} {et~al.}(2011){Kiuchi}, {Shibata}, {Montero}, \&
  {Font}}]{2011PhRvL.106y1102K}
{Kiuchi}, K., {Shibata}, M., {Montero}, P.~J., \& {Font}, J.~A. 2011, \prl,
  106, 251102

\bibitem[{{Kobayashi} {et~al.}(2004){Kobayashi}, {Laguna}, {Phinney}, \&
  {M{\'e}sz{\'a}ros}}]{2004ApJ...615..855K}
{Kobayashi}, S., {Laguna}, P., {Phinney}, E.~S., \& {M{\'e}sz{\'a}ros}, P.
  2004, \apj, 615, 855

\bibitem[{{Komossa} \& {Greiner}(1999)}]{1999A&A...349L..45K}
{Komossa}, S., \& {Greiner}, J. 1999, \aap, 349, L45

\bibitem[{{Komossa} {et~al.}(2004){Komossa}, {Halpern}, {Schartel}, {Hasinger},
  {Santos-Lleo}, \& {Predehl}}]{2004ApJ...603L..17K}
{Komossa}, S., {Halpern}, J., {Schartel}, N., {et~al.} 2004, \apjl, 603, L17

\bibitem[{{Komossa} {et~al.}(2008){Komossa}, {Zhou}, {Wang}, {Ajello}, {Ge},
  {Greiner}, {Lu}, {Salvato}, {Saxton}, {Shan}, {Xu}, \&
  {Yuan}}]{2008ApJ...678L..13K}
{Komossa}, S., {Zhou}, H., {Wang}, T., {et~al.} 2008, \apjl, 678, L13

\bibitem[{{Komossa} {et~al.}(2009){Komossa}, {Zhou}, {Rau}, {Dopita},
  {Gal-Yam}, {Greiner}, {Zuther}, {Salvato}, {Xu}, {Lu}, {Saxton}, \&
  {Ajello}}]{2009ApJ...701..105K}
{Komossa}, S., {Zhou}, H., {Rau}, A., {et~al.} 2009, \apj, 701, 105

\bibitem[{{Kormendy} \& {Ho}(2013)}]{2013ARA&A..51..511K}
{Kormendy}, J., \& {Ho}, L.~C. 2013, \araa, 51, 511

\bibitem[{{Kosovichev} \& {Novikov}(1992)}]{1992MNRAS.258..715K}
{Kosovichev}, A.~G., \& {Novikov}, I.~D. 1992, \mnras, 258, 715

\bibitem[{{Kotera} \& {Olinto}(2011)}]{2011ARAA..49..119K}
{Kotera}, K., \& {Olinto}, A.~V. 2011, \araa, 49, 119

\bibitem[{{Kremer} {et~al.}(2022{\natexlab{a}}){Kremer}, {Lombardi}, {Lu},
  {Piro}, \& {Rasio}}]{2022ApJ...933..203K}
{Kremer}, K., {Lombardi}, J.~C., {Lu}, W., {Piro}, A.~L., \& {Rasio}, F.~A.
  2022{\natexlab{a}}, \apj, 933, 203

\bibitem[{{Kremer} {et~al.}(2022{\natexlab{b}}){Kremer}, {Lombardi}, {Lu},
  {Piro}, \& {Rasio}}]{2022arXiv220112368K}
---. 2022{\natexlab{b}}, \apj, 933, 203

\bibitem[{{Kremer} {et~al.}(2021){Kremer}, {Lu}, {Piro}, {Chatterjee}, {Rasio},
  \& {Ye}}]{2021ApJ...911..104K}
{Kremer}, K., {Lu}, W., {Piro}, A.~L., {et~al.} 2021, \apj, 911, 104

\bibitem[{{Kremer} {et~al.}(2019){Kremer}, {Lu}, {Rodriguez}, {Lachat}, \&
  {Rasio}}]{2019ApJ...881...75K}
{Kremer}, K., {Lu}, W., {Rodriguez}, C.~L., {Lachat}, M., \& {Rasio}, F.~A.
  2019, \apj, 881, 75

\bibitem[{{Krolik} {et~al.}(2020{\natexlab{a}}){Krolik}, {Piran}, \&
  {Ryu}}]{2020ApJ...904...68K}
{Krolik}, J., {Piran}, T., \& {Ryu}, T. 2020{\natexlab{a}}, \apj, 904, 68

\bibitem[{{Krolik} {et~al.}(2016){Krolik}, {Piran}, {Svirski}, \&
  {Cheng}}]{2016ApJ...827..127K}
{Krolik}, J., {Piran}, T., {Svirski}, G., \& {Cheng}, R.~M. 2016, \apj, 827,
  127

\bibitem[{{Krolik} {et~al.}(2020{\natexlab{b}}){Krolik}, {Armitage}, {Jiang},
  \& {Lodato}}]{2020SSRv..216...88K}
{Krolik}, J.~H., {Armitage}, P.~J., {Jiang}, Y., \& {Lodato}, G.
  2020{\natexlab{b}}, \ssr, 216, 88

\bibitem[{{Kumar} {et~al.}(2013){Kumar}, {Barniol Duran}, {Bo{\v{s}}njak}, \&
  {Piran}}]{2013MNRAS.434.3078K}
{Kumar}, P., {Barniol Duran}, R., {Bo{\v{s}}njak}, {\v{Z}}., \& {Piran}, T.
  2013, \mnras, 434, 3078

\bibitem[{{Lacy} {et~al.}(1982){Lacy}, {Townes}, \&
  {Hollenbach}}]{1982ApJ...262..120L}
{Lacy}, J.~H., {Townes}, C.~H., \& {Hollenbach}, D.~J. 1982, \apj, 262, 120

\bibitem[{{Lattimer} \& {Schramm}(1976)}]{1976ApJ...210..549L}
{Lattimer}, J.~M., \& {Schramm}, D.~N. 1976, \apj, 210, 549

\bibitem[{{Law-Smith} {et~al.}(2019){Law-Smith}, {Guillochon}, \&
  {Ramirez-Ruiz}}]{2019ApJ...882L..25L}
{Law-Smith}, J., {Guillochon}, J., \& {Ramirez-Ruiz}, E. 2019, \apjl, 882, L25

\bibitem[{{Law-Smith} {et~al.}(2020){Law-Smith}, {Coulter}, {Guillochon},
  {Mockler}, \& {Ramirez-Ruiz}}]{2020ApJ...905..141L}
{Law-Smith}, J. A.~P., {Coulter}, D.~A., {Guillochon}, J., {Mockler}, B., \&
  {Ramirez-Ruiz}, E. 2020, \apj, 905, 141

\bibitem[{{Leloudas} {et~al.}(2016){Leloudas}, {Fraser}, {Stone}, {van Velzen},
  {Jonker}, {Arcavi}, {Fremling}, {Maund}, {Smartt}, {Kr{\`\i}hler},
  {Miller-Jones}, {Vreeswijk}, {Gal-Yam}, {Mazzali}, {De Cia}, {Howell},
  {Inserra}, {Patat}, {de Ugarte Postigo}, {Yaron}, {Ashall}, {Bar},
  {Campbell}, {Chen}, {Childress}, {Elias-Rosa}, {Harmanen}, {Hosseinzadeh},
  {Johansson}, {Kangas}, {Kankare}, {Kim}, {Kuncarayakti}, {Lyman}, {Magee},
  {Maguire}, {Malesani}, {Mattila}, {McCully}, {Nicholl}, {Prentice},
  {Romero-Ca{\~n}izales}, {Schulze}, {Smith}, {Sollerman}, {Sullivan},
  {Tucker}, {Valenti}, {Wheeler}, \& {Young}}]{2016NatAs...1E...2L}
{Leloudas}, G., {Fraser}, M., {Stone}, N.~C., {et~al.} 2016, Nature Astronomy,
  1, 0002

\bibitem[{{Leloudas} {et~al.}(2019){Leloudas}, {Dai}, {Arcavi}, {Vreeswijk},
  {Mockler}, {Roy}, {Malesani}, {Schulze}, {Wevers}, {Fraser}, {Ramirez-Ruiz},
  {Auchettl}, {Burke}, {Cannizzaro}, {Charalampopoulos}, {Chen}, {Cikota},
  {Della Valle}, {Galbany}, {Gromadzki}, {Heintz}, {Hiramatsu}, {Jonker},
  {Kostrzewa-Rutkowska}, {Maguire}, {Mandel}, {Nicholl}, {Onori}, {Roth},
  {Smartt}, {Wyrzykowski}, \& {Young}}]{2019ApJ...887..218L}
{Leloudas}, G., {Dai}, L., {Arcavi}, I., {et~al.} 2019, \apj, 887, 218

\bibitem[{{Leloudas} {et~al.}(2022){Leloudas}, {Bulla}, {Cikota}, {Dai},
  {Thomsen}, {Maund}, {Charalampopoulos}, {Roth}, {Arcavi}, {Auchettl},
  {Malesani}, {Nicholl}, \& {Ramirez-Ruiz}}]{2022NatAs...6.1193L}
{Leloudas}, G., {Bulla}, M., {Cikota}, A., {et~al.} 2022, Nature Astronomy, 6,
  1193

\bibitem[{{Levan} {et~al.}(2011){Levan}, {Tanvir}, {Cenko}, {Perley},
  {Wiersema}, {Bloom}, {Fruchter}, {de Ugarte Postigo}, {O'Brien}, {Butler},
  {van der Horst}, {Leloudas}, {Morgan}, {Misra}, {Bower}, {Farihi},
  {Tunnicliffe}, {Modjaz}, {Silverman}, {Hjorth}, {Th{\"o}ne}, {Cucchiara},
  {Cer{\'o}n}, {Castro-Tirado}, {Arnold}, {Bremer}, {Brodie}, {Carroll},
  {Cooper}, {Curran}, {Cutri}, {Ehle}, {Forbes}, {Fynbo}, {Gorosabel},
  {Graham}, {Hoffman}, {Guziy}, {Jakobsson}, {Kamble}, {Kerr}, {Kasliwal},
  {Kouveliotou}, {Kocevski}, {Law}, {Nugent}, {Ofek}, {Poznanski}, {Quimby},
  {Rol}, {Romanowsky}, {S{\'a}nchez-Ram{\'\i}rez}, {Schulze}, {Singh}, {van
  Spaandonk}, {Starling}, {Strom}, {Tello}, {Vaduvescu}, {Wheatley}, {Wijers},
  {Winters}, \& {Xu}}]{2011Sci...333..199L}
{Levan}, A.~J., {Tanvir}, N.~R., {Cenko}, S.~B., {et~al.} 2011, Science, 333,
  199

\bibitem[{{Li} {et~al.}(2002){Li}, {Narayan}, \& {Menou}}]{2002ApJ...576..753L}
{Li}, L.-X., {Narayan}, R., \& {Menou}, K. 2002, \apj, 576, 753

\bibitem[{{Lightman} \& {Shapiro}(1977)}]{1977ApJ...211..244L}
{Lightman}, A.~P., \& {Shapiro}, S.~L. 1977, \apj, 211, 244

\bibitem[{{Lin} {et~al.}(2022){Lin}, {Jiang}, {Kong}, {Huang}, {Lin}, {Zhu}, \&
  {Wang}}]{2022arXiv221014950L}
{Lin}, Z., {Jiang}, N., {Kong}, X., {et~al.} 2022, arXiv e-prints,
  arXiv:2210.14950

\bibitem[{{Liu} {et~al.}(2023){Liu}, {Malyali}, {Krumpe}, {Homan}, {Goodwin},
  {Grotova}, {Kawka}, {Rau}, {Merloni}, {Anderson}, {Miller-Jones},
  {Markowitz}, {Ciroi}, {Di Mille}, {Schramm}, {Tang}, {Buckley}, {Gromadzki},
  {Jin}, \& {Buchner}}]{Liu:23}
{Liu}, Z., {Malyali}, A., {Krumpe}, M., {et~al.} 2023, \aap, 669, A75

\bibitem[{{Lodato} {et~al.}(2009){Lodato}, {King}, \&
  {Pringle}}]{2009MNRAS.392..332L}
{Lodato}, G., {King}, A.~R., \& {Pringle}, J.~E. 2009, \mnras, 392, 332

\bibitem[{{Lopez} {et~al.}(2019){Lopez}, {Batta}, {Ramirez-Ruiz}, {Martinez},
  \& {Samsing}}]{2019ApJ...877...56L}
{Lopez}, Martin, J., {Batta}, A., {Ramirez-Ruiz}, E., {Martinez}, I., \&
  {Samsing}, J. 2019, \apj, 877, 56

\bibitem[{{Lu} {et~al.}(2016){Lu}, {Kumar}, \& {Evans}}]{2016MNRAS.458..575L}
{Lu}, W., {Kumar}, P., \& {Evans}, N.~J. 2016, \mnras, 458, 575

\bibitem[{{Luminet} \& {Carter}(1986)}]{1986ApJS...61..219L}
{Luminet}, J.~P., \& {Carter}, B. 1986, \apjs, 61, 219

\bibitem[{{Luo} {et~al.}(2016){Luo}, {Chen}, {Duan}, {Gong}, {Hu}, {Ji}, {Liu},
  {Mei}, {Milyukov}, {Sazhin}, {Shao}, {Toth}, {Tu}, {Wang}, {Wang}, {Yeh},
  {Zhan}, {Zhang}, {Zharov}, \& {Zhou}}]{2016CQGra..33c5010L}
{Luo}, J., {Chen}, L.-S., {Duan}, H.-Z., {et~al.} 2016, Classical and Quantum
  Gravity, 33, 035010

\bibitem[{Maggiore(????)}]{maggiore2018}
Maggiore, M. ????, Gravitational Waves – Volume 2: Astrophysics and
  Cosmology, doi:10.1093/oso/9780198570899.001.0001

\bibitem[{{Magorrian} \& {Tremaine}(1999)}]{1999MNRAS.309..447M}
{Magorrian}, J., \& {Tremaine}, S. 1999, \mnras, 309, 447

\bibitem[{{Mainetti} {et~al.}(2017){Mainetti}, {Lupi}, {Campana}, {Colpi},
  {Coughlin}, {Guillochon}, \& {Ramirez-Ruiz}}]{2017A&A...600A.124M}
{Mainetti}, D., {Lupi}, A., {Campana}, S., {et~al.} 2017, \aap, 600, A124

\bibitem[{{Mainzer} {et~al.}(2014){Mainzer}, {Bauer}, {Cutri}, {Grav},
  {Masiero}, {Beck}, {Clarkson}, {Conrow}, {Dailey}, {Eisenhardt}, {Fabinsky},
  {Fajardo-Acosta}, {Fowler}, {Gelino}, {Grillmair}, {Heinrichsen}, {Kendall},
  {Kirkpatrick}, {Liu}, {Masci}, {McCallon}, {Nugent}, {Papin}, {Rice},
  {Royer}, {Ryan}, {Sevilla}, {Sonnett}, {Stevenson}, {Thompson}, {Wheelock},
  {Wiemer}, {Wittman}, {Wright}, \& {Yan}}]{2014ApJ...792...30M}
{Mainzer}, A., {Bauer}, J., {Cutri}, R.~M., {et~al.} 2014, \apj, 792, 30

\bibitem[{{Maksym} {et~al.}(2014){Maksym}, {Lin}, \&
  {Irwin}}]{2014ApJ...792L..29M}
{Maksym}, W.~P., {Lin}, D., \& {Irwin}, J.~A. 2014, \apjl, 792, L29

\bibitem[{{Malyali} {et~al.}(2023){Malyali}, {Liu}, {Rau}, {Grotova},
  {Merloni}, {Goodwin}, {Anderson}, {Miller-Jones}, {Kawka}, {Arcodia},
  {Buchner}, {Nandra}, {Homan}, \& {Krumpe}}]{Malyali:23}
{Malyali}, A., {Liu}, Z., {Rau}, A., {et~al.} 2023, \mnras, 520, 3549

\bibitem[{{Mapelli} {et~al.}(2021){Mapelli}, {Santoliquido}, {Bouffanais},
  {Arca Sedda}, {Artale}, \& {Ballone}}]{2021Symm...13.1678M}
{Mapelli}, M., {Santoliquido}, F., {Bouffanais}, Y., {et~al.} 2021, Symmetry,
  13, 1678

\bibitem[{{Matsumoto} {et~al.}(2022){Matsumoto}, {Piran}, \&
  {Krolik}}]{2022MNRAS.511.5085M}
{Matsumoto}, T., {Piran}, T., \& {Krolik}, J.~H. 2022, \mnras, 511, 5085

\bibitem[{{McKernan} {et~al.}(2022){McKernan}, {Ford}, {Cantiello}, {Graham},
  {Jermyn}, {Leigh}, {Ryu}, \& {Stern}}]{2022MNRAS.514.4102M}
{McKernan}, B., {Ford}, K.~E.~S., {Cantiello}, M., {et~al.} 2022, \mnras, 514,
  4102

\bibitem[{{M{\'e}sz{\'a}ros}(2017)}]{2017ARNPS..67...45M}
{M{\'e}sz{\'a}ros}, P. 2017, Annual Review of Nuclear and Particle Science, 67,
  45

\bibitem[{{Metzger} {et~al.}(2012){Metzger}, {Giannios}, \&
  {Mimica}}]{2012MNRAS.420.3528M}
{Metzger}, B.~D., {Giannios}, D., \& {Mimica}, P. 2012, \mnras, 420, 3528

\bibitem[{{Metzger} {et~al.}(2022){Metzger}, {Stone}, \&
  {Gilbaum}}]{2022ApJ...926..101M}
{Metzger}, B.~D., {Stone}, N.~C., \& {Gilbaum}, S. 2022, \apj, 926, 101

\bibitem[{{Michaely} {et~al.}(2016){Michaely}, {Ginzburg}, \&
  {Perets}}]{2016arXiv161000593M}
{Michaely}, E., {Ginzburg}, D., \& {Perets}, H.~B. 2016, arXiv e-prints,
  arXiv:1610.00593

\bibitem[{{Miller} \& {Lauburg}(2009)}]{2009ApJ...692..917M}
{Miller}, M.~C., \& {Lauburg}, V.~M. 2009, \apj, 692, 917

\bibitem[{{Mimica} {et~al.}(2015){Mimica}, {Giannios}, {Metzger}, \&
  {Aloy}}]{2015MNRAS.450.2824M}
{Mimica}, P., {Giannios}, D., {Metzger}, B.~D., \& {Aloy}, M.~A. 2015, \mnras,
  450, 2824

\bibitem[{{Miniutti} {et~al.}(2019){Miniutti}, {Saxton}, {Giustini}, {Alexand
  er}, {Fender}, {Heywood}, {Monageng}, {Coriat}, {Tzioumis}, {Read}, {Knigge},
  {Gandhi}, {Pretorius}, \& {Ag{\'\i}s-Gonz{\'a}lez}}]{Miniutti:19}
{Miniutti}, G., {Saxton}, R.~D., {Giustini}, M., {et~al.} 2019, \nat, 573, 381

\bibitem[{{Miralda-Escud{\'e}} \& {Kollmeier}(2005)}]{2005ApJ...619...30M}
{Miralda-Escud{\'e}}, J., \& {Kollmeier}, J.~A. 2005, \apj, 619, 30

\bibitem[{{Mohan} {et~al.}(2022){Mohan}, {An}, {Zhang}, {Yang}, {Yang}, \&
  {Wang}}]{2022ApJ...927...74M}
{Mohan}, P., {An}, T., {Zhang}, Y., {et~al.} 2022, \apj, 927, 74

\bibitem[{{Morscher} {et~al.}(2015){Morscher}, {Pattabiraman}, {Rodriguez},
  {Rasio}, \& {Umbreit}}]{2015ApJ...800....9M}
{Morscher}, M., {Pattabiraman}, B., {Rodriguez}, C., {Rasio}, F.~A., \&
  {Umbreit}, S. 2015, \apj, 800, 9

\bibitem[{{Mummery}(2021)}]{2021arXiv210406212M}
{Mummery}, A. 2021, arXiv e-prints, arXiv:2104.06212

\bibitem[{{Murase} {et~al.}(2020){Murase}, {Kimura}, {Zhang}, {Oikonomou}, \&
  {Petropoulou}}]{2020ApJ...902..108M}
{Murase}, K., {Kimura}, S.~S., {Zhang}, B.~T., {Oikonomou}, F., \&
  {Petropoulou}, M. 2020, \apj, 902, 108

\bibitem[{{Nealon} {et~al.}(2018){Nealon}, {Price}, {Bonnerot}, \&
  {Lodato}}]{2018MNRAS.474.1737N}
{Nealon}, R., {Price}, D.~J., {Bonnerot}, C., \& {Lodato}, G. 2018, \mnras,
  474, 1737

\bibitem[{{Nicholl} {et~al.}(2022){Nicholl}, {Lanning}, {Ramsden}, {Mockler},
  {Lawrence}, {Short}, \& {Ridley}}]{2022MNRAS.515.5604N}
{Nicholl}, M., {Lanning}, D., {Ramsden}, P., {et~al.} 2022, \mnras, 515, 5604

\bibitem[{{Nicholl} {et~al.}(2019){Nicholl}, {Blanchard}, {Berger}, {Gomez},
  {Margutti}, {Alexander}, {Guillochon}, {Leja}, {Chornock}, {Snios},
  {Auchettl}, {Bruce}, {Challis}, {D'Orazio}, {Drout}, {Eftekhari}, {Foley},
  {Graur}, {Kilpatrick}, {Lawrence}, {Piro}, {Rojas-Bravo}, {Ross}, {Short},
  {Smartt}, {Smith}, \& {Stalder}}]{2019MNRAS.488.1878N}
{Nicholl}, M., {Blanchard}, P.~K., {Berger}, E., {et~al.} 2019, \mnras, 488,
  1878

\bibitem[{{Nicholl} {et~al.}(2020){Nicholl}, {Wevers}, {Oates}, {Alexander},
  {Leloudas}, {Onori}, {Jerkstrand}, {Gomez}, {Campana}, {Arcavi},
  {Charalampopoulos}, {Gromadzki}, {Ihanec}, {Jonker}, {Lawrence}, {Mandel},
  {Schulze}, {Short}, {Burke}, {McCully}, {Hiramatsu}, {Howell}, {Pellegrino},
  {Abbot}, {Anderson}, {Berger}, {Blanchard}, {Cannizzaro}, {Chen},
  {Dennefeld}, {Galbany}, {Gonz{\'a}lez-Gait{\'a}n}, {Hosseinzadeh}, {Inserra},
  {Irani}, {Kuin}, {M{\"u}ller-Bravo}, {Pineda}, {Ross}, {Roy}, {Smartt},
  {Smith}, {Tucker}, {Wyrzykowski}, \& {Young}}]{2020MNRAS.499..482N}
{Nicholl}, M., {Wevers}, T., {Oates}, S.~R., {et~al.} 2020, \mnras, 499, 482

\bibitem[{{Nordin} {et~al.}(2019{\natexlab{a}}){Nordin}, {Brinnel}, {Giomi},
  {Santen}, {Gal-Yam}, {Yaron}, \& {Schulze}}]{2019TNSTR.615....1N}
{Nordin}, J., {Brinnel}, V., {Giomi}, M., {et~al.} 2019{\natexlab{a}},
  Transient Name Server Discovery Report, 2019-615, 1

\bibitem[{{Nordin} {et~al.}(2019{\natexlab{b}}){Nordin}, {Brinnel}, {Giomi},
  {Santen}, {Gal-Yam}, {Yaron}, \& {Schulze}}]{2019TNSTR.771....1N}
---. 2019{\natexlab{b}}, Transient Name Server Discovery Report, 2019-771, 1

\bibitem[{{Novikov} {et~al.}(1992){Novikov}, {Pethick}, \&
  {Polnarev}}]{1992MNRAS.255..276N}
{Novikov}, I.~D., {Pethick}, C.~J., \& {Polnarev}, A.~G. 1992, \mnras, 255, 276

\bibitem[{{Onori} {et~al.}(2022){Onori}, {Cannizzaro}, {Jonker}, {Kim},
  {Nicholl}, {Mattila}, {Reynolds}, {Fraser}, {Wevers}, {Brocato}, {Anderson},
  {Carini}, {Charalampopoulos}, {Clark}, {Gromadzki}, {Guti{\'e}rrez},
  {Ihanec}, {Inserra}, {Lawrence}, {Leloudas}, {Lundqvist}, {M{\"u}ller-Bravo},
  {Piranomonte}, {Pursiainen}, {Rybicki}, {Somero}, {Young}, {Chambers}, {Gao},
  {de Boer}, \& {Magnier}}]{2022MNRAS.517...76O}
{Onori}, F., {Cannizzaro}, G., {Jonker}, P.~G., {et~al.} 2022, \mnras, 517, 76

\bibitem[{{Papaloizou} \& {Stanley}(1986)}]{1986MNRAS.220..593P}
{Papaloizou}, J.~C.~B., \& {Stanley}, G.~Q.~G. 1986, \mnras, 220, 593

\bibitem[{{Parkinson} {et~al.}(2020){Parkinson}, {Knigge}, {Long}, {Matthews},
  {Higginbottom}, {Sim}, \& {Hewitt}}]{2020MNRAS.494.4914P}
{Parkinson}, E.~J., {Knigge}, C., {Long}, K.~S., {et~al.} 2020, \mnras, 494,
  4914

\bibitem[{{Pasham} \& {van Velzen}(2018)}]{2018ApJ...856....1P}
{Pasham}, D.~R., \& {van Velzen}, S. 2018, \apj, 856, 1

\bibitem[{{Pasham} {et~al.}(2022){Pasham}, {Lucchini}, {Laskar}, {Gompertz},
  {Srivastav}, {Nicholl}, {Smartt}, {Miller-Jones}, {Alexander}, {Fender},
  {Smith}, {Fulton}, {Dewangan}, {Gendreau}, {Coughlin}, {Rhodes}, {Horesh},
  {van Velzen}, {Sfaradi}, {Guolo}, {Castro Segura}, {Aamer}, {Anderson},
  {Arcavi}, {Brennan}, {Chambers}, {Charalampopoulos}, {Chen}, {Clocchiatti},
  {de Boer}, {Dennefeld}, {Ferrara}, {Galbany}, {Gao}, {Gillanders}, {Goodwin},
  {Gromadzki}, {Huber}, {Jonker}, {Joshi}, {Kara}, {Killestein}, {Kosec},
  {Kocevski}, {Leloudas}, {Lin}, {Margutti}, {Mattila}, {Moore},
  {M{\"u}ller-Bravo}, {Ngeow}, {Oates}, {Onori}, {Pan}, {Perez-Torres}, {Rani},
  {Remillard}, {Ridley}, {Schulze}, {Sheng}, {Shingles}, {Smith}, {Steiner},
  {Wainscoat}, {Wevers}, \& {Yang}}]{2022NatAs.tmp..252P}
{Pasham}, D.~R., {Lucchini}, M., {Laskar}, T., {et~al.} 2022, Nature Astronomy,
  arXiv:2211.16537

\bibitem[{{Patra} {et~al.}(2022){Patra}, {Lu}, {Brink}, {Yang}, {Filippenko},
  \& {Vasylyev}}]{2022MNRAS.515..138P}
{Patra}, K.~C., {Lu}, W., {Brink}, T.~G., {et~al.} 2022, \mnras, 515, 138

\bibitem[{{Paxton} {et~al.}(2011){Paxton}, {Bildsten}, {Dotter}, {Herwig},
  {Lesaffre}, \& {Timmes}}]{2011ApJS..192....3P}
{Paxton}, B., {Bildsten}, L., {Dotter}, A., {et~al.} 2011, \apjs, 192, 3

\bibitem[{{Payne} {et~al.}(2021){Payne}, {Shappee}, {Hinkle}, {Vallely},
  {Kochanek}, {Holoien}, {Auchettl}, {Stanek}, {Thompson}, {Neustadt},
  {Tucker}, {Armstrong}, {Brimacombe}, {Cacella}, {Cornect}, {Denneau},
  {Fausnaugh}, {Flewelling}, {Grupe}, {Heinze}, {Lopez}, {Monard}, {Prieto},
  {Schneider}, {Sheppard}, {Tonry}, \& {Weiland}}]{Payne:21}
{Payne}, A.~V., {Shappee}, B.~J., {Hinkle}, J.~T., {et~al.} 2021, \apj, 910,
  125

\bibitem[{{Perets} {et~al.}(2016){Perets}, {Li}, {Lombardi}, \&
  {Milcarek}}]{2016ApJ...823..113P}
{Perets}, H.~B., {Li}, Z., {Lombardi}, James~C., J., \& {Milcarek}, Stephen~R.,
  J. 2016, \apj, 823, 113

\bibitem[{{Pfister} {et~al.}(2022){Pfister}, {Toscani}, {Wong}, {Dai},
  {Lodato}, \& {Rossi}}]{2022MNRAS.510.2025P}
{Pfister}, H., {Toscani}, M., {Wong}, T. H.~T., {et~al.} 2022, \mnras, 510,
  2025

\bibitem[{{Phinney}(1989)}]{1989IAUS..136..543P}
{Phinney}, E.~S. 1989, in The Center of the Galaxy, ed. M.~{Morris}, Vol. 136,
  543

\bibitem[{{Piran} {et~al.}(2015){Piran}, {Svirski}, {Krolik}, {Cheng}, \&
  {Shiokawa}}]{2015ApJ...806..164P}
{Piran}, T., {Svirski}, G., {Krolik}, J., {Cheng}, R.~M., \& {Shiokawa}, H.
  2015, \apj, 806, 164

\bibitem[{{Press} \& {Teukolsky}(1977)}]{1977ApJ...213..183P}
{Press}, W.~H., \& {Teukolsky}, S.~A. 1977, \apj, 213, 183

\bibitem[{{Rauch}(1995)}]{1995MNRAS.275..628R}
{Rauch}, K.~P. 1995, \mnras, 275, 628

\bibitem[{{Rees}(1988)}]{1988Natur.333..523R}
{Rees}, M.~J. 1988, \nat, 333, 523

\bibitem[{{Reines} \& {Volonteri}(2015)}]{2015ApJ...813...82R}
{Reines}, A.~E., \& {Volonteri}, M. 2015, \apj, 813, 82

\bibitem[{{Remillard} \& {McClintock}(2006)}]{2006ARA&A..44...49R}
{Remillard}, R.~A., \& {McClintock}, J.~E. 2006, \araa, 44, 49

\bibitem[{{Reusch} {et~al.}(2022){Reusch}, {Stein}, {Kowalski}, {van Velzen},
  {Franckowiak}, {Lunardini}, {Murase}, {Winter}, {Miller-Jones}, {Kasliwal},
  {Gilfanov}, {Garrappa}, {Paliya}, {Ahumada}, {Anand}, {Barbarino}, {Bellm},
  {Brinnel}, {Buson}, {Cenko}, {Coughlin}, {De}, {Dekany}, {Frederick},
  {Gal-Yam}, {Gezari}, {Giroletti}, {Graham}, {Karambelkar}, {Kimura}, {Kong},
  {Kool}, {Laher}, {Medvedev}, {Necker}, {Nordin}, {Perley}, {Rigault},
  {Rusholme}, {Schulze}, {Schweyer}, {Singer}, {Sollerman}, {Strotjohann},
  {Sunyaev}, {van Santen}, {Walters}, {Zhang}, \&
  {Zimmerman}}]{2022PhRvL.128v1101R}
{Reusch}, S., {Stein}, R., {Kowalski}, M., {et~al.} 2022, \prl, 128, 221101

\bibitem[{{Rodriguez} {et~al.}(2016){Rodriguez}, {Chatterjee}, \&
  {Rasio}}]{2016PhRvD..93h4029R}
{Rodriguez}, C.~L., {Chatterjee}, S., \& {Rasio}, F.~A. 2016, \prd, 93, 084029

\bibitem[{{Rossi} {et~al.}(2020){Rossi}, {Stone}, {Law-Smith}, {MacLeod},
  {Lodato}, {Dai}, \& {Mandel}}]{2020arXiv200512528R}
{Rossi}, E.~M., {Stone}, N.~C., {Law-Smith}, J. A.~P., {et~al.} 2020, arXiv
  e-prints, arXiv:2005.12528

\bibitem[{{Ryu} {et~al.}(2020{\natexlab{a}}){Ryu}, {Krolik}, \&
  {Piran}}]{2020ApJ...904...73R}
{Ryu}, T., {Krolik}, J., \& {Piran}, T. 2020{\natexlab{a}}, \apj, 904, 73

\bibitem[{{Ryu} {et~al.}(2020{\natexlab{b}}){Ryu}, {Krolik}, {Piran}, \&
  {Noble}}]{2020ApJ...904...98R}
{Ryu}, T., {Krolik}, J., {Piran}, T., \& {Noble}, S.~C. 2020{\natexlab{b}},
  \apj, 904, 98

\bibitem[{{Ryu} {et~al.}(2020{\natexlab{c}}){Ryu}, {Krolik}, {Piran}, \&
  {Noble}}]{2020ApJ...904...99R}
---. 2020{\natexlab{c}}, \apj, 904, 99

\bibitem[{{Ryu} {et~al.}(2020{\natexlab{d}}){Ryu}, {Krolik}, {Piran}, \&
  {Noble}}]{2020ApJ...904..100R}
---. 2020{\natexlab{d}}, \apj, 904, 100

\bibitem[{{Ryu} {et~al.}(2020{\natexlab{e}}){Ryu}, {Krolik}, {Piran}, \&
  {Noble}}]{2020ApJ...904..101R}
---. 2020{\natexlab{e}}, \apj, 904, 101

\bibitem[{{Ryu} {et~al.}(2023{\natexlab{a}}){Ryu}, {Perna}, {Pakmor}, {Ma},
  {Farmer}, \& {de Mink}}]{2023MNRAS.519.5787R}
{Ryu}, T., {Perna}, R., {Pakmor}, R., {et~al.} 2023{\natexlab{a}}, \mnras, 519,
  5787

\bibitem[{{Ryu} {et~al.}(2023{\natexlab{b}}){Ryu}, {Perna}, {Pakmor}, {Ma},
  {Farmer}, \& {de Mink}}]{2023MNRAS.tmp...76R}
---. 2023{\natexlab{b}}, \mnras, arXiv:2211.02734

\bibitem[{{Ryu} {et~al.}(2022){Ryu}, {Perna}, \& {Wang}}]{2022MNRAS.516.2204R}
{Ryu}, T., {Perna}, R., \& {Wang}, Y.-H. 2022, \mnras, 516, 2204

\bibitem[{{Ryu} {et~al.}(2023{\natexlab{c}}){Ryu}, {Valli}, {Pakmor}, {Perna},
  {de Mink}, \& {Springel}}]{2023MNRAS.tmp.1885R}
{Ryu}, T., {Valli}, R., {Pakmor}, R., {et~al.} 2023{\natexlab{c}}, \mnras,
  arXiv:2304.01792

\bibitem[{{Ryu} {et~al.}(2023{\natexlab{d}}){Ryu}, {Valli}, {Pakmor}, {Perna},
  {de Mink}, \& {Springel}}]{2023arXiv230703097R}
---. 2023{\natexlab{d}}, \mnras, 525, 5752

\bibitem[{{Samsing} {et~al.}(2019){Samsing}, {Hamers}, \&
  {Tyles}}]{2019PhRvD.100d3010S}
{Samsing}, J., {Hamers}, A.~S., \& {Tyles}, J.~G. 2019, \prd, 100, 043010

\bibitem[{{Saxton} {et~al.}(2021){Saxton}, {Komossa}, {Auchettl}, \&
  {Jonker}}]{2021SSRv..217...18S}
{Saxton}, R., {Komossa}, S., {Auchettl}, K., \& {Jonker}, P.~G. 2021, \ssr,
  217, 18

\bibitem[{{Saxton} {et~al.}(2008){Saxton}, {Read}, {Esquej}, {Freyberg},
  {Altieri}, \& {Bermejo}}]{2008A&A...480..611S}
{Saxton}, R.~D., {Read}, A.~M., {Esquej}, P., {et~al.} 2008, \aap, 480, 611

\bibitem[{{Saxton} {et~al.}(2012){Saxton}, {Read}, {Esquej}, {Komossa},
  {Dougherty}, {Rodriguez-Pascual}, \& {Barrado}}]{2012A&A...541A.106S}
---. 2012, \aap, 541, A106

\bibitem[{{Sazonov} {et~al.}(2021){Sazonov}, {Gilfanov}, {Medvedev}, {Yao},
  {Khorunzhev}, {Semena}, {Sunyaev}, {Burenin}, {Lyapin}, {Meshcheryakov},
  {Uskov}, {Zaznobin}, {Postnov}, {Dodin}, {Belinski}, {Cherepashchuk},
  {Eselevich}, {Dodonov}, {Grokhovskaya}, {Kotov}, {Bikmaev}, {Zhuchkov},
  {Gumerov}, {van Velzen}, \& {Kulkarni}}]{2021MNRAS.508.3820S}
{Sazonov}, S., {Gilfanov}, M., {Medvedev}, P., {et~al.} 2021, \mnras, 508, 3820

\bibitem[{{Senno} {et~al.}(2017){Senno}, {Murase}, \&
  {M{\'e}sz{\'a}ros}}]{2017ApJ...838....3S}
{Senno}, N., {Murase}, K., \& {M{\'e}sz{\'a}ros}, P. 2017, \apj, 838, 3

\bibitem[{{Seto} \& {Kyutoku}(2017)}]{2017PhRvL.118o1101S}
{Seto}, N., \& {Kyutoku}, K. 2017, \prl, 118, 151101

\bibitem[{{Shappee} {et~al.}(2014){Shappee}, {Prieto}, {Grupe}, {Kochanek},
  {Stanek}, {De Rosa}, {Mathur}, {Zu}, {Peterson}, {Pogge}, {Komossa}, {Im},
  {Jencson}, {Holoien}, {Basu}, {Beacom}, {Szczygie{\l}}, {Brimacombe},
  {Adams}, {Campillay}, {Choi}, {Contreras}, {Dietrich}, {Dubberley},
  {Elphick}, {Foale}, {Giustini}, {Gonzalez}, {Hawkins}, {Howell}, {Hsiao},
  {Koss}, {Leighly}, {Morrell}, {Mudd}, {Mullins}, {Nugent}, {Parrent},
  {Phillips}, {Pojmanski}, {Rosing}, {Ross}, {Sand}, {Terndrup}, {Valenti},
  {Walker}, \& {Yoon}}]{2014ApJ...788...48S}
{Shappee}, B.~J., {Prieto}, J.~L., {Grupe}, D., {et~al.} 2014, \apj, 788, 48

\bibitem[{{Shiokawa} {et~al.}(2015){Shiokawa}, {Krolik}, {Cheng}, {Piran}, \&
  {Noble}}]{2015ApJ...804...85S}
{Shiokawa}, H., {Krolik}, J.~H., {Cheng}, R.~M., {Piran}, T., \& {Noble}, S.~C.
  2015, \apj, 804, 85

\bibitem[{{Short} {et~al.}(2020){Short}, {Nicholl}, {Lawrence}, {Gomez},
  {Arcavi}, {Wevers}, {Leloudas}, {Schulze}, {Anderson}, {Berger}, {Blanchard},
  {Burke}, {Segura}, {Charalampopoulos}, {Chornock}, {Galbany}, {Gromadzki},
  {Herzog}, {Hiramatsu}, {Horne}, {Hosseinzadeh}, {Howell}, {Ihanec},
  {Inserra}, {Kankare}, {Maguire}, {McCully}, {M{\"u}ller Bravo}, {Onori},
  {Sollerman}, \& {Young}}]{2020MNRAS.498.4119S}
{Short}, P., {Nicholl}, M., {Lawrence}, A., {et~al.} 2020, \mnras, 498, 4119

\bibitem[{{Sigurdsson} \& {Rees}(1997)}]{1997MNRAS.284..318S}
{Sigurdsson}, S., \& {Rees}, M.~J. 1997, \mnras, 284, 318

\bibitem[{{Soszy{\'n}ski} {et~al.}(2004){Soszy{\'n}ski}, {Udalski}, {Kubiak},
  {Szyma{\'n}ski}, {Pietrzy{\'n}ski}, {{\.Z}ebru{\'n}}, {Szewczyk}, \&
  {Wyrzykowski}}]{2004AcA....54..129S}
{Soszy{\'n}ski}, I., {Udalski}, A., {Kubiak}, M., {et~al.} 2004, \actaa, 54,
  129

\bibitem[{{Stein} {et~al.}(2021){Stein}, {Velzen}, {Kowalski}, {Franckowiak},
  {Gezari}, {Miller-Jones}, {Frederick}, {Sfaradi}, {Bietenholz}, {Horesh},
  {Fender}, {Garrappa}, {Ahumada}, {Andreoni}, {Belicki}, {Bellm},
  {B{\"o}ttcher}, {Brinnel}, {Burruss}, {Cenko}, {Coughlin}, {Cunningham},
  {Drake}, {Farrar}, {Feeney}, {Foley}, {Gal-Yam}, {Golkhou}, {Goobar},
  {Graham}, {Hammerstein}, {Helou}, {Hung}, {Kasliwal}, {Kilpatrick}, {Kong},
  {Kupfer}, {Laher}, {Mahabal}, {Masci}, {Necker}, {Nordin}, {Perley},
  {Rigault}, {Reusch}, {Rodriguez}, {Rojas-Bravo}, {Rusholme}, {Shupe},
  {Singer}, {Sollerman}, {Soumagnac}, {Stern}, {Taggart}, {van Santen}, {Ward},
  {Woudt}, \& {Yao}}]{2021NatAs...5..510S}
{Stein}, R., {Velzen}, S.~v., {Kowalski}, M., {et~al.} 2021, Nature Astronomy,
  5, 510

\bibitem[{{Stone} {et~al.}(2013){Stone}, {Sari}, \&
  {Loeb}}]{2013MNRAS.435.1809S}
{Stone}, N., {Sari}, R., \& {Loeb}, A. 2013, \mnras, 435, 1809

\bibitem[{{Stone} \& {Metzger}(2016)}]{2016MNRAS.455..859S}
{Stone}, N.~C., \& {Metzger}, B.~D. 2016, \mnras, 455, 859

\bibitem[{{Svirski} {et~al.}(2017){Svirski}, {Piran}, \&
  {Krolik}}]{2017MNRAS.467.1426S}
{Svirski}, G., {Piran}, T., \& {Krolik}, J. 2017, \mnras, 467, 1426

\bibitem[{{Thorne}(1998)}]{1998bhrs.conf...41T}
{Thorne}, K.~S. 1998, in Black Holes and Relativistic Stars, ed. R.~M. {Wald},
  41

\bibitem[{{Toscani} {et~al.}(2019){Toscani}, {Lodato}, \&
  {Nealon}}]{2019MNRAS.489..699T}
{Toscani}, M., {Lodato}, G., \& {Nealon}, R. 2019, \mnras, 489, 699

\bibitem[{{Toscani} {et~al.}(2020){Toscani}, {Rossi}, \&
  {Lodato}}]{2020MNRAS.498..507T}
{Toscani}, M., {Rossi}, E.~M., \& {Lodato}, G. 2020, \mnras, 498, 507

\bibitem[{{van Velzen}(2018)}]{2018ApJ...852...72V}
{van Velzen}, S. 2018, \apj, 852, 72

\bibitem[{{van Velzen} \& {Farrar}(2014)}]{2014ApJ...792...53V}
{van Velzen}, S., \& {Farrar}, G.~R. 2014, \apj, 792, 53

\bibitem[{{van Velzen} {et~al.}(2013){van Velzen}, {Frail}, {K{\"o}rding}, \&
  {Falcke}}]{2013A&A...552A...5V}
{van Velzen}, S., {Frail}, D.~A., {K{\"o}rding}, E., \& {Falcke}, H. 2013,
  \aap, 552, A5

\bibitem[{{van Velzen} {et~al.}(2020){van Velzen}, {Holoien}, {Onori}, {Hung},
  \& {Arcavi}}]{2020SSRv..216..124V}
{van Velzen}, S., {Holoien}, T. W.~S., {Onori}, F., {Hung}, T., \& {Arcavi}, I.
  2020, \ssr, 216, 124

\bibitem[{{van Velzen} {et~al.}(2016){van Velzen}, {Mendez}, {Krolik}, \&
  {Gorjian}}]{2016ApJ...829...19V}
{van Velzen}, S., {Mendez}, A.~J., {Krolik}, J.~H., \& {Gorjian}, V. 2016,
  \apj, 829, 19

\bibitem[{{van Velzen} {et~al.}(2019){van Velzen}, {Stone}, {Metzger},
  {Gezari}, {Brown}, \& {Fruchter}}]{2019ApJ...878...82V}
{van Velzen}, S., {Stone}, N.~C., {Metzger}, B.~D., {et~al.} 2019, \apj, 878,
  82

\bibitem[{{van Velzen} {et~al.}(2011){van Velzen}, {Farrar}, {Gezari},
  {Morrell}, {Zaritsky}, {{\"O}stman}, {Smith}, {Gelfand}, \&
  {Drake}}]{2011ApJ...741...73V}
{van Velzen}, S., {Farrar}, G.~R., {Gezari}, S., {et~al.} 2011, \apj, 741, 73

\bibitem[{{van Velzen} {et~al.}(2021{\natexlab{a}}){van Velzen}, {Stein},
  {Gilfanov}, {Kowalski}, {Hayasaki}, {Reusch}, {Yao}, {Garrappa},
  {Franckowiak}, {Gezari}, {Nordin}, {Fremling}, {Sharma}, {Yan}, {Kool},
  {Sollerman}, {Medvedev}, {Sunyaev}, {Bellm}, {Dekany}, {Duev}, {Graham},
  {Kasliwal}, {Laher}, {Riddle}, \& {Rusholme}}]{2021arXiv211109391V}
{van Velzen}, S., {Stein}, R., {Gilfanov}, M., {et~al.} 2021{\natexlab{a}},
  arXiv e-prints, arXiv:2111.09391

\bibitem[{{van Velzen} {et~al.}(2021{\natexlab{b}}){van Velzen}, {Gezari},
  {Hammerstein}, {Roth}, {Frederick}, {Ward}, {Hung}, {Cenko}, {Stein},
  {Perley}, {Taggart}, {Foley}, {Sollerman}, {Blagorodnova}, {Andreoni},
  {Bellm}, {Brinnel}, {De}, {Dekany}, {Feeney}, {Fremling}, {Giomi}, {Golkhou},
  {Graham}, {Ho}, {Kasliwal}, {Kilpatrick}, {Kulkarni}, {Kupfer}, {Laher},
  {Mahabal}, {Masci}, {Miller}, {Nordin}, {Riddle}, {Rusholme}, {van Santen},
  {Sharma}, {Shupe}, \& {Soumagnac}}]{2021ApJ...908....4V}
{van Velzen}, S., {Gezari}, S., {Hammerstein}, E., {et~al.} 2021{\natexlab{b}},
  \apj, 908, 4

\bibitem[{{Wang} \& {Merritt}(2004)}]{2004ApJ...600..149W}
{Wang}, J., \& {Merritt}, D. 2004, \apj, 600, 149

\bibitem[{{Wang} {et~al.}(2012){Wang}, {Zhou}, {Komossa}, {Wang}, {Yuan}, \&
  {Yang}}]{2012ApJ...749..115W}
{Wang}, T.-G., {Zhou}, H.-Y., {Komossa}, S., {et~al.} 2012, \apj, 749, 115

\bibitem[{{Wang} {et~al.}(2011{\natexlab{a}}){Wang}, {Zhou}, {Wang}, {Lu}, \&
  {Xu}}]{2011ApJ...740...85W}
{Wang}, T.-G., {Zhou}, H.-Y., {Wang}, L.-F., {Lu}, H.-L., \& {Xu}, D.
  2011{\natexlab{a}}, \apj, 740, 85

\bibitem[{{Wang} \& {Liu}(2016)}]{2016PhRvD..93h3005W}
{Wang}, X.-Y., \& {Liu}, R.-Y. 2016, \prd, 93, 083005

\bibitem[{{Wang} {et~al.}(2011{\natexlab{b}}){Wang}, {Liu}, {Dai}, \&
  {Cheng}}]{2011PhRvD..84h1301W}
{Wang}, X.-Y., {Liu}, R.-Y., {Dai}, Z.-G., \& {Cheng}, K.~S.
  2011{\natexlab{b}}, \prd, 84, 081301

\bibitem[{{Weatherford} {et~al.}(2020){Weatherford}, {Chatterjee}, {Kremer}, \&
  {Rasio}}]{2020ApJ...898..162W}
{Weatherford}, N.~C., {Chatterjee}, S., {Kremer}, K., \& {Rasio}, F.~A. 2020,
  \apj, 898, 162

\bibitem[{{Weinberg}(1972)}]{1972gcpa.book.....W}
{Weinberg}, S. 1972, {Gravitation and Cosmology: Principles and Applications of
  the General Theory of Relativity}

\bibitem[{{Weisskopf} {et~al.}(2000){Weisskopf}, {Tananbaum}, {Van Speybroeck},
  \& {O'Dell}}]{2000SPIE.4012....2W}
{Weisskopf}, M.~C., {Tananbaum}, H.~D., {Van Speybroeck}, L.~P., \& {O'Dell},
  S.~L. 2000, in Society of Photo-Optical Instrumentation Engineers (SPIE)
  Conference Series, Vol. 4012, X-Ray Optics, Instruments, and Missions III,
  ed. J.~E. {Truemper} \& B.~{Aschenbach}, 2--16

\bibitem[{{Wevers}(2020)}]{2020MNRAS.497L...1W}
{Wevers}, T. 2020, \mnras, 497, L1

\bibitem[{{Wevers} {et~al.}(2019){Wevers}, {Pasham}, {van Velzen}, {Leloudas},
  {Schulze}, {Miller-Jones}, {Jonker}, {Gromadzki}, {Kankare}, {Hodgkin},
  {Wyrzykowski}, {Kostrzewa-Rutkowska}, {Moran}, {Berton}, {Maguire}, {Onori},
  {Mattila}, \& {Nicholl}}]{2019MNRAS.488.4816W}
{Wevers}, T., {Pasham}, D.~R., {van Velzen}, S., {et~al.} 2019, \mnras, 488,
  4816

\bibitem[{{Wevers} {et~al.}(2021){Wevers}, {Pasham}, {van Velzen},
  {Miller-Jones}, {Uttley}, {Gendreau}, {Remillard}, {Arzoumanian},
  {L{\"o}wenstein}, \& {Chiti}}]{2021ApJ...912..151W}
---. 2021, \apj, 912, 151

\bibitem[{{Wevers} {et~al.}(2022){Wevers}, {Nicholl}, {Guolo},
  {Charalampopoulos}, {Gromadzki}, {Reynolds}, {Kankare}, {Leloudas},
  {Anderson}, {Arcavi}, {Cannizzaro}, {Chen}, {Ihanec}, {Inserra},
  {Guti{\'e}rrez}, {Jonker}, {Lawrence}, {Magee}, {M{\"u}ller-Bravo}, {Onori},
  {Ridley}, {Schulze}, {Short}, {Hiramatsu}, {Newsome}, {Terwel}, {Yang}, \&
  {Young}}]{2022A&A...666A...6W}
{Wevers}, T., {Nicholl}, M., {Guolo}, M., {et~al.} 2022, \aap, 666, A6

\bibitem[{{Wevers} {et~al.}(2023){Wevers}, {Coughlin}, {Pasham}, {Guolo},
  {Sun}, {Wen}, {Jonker}, {Zabludoff}, {Malyali}, {Arcodia}, {Liu}, {Merloni},
  {Rau}, {Grotova}, {Short}, \& {Cao}}]{Wevers:23}
{Wevers}, T., {Coughlin}, E.~R., {Pasham}, D.~R., {et~al.} 2023, \apjl, 942,
  L33

\bibitem[{{Winter} \& {Lunardini}(2021)}]{2021NatAs...5..472W}
{Winter}, W., \& {Lunardini}, C. 2021, Nature Astronomy, 5, 472

\bibitem[{{Yang} {et~al.}(2021){Yang}, {Bartos}, {Fragione}, {Haiman},
  {Kowalski}, {Marka}, {Perna}, \& {Tagawa}}]{2021arXiv210502342Y}
{Yang}, Y., {Bartos}, I., {Fragione}, G., {et~al.} 2021, arXiv e-prints,
  arXiv:2105.02342

\bibitem[{{Zauderer} {et~al.}(2013){Zauderer}, {Berger}, {Margutti}, {Pooley},
  {Sari}, {Soderberg}, {Brunthaler}, \& {Bietenholz}}]{2013ApJ...767..152Z}
{Zauderer}, B.~A., {Berger}, E., {Margutti}, R., {et~al.} 2013, \apj, 767, 152

\bibitem[{{Zauderer} {et~al.}(2011){Zauderer}, {Berger}, {Soderberg}, {Loeb},
  {Narayan}, {Frail}, {Petitpas}, {Brunthaler}, {Chornock}, {Carpenter},
  {Pooley}, {Mooley}, {Kulkarni}, {Margutti}, {Fox}, {Nakar}, {Patel},
  {Volgenau}, {Culverhouse}, {Bietenholz}, {Rupen}, {Max-Moerbeck}, {Readhead},
  {Richards}, {Shepherd}, {Storm}, \& {Hull}}]{2011Natur.476..425Z}
{Zauderer}, B.~A., {Berger}, E., {Soderberg}, A.~M., {et~al.} 2011, \nat, 476,
  425

\bibitem[{{Zhang} {et~al.}(2017){Zhang}, {Murase}, {Oikonomou}, \&
  {Li}}]{2017PhRvD..96f3007Z}
{Zhang}, B.~T., {Murase}, K., {Oikonomou}, F., \& {Li}, Z. 2017, \prd, 96,
  063007

\bibitem[{{Zhong} {et~al.}(2022){Zhong}, {Li}, {Berczik}, \&
  {Spurzem}}]{2022ApJ...933...96Z}
{Zhong}, S., {Li}, S., {Berczik}, P., \& {Spurzem}, R. 2022, \apj, 933, 96

\bibitem[{{Zrake} {et~al.}(2021){Zrake}, {Tiede}, {MacFadyen}, \&
  {Haiman}}]{2021ApJ...909L..13Z}
{Zrake}, J., {Tiede}, C., {MacFadyen}, A., \& {Haiman}, Z. 2021, \apjl, 909,
  L13

\end{thebibliography}
\end{document}